\documentclass[aps,showpacs,twocolumn,amsmath,amssymb]{revtex4-1}
\usepackage{graphicx}
\usepackage{epstopdf}
\usepackage{bm}% bold math
\usepackage{subfigure}
\usepackage{sidecap}

\usepackage{graphicx}
\usepackage{epstopdf}
\usepackage{amssymb}
\usepackage{amsmath}

\usepackage{bm}% bold math
\graphicspath{{Figures/}}
\usepackage{subfigure}
\usepackage{sidecap}
\usepackage {times}
\usepackage{color}

\newcommand\redout{\bgroup\markoverwith{\textcolor{red}{\rule[.5ex]{2pt}{0.4pt}}}\ULon}

\usepackage[colorlinks=true,citecolor=blue]{hyperref}
\hypersetup{colorlinks=true,citecolor=blue,linkcolor=red,urlcolor=blue}
\newcommand{\bsig}{{\bm\sigma}}
\newcommand{\bkap}{{\bm\kappa}}
\newcommand{\balp}{{\bm\alpha}}

\newcommand{\be}{\begin{equation}}
\newcommand{\ee}{\end{equation}}

\newcommand{\bea}{\begin{eqnarray}}
\newcommand{\eea}{\end{eqnarray}}
\newcommand{\bd}{\begin{displaymath}}
\newcommand{\ed}{\end{displaymath}}
\newcommand{\ba}{\begin{array}}
\newcommand{\ea}{\end{array}}
\newcommand{\bi}{\begin{itemize}}
\newcommand{\ei}{\end{itemize}}
\newcommand{\bc}{\begin{center}}
\newcommand{\ec}{\end{center}}
\newcommand{\bfl}{\begin{flushleft}}
\newcommand{\efl}{\end{flushleft}}
\newcommand{\bfr}{\begin{flushright}}
\newcommand{\efr}{\end{flushright}}
\newcommand{\no}{\nonumber}

\newcommand{\bl}{\begin{aligned}}
\newcommand{\el}{\end{aligned}}

\newcommand{\hatt}{\hat{t}}

\newcommand{\hG}{\hat{G}}

\newcommand{\hV}{\hat{V}}

\newcommand{\hE}{\hat{E}}

\newcommand{\hk}{\hat{k}}
\newcommand{\hbk}{\hat{{\bf k}}}
\newcommand{\hbq}{\hat{{\bf q}}}

\newcommand{\hbz}{\hat{{\bf z}}}

\newcommand{\tA}{\tilde{A}}
\newcommand{\tB}{\tilde{B}}

\newcommand{\tE}{\tilde{E}}
\newcommand{\tG}{\tilde{G}}

\newcommand{\tg}{\tilde{g}}
\newcommand{\tilh}{\tilde{h}}
\newcommand{\tilH}{\tilde{H}}
\newcommand{\tilt}{\tilde{t}}

\newcommand{\tps}{\tilde{\psi}}

\newcommand{\fs}{\frac{1}{2}}

\newcommand{\fphik}{\frac{\phi_{{\bf k}}}{2}}

\newcommand{\om}{i\omega_n}

\newcommand{\cph}{\cos\frac{\phi_\bk}{2}}
\newcommand{\cpr}{\cos\frac{\phi_{\bk'}}{2}}
\newcommand{\sph}{\sin\frac{\phi_\bk}{2}}
\newcommand{\spr}{\sin\frac{\phi_{\bk'}}{2}}

\newcommand{\bse}{Bi$_2$Se$_3$}
\newcommand{\bte}{Bi$_2$Te$_3$}

\def\dg{^{\dagger}}

\def\bk{{\bf k}}  \def\bq{{\bf q}} 
 
\def\bp{{\bf p}} \def\hg{\rm\hat{g}}

 \def\bd{{\bf d}}  
  \def\hbz{\hat{{\bf z}}}

\def\da{\downarrow} \def\ua{\uparrow}
 
\def\dg{\dagger}

\def\bra{\langle}
\def\ket{\rangle}
\begin{document}
\date{\today}
\title{Gapped Dirac cones and spin texture in thin film topological insulator}
\author{Peter Thalmeier$^1$}
\author{Alireza Akbari$^{1,2,3}$}
\email{akbari@postech.ac.kr}
\affiliation{$^1$Max Planck Institute for the  Chemical Physics of Solids, D-01187 Dresden, Germany}
\affiliation{$^2$Max Planck POSTECH Center for Complex Phase Materials, POSTECH, Pohang 790-784, Korea} 
\affiliation{$^3$Department of Physics, POSTECH, Pohang, Gyeongbuk 790-784, Korea}
\begin{abstract}
The protected surface states of topological insulators (TIs) form gapless Dirac cones corresponding non-degenerate eigenstates with helical spin polarisation. The presence of a warping term deforms the isotropic cone of the most simple model into snowflake Fermi surfaces as in \bse~and \bte. Their features have been identified in STM quasiparticle interference (QPI) experiments on isolated surfaces. Here we investigate the QPI spectrum for the TI thin film geometry with a finite tunnelling between the surface states. This leads to a dramatic change of spectrum due to gapping and a change in spin texture that should leave distinct signatures in the QPI pattern. We consider both normal and magnetic exchange scattering from the surface impurities and obtain the scattering t-matrix in Born approximation as well as the general closed solution. We show the expected systematic variation of QPI snowflake intensity features by varying film thickness and study in particular the influence on back scattering processes. We predict the variation of QPI spectrum for \bse~thin films using the observed gap dependence from ARPES results.
\end{abstract}

\maketitle

\section{Introduction}

The most dramatic manifestation in a topological insulator (TI) is the presence of protected surface states which are helical spin locked nondegenerate eigenstates with a gapless Dirac dispersion. This has been manifestly demonstrated e.g. by countless (spin-polarised) ARPES experiments~\cite{hsieh:09,kuroda:10,hoefer:14} and STM-QPI~\cite{roushan:09,zhang:09,gomes:09,alpichshev:10, okada:11,cheng:12,kohsaka:17} investigations on isolated surfaces, in particular in the canonical examples \bse~and \bte. The theory of QPI spectra on single surfaces has been presented in numerous effective model investigations~\cite{lee:09,zhou:09,guo:10,Liu:2012aa,kohsaka:17} and recently also with an ab-initio approach~\cite{ruessmann:20}. Due to their helical nature these surface states have forbidden backsckattering by normal (scalar) impurities leading to the well-know weak anti-localization (WAL) effects  in the surface  magnetoconductance experiments~\cite{taskin:12,lu:11}. However the strict WAL due to the destructive interference caused by a Berry phase $\mp\pi$  applies only to isolated surfaces. When we consider thin films there will be a tunnelling between opposite surface states due to their finite decay lenghts into the bulk. The tunnelling  mixes the surfaces states leading to a finite gap at the Dirac point and it lifts their protection as witnessed by the suppression of helical polarisation and modification of the Berry phase for states close to the gap energy~\cite{lu:11b}. This has been directly observed in magnetotransport experiments in \bse~where the weak anti-localization as indicator of the ungapped states rather suddenly breaks down below a thickness of five quintuple layers (QL)~\cite{taskin:12} (In \bse~1QL =c/3 =9.55\AA; c= lattice constant). The gapping of the Dirac cones in thin films has also been observed directly with the ARPES technique\cite{zhang:10,neupane:14}.\\

Here we want to investigate theoretically what kind of influence the inter-surface tunnelling in TI thin-films may have on the QPI pattern in scanning STM experiments. In the known QPI experiments on single surfaces the Fourier transformed scanning image is dominated by the scattering vectors that interconnect the characteristic points of the real `snowflake' type Fermi surfaces, with those of the backscattering characteristically missing. Therefore our investigation demands that we take into account not only the inter-surface tunnelling but also includes the higher order intra-surface warping terms that deform the isotropic Dirac cones with Fermi surface (FS) circles into the observed snowflake geometry. The QPI pattern of the isolated surfaces should be dramatically modified for small bias voltages such that $\omega = { eV} \approx \pm |t|$ where $V$ is the STM bias voltage and t is the inter-surface tunnelling leading to a gap energy $2|t|$. Furthermore the QPI spectrum due to scattering from magnetic impurities should also be sensitive to the change of spin texture close to the gap threshold 
which is induced by the tunnelling consisting primarily in a suppression of helical polarisation for small energies.
Most importantly  the intensity variation of QPI peaks on the constant frequency cuts should carry an imprint of the reemergence of backscattering  for the topologically trivial surface states close to the hybridisation gap edge.  In this work we intend to give a detailed theoretical analysis of the tunnelling effects in TI thin films that one may expect to see as a guidance to QPI experiments with systematic variation of film thickness. We also present the expected variation of the QPI spectrum using the thickness dependent surface state gap obtained from ARPES.  

These issues are central to understand properties of thin film topological  surface states. Our theoretical approach to them is entirely analytical t-matrix theory  \cite{capriotti:03,balatsky:06} because we want to study in detail the combined effect of in-plane warping and inter-surface hybridisation on QPI images. This allows a transparent interpretation of their  spin texture, Berry phase and gradual change of backscattering characteristics with inter-surface tunnelling which are at the heart of the TI thin film problem.
Naturally this demands a restriction to indispensable ingredients of the TI thin film model, e.g. to only four basis states and a momentum independent  
intra-layer impurity scattering. Otherwise a fully numerical approach would be necessary which is not the aim of this work.
We will derive the QPI spectrum in Born approximation as well as in closed fully analytical t-matrix form, using both normal and magnetic example of (momentum-isotropic) impurity scattering.

The paper is organized as follows: In Secs.~\ref{sec:model}, \ref{sec:tunnelling} we define the model of warped and inter-surface coupled thin film TI states and discuss their spin texture in Sec.~\ref{sec:spintext}. In the main formal part of Sec.~\ref{sec:tmat} we derive the full t-matrix for the assumed impurity scattering potential and discuss the simplified Born approximation. These results are then applied to calculate the QPI spectrum in Sec.~\ref{sec:QPI} in the normal and magnetic scattering cases with full t-matrix approach. An extended discussion of the numerical results as function of bias voltage and film thickness is presented in Sec.~\ref{sec:numerical}. Finally Sec.~\ref{sec:conc} gives the conclusions.

\section{Model for warped topological insulator Dirac cones}
\label{sec:model}
The protected surface states in topological insulators like e.g. \bse~and \bte~ may be derived from a $\bk\cdot\bp$-type Hamiltonian model for the 3D bulk states~\cite{lu:10,asmar:18} using four strongly spin-orbit coupled basis states~with angular momentum (pseudospin) components $m_j=\pm \fs$ and parity $P=\pm$. Because the crystal-field splitting is much larger than the spin-orbit coupling these states are dominated by $p_z$ orbitals from Se(4p) and Bi(6p) and therefore the $m_j$ pseudospin  is proportional to the real electron spin $\sigma_z$ \cite{zhang:09a,zhang:12} which will be used for simplicity in the following. The surface states are obtained by solving the 1D Dirac equation along the surface normal direction ($\hbz$) under appropriate boundary conditions~\cite{lu:10,asmar:18}. The effective low energy Hamiltonian for the 2D topological surface states  can then be parametrized in the well-known form~\cite{fu:09,lee:09,thalmeier:11,wang:13}
%
%---------------------------------------------------------------------------------------------------
\be
{\cal H}=\sum_\bk h_\bk =\sum_\bk\bigl[v(\bk\times\bsig)\cdot\hbz+\lambda k^3\cos 3\theta_\bk\sigma_z\bigr].
\label{eqn:hamsig}
\ee
%---------------------------------------------------------------------------------------------------
%
Here the first term describes the massless Dirac states forming an isotropic cone dispersion in the 2D surface Brillouin zone (SBZ) where the velocity $v$ is the slope of the cone and its position is the SBZ projection of a bulk time reversal invariant (TRI) point of the bulk Brillouin zone, e.g., the $\Gamma$ point for the above compounds. The second `warping term' distorts the cone anisotropically in accordance with crystal symmetry in such a way that the 2D Fermi surfaces (cuts through the cone)  evolve from circular shapes to hexagons to  six-pronged  `snowflakes' \cite{fu:09,bansil:16} when the Fermi energy increases. The amount of distortion is determined by the strength $\lambda$ of the warping term. The spin degeneracy of states at the projected TRI points is lifted due to spin orbit coupling when moving away from them. Furthermore the surface wave vector $\bk =(k_x,k_y) =k\hbk =k(\cos\theta_\bk,\sin\theta_\bk) $ has the polar presentation with magnitude  $k=(k_x^2+k_y^2)^\fs$ and the azimuthal angle $\theta_\bk=\tan^{-1}(k_y/k_x)$. The above $2\times 2$ Hamiltionian can be explicitly written as 
%
%---------------------------------------------------------------------------------------------------
\be
\bl
h_{\bk}&=\epsilon_\bk
\left(
 \begin{array}{cc}
\hk^2\cos 3\theta_\bk& -ie^{-i\theta_\bk}
\\
ie^{i\theta_\bk} & -\hk^2\cos 3\theta_\bk
\end{array}
\right),
\label{eqn:hamat}
\el
\ee
%---------------------------------------------------------------------------------------------------
%
where $\epsilon_\bk=E^*\hk=vk$ is the isotropic $(\lambda =0)$ Dirac dispersion. An overall parabolic term $\epsilon_{0\bk}=\bk^2/2m^*$ that breaks particle hole-symmetry for larger energies~\cite{thalmeier:11} will be neglected.   We also introduced a momentum scale $k_c=\sqrt{v/\lambda}$ and energy scale $E^*=vk_c$ to define dimensionless momentum variables $\hk=k/k_c$ and energies $\hat{\epsilon}_\bk =\epsilon_\bk/E^*=\hk$, see also Table~\ref{tbl:table1}. The single surface Hamiltonian may be diagonalized by a unitary transformation according to
%
%
%%%%%%%%%%%%%%%%%%%%%% figure %%%%%%%%%%%%%%%%%%%%%%%%%%%%%%
\begin{figure}[t]
\vspace{0.cm}
\begin{center}
\includegraphics[width=0.99\columnwidth]{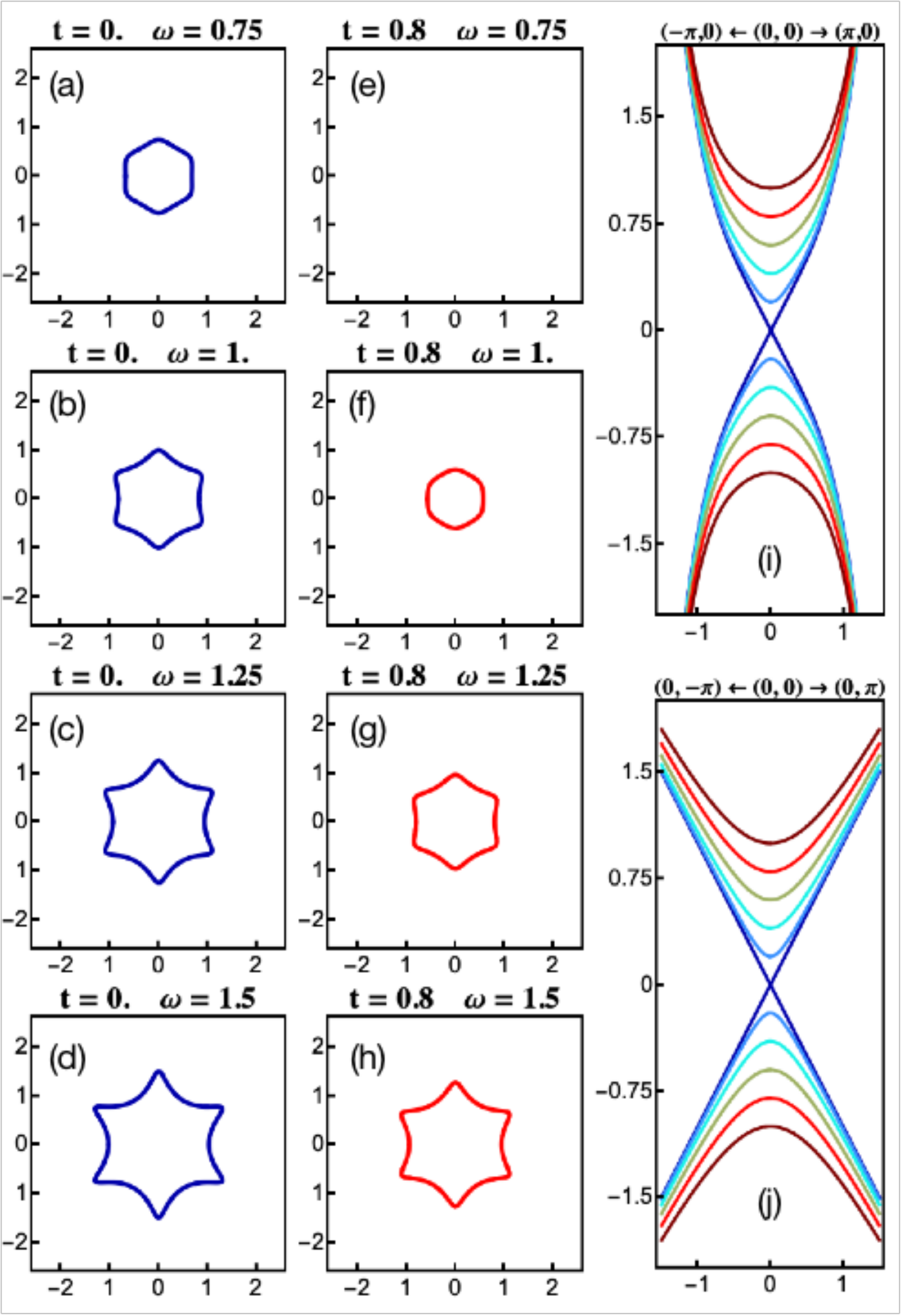}
\end{center}
\vspace{-0.5cm}
\caption{(a-h) Constant-energy $(\omega)$ surfaces of warped Dirac cones with tips at  $\theta_n=(2n+1)\frac{\pi}{6};\;(n=0-5)$ and dents in between. Left column corresponds to isolated surfaces $(t=0)$ and center column to coupled surfaces $(t=0.8)$. In the latter the surfaces are absent for small $\omega$ and  reappear for $\omega\simeq t$, becoming similar to uncoupled case for $\omega \gg |t|$. (i,j) Gap opening in warped Dirac cone dispersion due to tunnelling $t$ for momentum along dent (i) and tip (j) directions, 
where each curves corresponds to a different  inter-surface tunnelling values, as  $0\leq t \leq 1$ in steps of $0.2$, from blue to brown lines.
Here, and in the rest of paper we set the energy unit as $E^*$.}
\label{fig:FermiSurf}
\end{figure}
%%%%%%%%%%%%%%%%%%%%%%fig%%%%%%%%%%%%%%%%%%%%%%%%%%%%%%%%
%
%
%%%%%%%%%%%%%%%%%%%%%%%%%%%%%%%%%%%%%%%%
\begin{center}
\begin{table*}
\caption{Dirac cone parameters for \bse~and \bte~(Refs.~\onlinecite{kuroda:10, fu:09,thalmeier:11}). Here $v, \lambda$ are determined
from experiment, e.g. by ARPES. They also fix the intrinsic momentum and energy scales $k_c$ and  $E^*$ .}
\vspace{0.2cm}
\begin{center}
\begin{tabular}{l @{\hspace{4mm}} l @{\hspace{4mm}}  l @{\hspace{4mm}}  l }
\hline\hline
TI compound && \bse~[\onlinecite{kuroda:10}] & \bte~[\onlinecite{fu:09}] \\
quintuple layer & 1QL & 9.55 \AA & 10.16 \AA   \\
Fermi velocity&  $v$ & 3.55 eV \AA & 2.55 \text{eV}\AA     \\
intrinsic momentum scale & $k_c=\sqrt{v/\lambda}$ & 0.17 \AA$^{-1}\simeq (0.6\text{QL})^{-1}$ &0.10 \AA$^{-1}\simeq (1\text{QL})^{-1}$  \\
intrinisic energy scale & $E^*= vk_c$   & 0.59 eV & 0.25 eV \\
warping parameter &$\lambda=v/k_c^2$ &  $1.28\times 10^2$ eV \AA$^3$ &  $2.5\times 10^2$ eV \AA$^3$\\
 &$\lambda =v^3/E^{*2}=E^*/k_c^3$ & &\\
\hline\hline
\end{tabular}
\end{center}
\label{tbl:table1}
\end{table*}
\end{center}
%%%%%%%%%%%%%%%%%%%%%%%%%%%%%%%%%%%%%%%%%%
%
%---------------------------------------------------------------------------------------------------
\be
\tilh_\bk=S_\bk^\dg h_\bk S_\bk=E_\bk\kappa_z,
\ee
%---------------------------------------------------------------------------------------------------
%
with the warped Dirac cone energy now given by
%---------------------------------------------------------------------------------------------------
\be
\bl
E_\bk=[(vk)^2+(\lambda k^3\cos 3\theta)^2]^\fs=E^*\hk[1+(\hk^2\cos 3\theta_\bk)^2]^\fs,
\label{eqn:cone1}
\el
\ee
%---------------------------------------------------------------------------------------------------
%
or dimensionless $\hE_\bk=E_\bk/E^*$.
The dispersion of Eq.~(\ref{eqn:cone1}) shown in Figs.~\ref{fig:FermiSurf}(i~and~j) results in a warped  constant energy that surfaces have the
shape of six-pronged  `snowflakes'  (Fig.~\ref{fig:FermiSurf}).
The unitary transformation to the eigenstates is given by the matrix
%
%---------------------------------------------------------------------------------------------------
\be
\bl
S_{\bk}&=
\left(
 \begin{array}{cc}
\cos\fphik& i\sin\fphik e^{-i\theta_\bk} \\
 i\sin\fphik e^{i\theta_\bk}& \cos\fphik
\end{array}
\right).
\label{eqn:unmat}
\el
\ee
%---------------------------------------------------------------------------------------------------
%
The columns of $S_\bk$ are the eigenvectors   $|\psi_\kappa\ket$  (in spin $\ua\da$ basis) to the eigenvalues $\pm E_\bk$ $(\kappa=\pm)$. These are called
`chiral' or `helical' basis states because in the isotropic case $\lambda=0$ they are also eigenstates to the chirality operator defined by $\kappa_z=(\bsig\times\hbk)\cdot\hbz=\pm 1$. The spin mixing angle $\phi_\bk$ is given by
%
%---------------------------------------------------------------------------------------------------
\be
\bl
\tan\phi_\bk=k_c^2/(k^2\cos 3\theta_\bk)=1/(\hk^2\cos 3\theta_\bk).
\label{eqn:phiangle}
\el
\ee
%---------------------------------------------------------------------------------------------------
%
In the full circle $0\leq\theta_\bk<2\pi$, $\cos 3\theta_\bk$ changes sign six times at $\theta_n=(2n+1)\frac{\pi}{6};\;(n=0-5)$. To obtain a continuous variation of $\phi_\bk(\theta_\bk)$ at these boundaries and to have a well defined uniform convergence to the isotropic limit $\phi_\bk\rightarrow \frac{\pi}{2}$ for  vanishing warping 
$(\lambda\rightarrow 0)$
 we define 
 $\phi_\bk=tan^{-1}(1/\hk^2\cos 3\theta_\bk)$ for  $\cos 3\theta_\bk >0$
  and  
  $\phi_\bk=tan^{-1}(1/\hk^2\cos 3\theta_\bk)+\pi$
   for  $\cos 3\theta_\bk <0$. 
   This amounts to taking the second (upper) branch of $tan^{-1}$ in the latter case.
The $\hk$ and $\theta_\bk$ dependence of the spin mixing angle $\phi_\bk$ is shown in Fig.~\ref{fig:phipsi},
 and it indeed varies continuously with $\theta_\bk$ around the (half-)circle centered at the isotropic limit $\phi_\bk=\pi/2$.\\

For calculation of the QPI spectrum one needs the Green's function in spin representation. For the single surface problem it is given by
%
%---------------------------------------------------------------------------------------------------
\be
\bl
G_\bk
&=(\om-h_\bk)^{-1}
\\
&=
\left(
 \begin{array}{cc}
\om -\epsilon_\bk\hk^2\cos 3\theta_\bk&-i\epsilon_\bk e^{-i\theta_\bk} \\
i\epsilon_\bk e^{i\theta_\bk} & \om +\epsilon_\bk\hk^2\cos 3\theta_\bk
\end{array}
\right)^{-1}.
\label{eqn:Green_single1}
\el
\ee
%---------------------------------------------------------------------------------------------------
%
Defining the warping energy by 
$$\Delta_\bk=\epsilon_\bk(\hk^2\cos 3\theta_\bk)=\pm(E_\bk^2-\epsilon_\bk^2)^\frac{1}{2},$$
the Green's function is obtained as
%
%---------------------------------------------------------------------------------------------------
\be
\bl
G_\bk&=
\frac{1}{(\om)^2-E_\bk^2}
\left(
 \begin{array}{cc}
\om+\Delta_\bk&-i\epsilon_\bk e^{-i\theta_\bk} \\
i\epsilon_\bk e^{i\theta_\bk} & \om - \Delta_\bk
\end{array}
\right).
\label{eqn:Green_single2}
\el
\ee
%---------------------------------------------------------------------------------------------------
%
In the isotropic Dirac cone case one simply has to set $\Delta_\bk=0$ and $E_\bk=\epsilon_\bk$.

%
%%%%%%%%%%%%%%%%%%%%%% figure %%%%%%%%%%%%%%%%%%%%%%%%%%%%%%
\begin{figure}[t]
\begin{center}
\includegraphics[width=0.995\columnwidth]{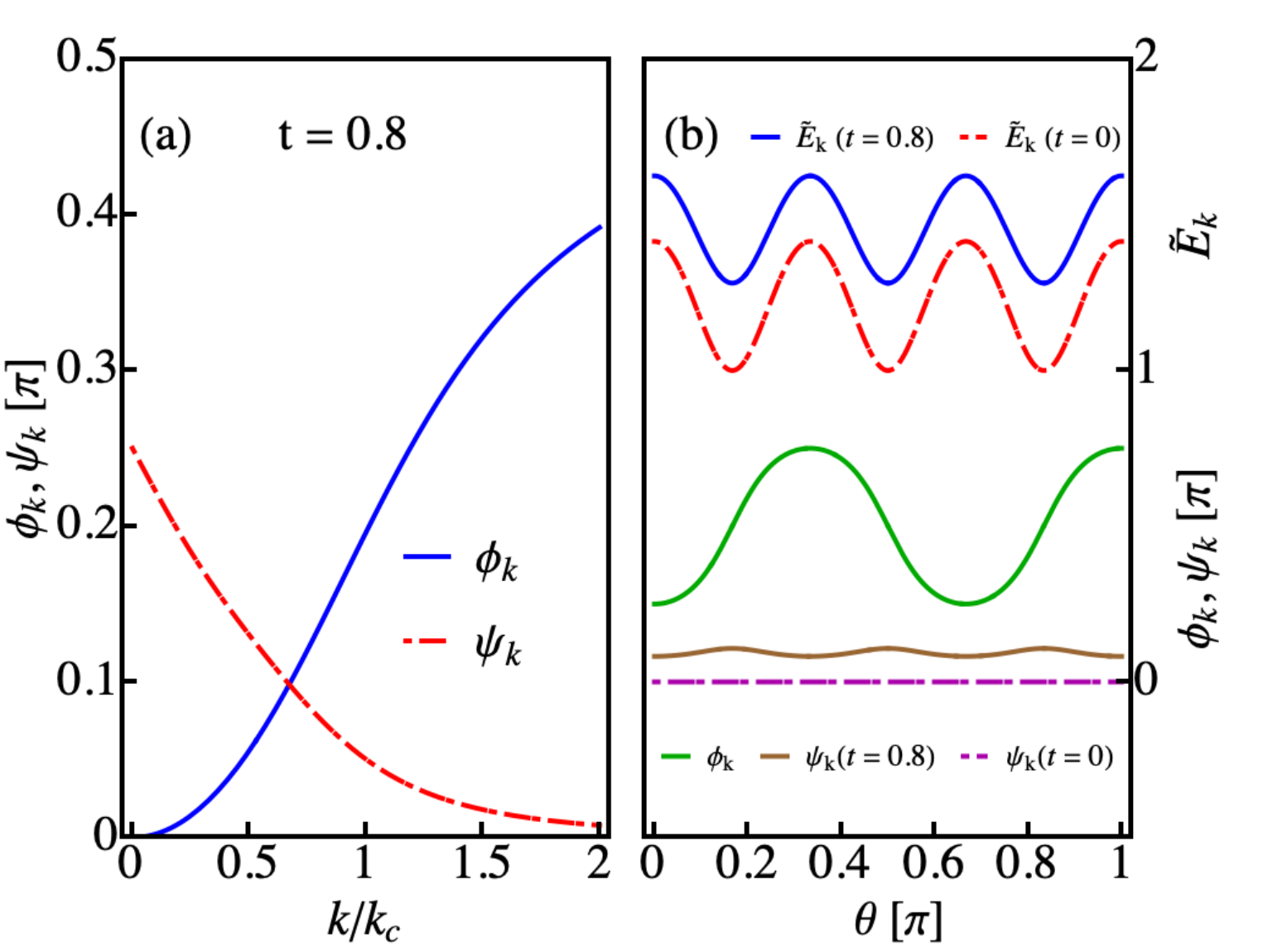}
\end{center}
\vspace{-0.5cm}
\caption{(a) Dependence of  spin mixing angle $\phi_\bk$ due to warping term and inter-surface state mixing angle  $\psi_\bk$ due to T,B tunnelling as function of wave number. Here $\theta_\bk$=0, i.e. the wave vector points 
to the dents in Fig.~\ref{fig:FermiSurf}. (b) Azimuthal $\theta_\bk$ angle dependence of warped and T,B hybridised cone energies $\hE=\tE_\bk/E^*$ and mixing angles $\phi_\bk, \psi_\bk$. Full lines for $\hatt=t/E^*=0.8$, broken lines for isolated T,B surfaces (${\hatt} = 0$). Here $\phi_\bk$ does not depend on $\hatt$ because the latter is diagonal in the helicities. }
\label{fig:phipsi}
\end{figure}
%%%%%%%%%%%%%%%%%%%%%%fig%%%%%%%%%%%%%%%%%%%%%%%%%%%%%%%%
%

\section{Splitting of Dirac cones by inter-surface tunnelling in TI thin films}
\label{sec:tunnelling}

The warped cones of single TI surfaces have been abundantly demonstrated by ARPES experiments~\cite{hasan:11,ando:13}.  As outlined in the introduction in thin films the tunnelling of helical Dirac states between the two surfaces introduces a gap in the spectrum signifyng the transition to topologically trivial surface states close to the gap threshold as function of thickness.
This has been observed again directly with ARPES~\cite{zhang:10,neupane:14} in agreement with magnetotransport results\cite{taskin:12}. Theoretically it has been studied in detail~\cite{lu:10,liu:10,asmar:18} by starting from the bulk thin film states and introducing the approriate boundary conditions. A nonmonotonic, even oscillating behaviour of surface state tunnelling and gap size  as function of thickness $d$ is possible although this has sofar not been observed in photoemission (Sec. \ref{sec:numerical}).

\begin{figure}[t]
\begin{center}
\includegraphics[width=1\columnwidth]{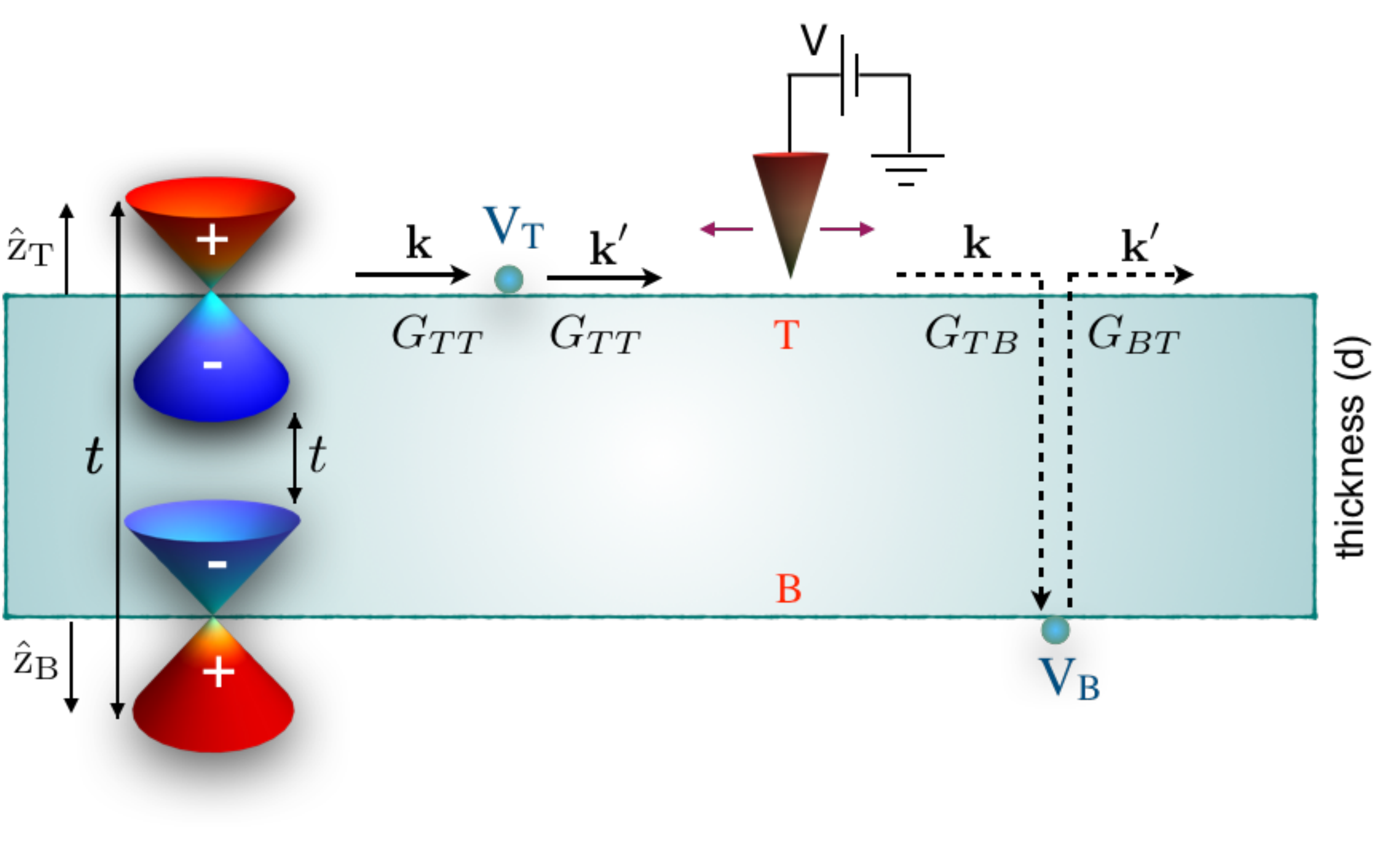}
\end{center}
\vspace{-0.75cm}
\caption{Sketch of surface scattering processes in QPI for thin film geometry. Surface normals $\hbz_B=-\hbz_T$ are opposite. Isolated T, B Dirac cones with helicities $\kappa=\pm 1$ are indicated. $t(d)$ is thickness $d$- dependent tunnelling energy between equal helicity states on T,B. Impurity scattering of (top) cone states with momentum $\bk$ to $\bk'=\bk+\bq$ is possible on both surfaces $(V_T,V_B)$ due to effect of tunnelling; simlar for the bottom states. Interference of waves with $\bk$ and $\bk'=\bk+\bq$ momenta leads to surface density oscillations $\delta N_T(\bq,\omega)$ described by Eq.~(\ref{eqn:QPI}) and scanned by the tip at bias voltage $eV=\omega$.}
\label{fig:film}
\end{figure}

For the purpose of investigating QPI signatures of the topological transition in thin films we employ a phenomenlogical model starting from the independent helical states on the two isolated surfaces.  They are coupled by the thickness dependent tunnelling matrix element $t(d)$ of top (T) and bottom (B) surface states with {\it equal helicities} (Fig.~\ref{fig:film}).
The $4 \times 4$ thin film Hamiltonian in the space with basis $|\kappa\alpha\ket$ (helicity $\kappa=\pm 1$ and surface index $\alpha=$ T,B) may be written as $2\times 2$ block matrix in T,B space according to
%
%---------------------------------------------------------------------------------------------------
\bea
\bl
H_\bk&=
\left(
 \begin{array}{cc}
h_\bk& t\sigma_0 \\
 t\sigma_0& h_\bk
\end{array}
\right)
\longrightarrow
\tilH_\bk=\left(
 \begin{array}{cc}
E_\bk\kappa_z& t\kappa_0 \\
 t\kappa_0& -E_\bk\kappa_z 
\end{array}
\right),
\label{eqn:filmmat}
\el
\eea
%---------------------------------------------------------------------------------------------------
%
in the spin space with basis $|\sigma\alpha\ket$ ( $\sigma=\ua,\da$) (left)  and helicity space with basis $|\kappa\alpha\ket$ ($\kappa=\pm 1$) (right). In both representations $\alpha=$ T,B is the surface index.
The minus sign in the last element appears because states of equal helicities on B,T surfaces belong to opposite half-cones (see Fig.~\ref{fig:film}). We use the convention that $\bsig, \bkap, \balp$ denote the vector of $2\times 2$ Pauli matrices in spin, helicity and surface layer (T,B) space, respectively. Furthermore the unit in each space is denoted by $\sigma_0,\kappa_0,\alpha_0$. The film Hamiltonian may be diagonalized by two transformations
for the $\kappa=\pm1$ helicities separately with the simple form
%
%---------------------------------------------------------------------------------------------------
\be
\bl
U^\pm_\bk&=
\left(
 \begin{array}{cc}
\cos\psi_\bk& \mp\sin\psi_\bk \\
\pm\sin\psi_\bk& \cos\psi_\bk 
\end{array}
\right);\;\;\;
\tan 2\psi_\bk=\frac{t}{E_\bk}
=\frac{\hat{t}}{\hE_\bk},
\label{eqn:filmtrans}
\el
\ee
%---------------------------------------------------------------------------------------------------
%
where $\hat{t}=t/E^*$. The total transformation matrix from spin single surface representation to helicity-film eigenstate representation that diagonalizes $H_\bk$ is then given by $W_\bk$. The column vectors of this $4\times 4$ matrix are the helicity eigenstates of the film (explicitly given in Appendix \ref{sec:heleigen}). They have the energies resulting from the diagonalization $W^\dg_\bk H_\bk W_\bk=\tilde{E_\bk}\kappa_0\otimes\tau_z$ with
%
%---------------------------------------------------------------------------------------------------
\be
\tE_\bk=[E_\bk^2+t^2]^\fs=
E^* \bigl[\hk^2[1+(\hk^2\cos 3\theta_\bk)^2]+\hat{t}^2\bigr]^\fs.
\label{eqn:filmdirac}
\ee
%---------------------------------------------------------------------------------------------------
%
Therefore at the Dirac point $\bk=0 $ a gap of size $\Delta_t=2|t|$ opens between T,B hybridised film states  $(\tau=1,2)$  each twofold degenerate due to time reversal symmetry with helicities $\kappa =\pm 1$.
%
%%%%%%%%%%%%%%%%%%%%%% figure %%%%%%%%%%%%%%%%%%%%%%%%%%%%%%
%%%%%%%%%%%%%%%%%%%%%%fig%%%%%%%%%%%%%%%%%%%%%%%%%%%%%%%%
%

\section{Mixed topological surface states and their spin texture and Berry phase}
\label{sec:spintext}

\begin{figure}[t]
\begin{center}
\includegraphics[width=0.92\linewidth]{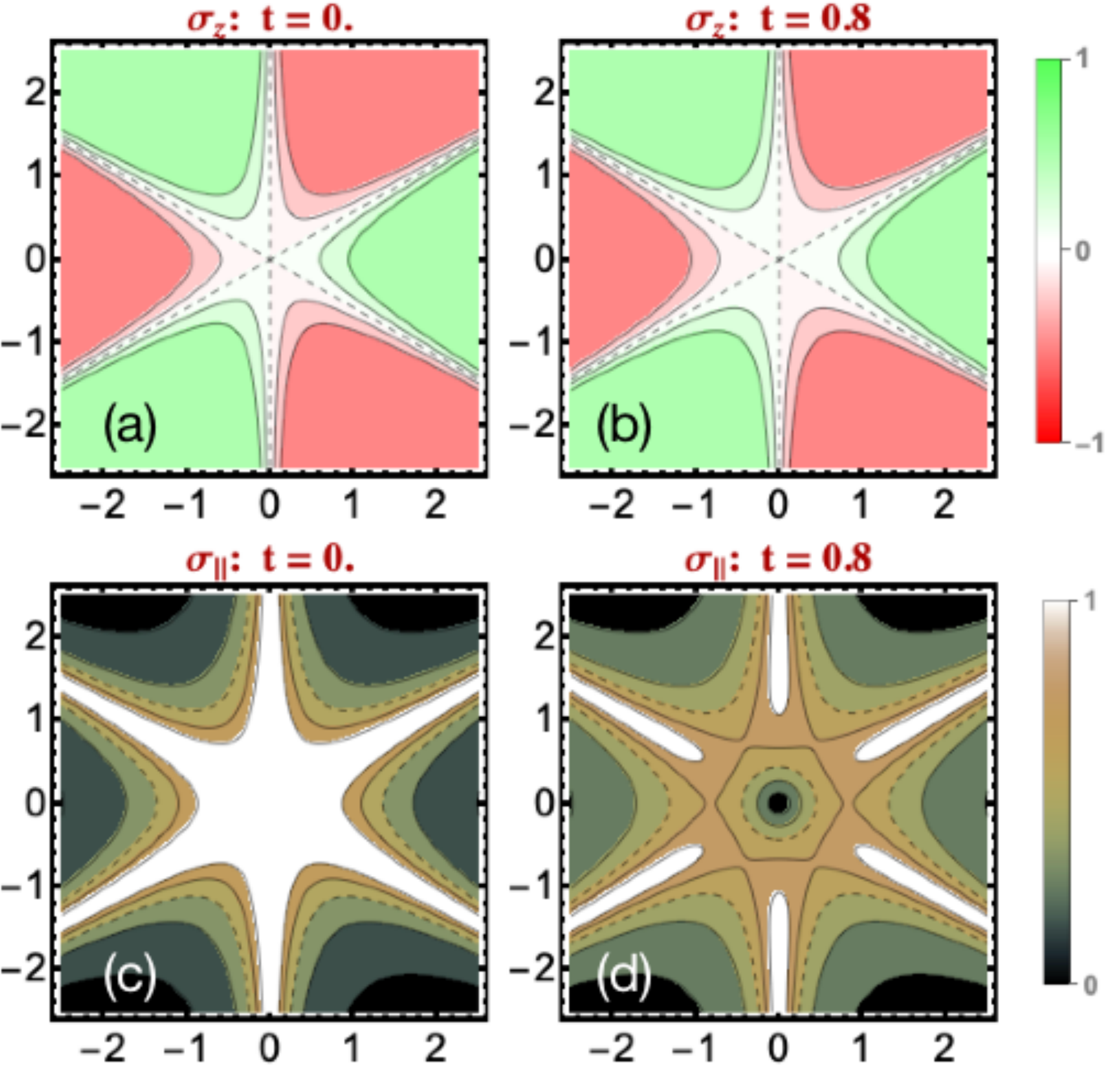}
\end{center}
\vspace{-0.5cm}
\caption{Contour plot of spin texture. Left (a,c) for $t=0$: In the cone center $\langle\sigma_z\rangle_T$ vanishes whereas $\bra\sigma_\parallel\ket_T\simeq 1$ (white region). Right (b,d) for coupled surfaces $t=0.8$ both  $\langle\sigma_{z}\rangle_T$, $\langle\sigma_{\parallel}\rangle_T$ vanish around $\bk\approx 0$. The sign of  $\langle\sigma_z\rangle_T$ alternates (green/red) when moving around the cone center. See also Fig.~\ref{fig:spin}.}
\label{fig:spintex}
\end{figure}

The explicit form of the surface states  is given by the columns of Eq.~(\ref{eqn:Wunitary}). For the upper and lower split half cones with energies $E_{(1,2)}(\bk)=\pm \tE_\bk$ the pairs of doubly degenerate $(\kappa=\pm)$ upper and lower $(\tau=1,2)$ band states $|\tilde{\psi}_{\tau\kappa}\ket$ can be written, respectively as
\be
\bl
|\tps_{1+}^\bk\ket&=\cos\psi_\bk |\psi_{T+}\ket +\sin\psi_\bk |\psi_{B+}\ket \;\; : E_{1\bk}=+\tE_\bk 
\\
|\tps_{1-}^\bk\ket&=\cos\psi_\bk |\psi_{B-}\ket +\sin\psi_\bk |\psi_{T-}\ket  \;\; : E_{1\bk}=+\tE_\bk 
\\[0.2cm]
|\tps_{2+}^\bk\ket&=\cos\psi_\bk |\psi_{B+}\ket -\sin\psi_\bk |\psi_{T+}\ket  \;\; : E_{2\bk}=-\tE_\bk 
\\
|\tps_{2-}^\bk\ket&=\cos\psi_\bk |\psi_{T-}\ket -\sin\psi_\bk |\psi_{B-}\ket  \;\; : E_{2\bk}=-\tE_\bk
\hspace{0.5cm}
\label{eqn:eigenvec}
\el
\ee
in terms of the isolated surface chiral states $|\psi_{\alpha\kappa}\ket$.  These eigenstates are related by combined time reversal $\pm\leftrightarrow \mp$ represented by $\Theta=i\sigma_yK$ ($K=$ complex conjugation) and interchange of surfaces $T,B \leftrightarrow B,T$ represented by the reflection $\Sigma=\alpha_x$. The pairwise degeneracy of  hybridised states in Eq.~(\ref{eqn:eigenvec}) is due to the symmetry $(\Theta\Sigma)|\tps^\bk_{\tau +}\ket=|\tps^\bk_{\tau -}\ket$. For states close to the gap with $\tE_\bk\geq t$ $(k\ll k_c)$ one has $\psi_\bk\simeq \frac{\pi}{4}$ meaning  these are true film states with equal amplitudes on B,T surfaces.
On the other hand far from the gap with $\tE_\bk\gg t$   $(k\gg k_c)$, we have $\psi_\bk\approx \frac{1}{2}\frac{t}{E_\bk} \ll \frac{\pi}{4}$ and consequently the above states are mostly localized on either T or B surface with only small amplitude on the opposite B or T surfaces. Therefore for large thickness when $t\rightarrow 0$ the  state in Eq.~(\ref{eqn:eigenvec}) decouple to isolated surface states on T,B such that $|\tps_{1\pm}^\bk\ket$ become states with opposite chirality on opposite surfaces with positive energies and likewise $|\tps_{2\pm}^\bk\ket$ with negative energies.
From these states we may now  calculate the surface spin textures $\bra\bsig\ket_{\tau\bk}^{T,B}$
on T or B of the model as function of the tunnelling matrix element $t$ between the isolated helicity surface states. The spin textures of states 1,2 will be opposite on the same surface and because of 
$\bra\bsig\ket_{1,2\bk}^B=-\bra\bsig\ket_{1,2\bk}^T$ identical on opposite surfaces. Therefore, restricting to the top surface we have for the spin expectation value of each pair corresponding to each half cone:
%
%---------------------------------------------------------------------------------------------------
\be
\bl
\bra\bsig\ket_{1,2\bk}^T
\!
&=
\cos^2\psi_\bk\bra\psi_{T\pm}|\bsig|\psi_{T\pm}\ket         
+\sin^2\psi_\bk\bra\psi_{T\mp}|\bsig|\psi_{T\mp}\ket 
\\
&=
\pm(-\sin\phi_\bk\sin\theta_\bk,\sin\phi_\bk\cos\theta_\bk,\cos\phi_\bk)\cos 2\psi_\bk.
\label{eqn:spinpol}
\el
\ee
%---------------------------------------------------------------------------------------------------
%
Using the previous expressions for the mixing angles $\phi_\bk, \psi_\bk$ in terms of the polar angle $\theta_\bk$ we get explicitly for in- and out-of-plane spin polarisation:
%
%---------------------------------------------------------------------------------------------------
\be
\bl
\bra\sigma_\parallel\ket_T
&=
\frac{\hk}{\{\hk^2[1+(\hk^2\cos 3\theta)^2]+\hat{t}^2\}^{\fs}};
\\
\bra\sigma_z\ket_T
&=
\frac{\hk^3\cos 3\theta}{\{\hk^2[1+(\hk^2\cos 3\theta)^2]+\hat{t}^2\}^{\fs}}.
\el
\ee
%---------------------------------------------------------------------------------------------------
%
Here $\bra\sigma_\parallel\ket_T=(\bra\sigma_x\ket_T^2+\bra\sigma_y\ket_T^2)^\fs$ and therefore the total 
length of the spin on T is
%
%---------------------------------------------------------------------------------------------------
\be
\bra\sigma_{tot}\ket^2_T
=(\bra\sigma_z\ket_T^2+\bra\sigma_\parallel\ket_T^2)
=\cos^2 2\psi_\bk=\Bigl(\frac{E_\bk}{\tE_\bk}\Bigr)^2.
\ee
%---------------------------------------------------------------------------------------------------
%
When the tunnelling between surfaces becomes negligible $\psi_\bk\rightarrow 0$ and $\bra\sigma_{tot}\ket^2_T\rightarrow 1$. Contour plots of $\bra\sigma_\parallel\ket_T$ and $\bra\sigma_z\ket_T$ are shown in Fig.~\ref{fig:spintex} for isolated and coupled surfaces. In particular the in-plane component which is maximum $(\leq 1)$ at the $\Gamma$ point and along the tips (white region) in Fig.~\ref{fig:spintex}(c) is suppressed to zero by the tunnelling at the $\Gamma$ point in (d) (dark green area). We note that ab-initio calculations of the spin-textures for thick slabs have also been perfomed~\cite{zhu:13}.\\

%
%%%%%%%%%%%%%%%%%%%%%% figure %%%%%%%%%%%%%%%%%%%%%%%%%%%%%%
%%%%%%%%%%%%%%%%%%%%%%fig%%%%%%%%%%%%%%%%%%%%%%%%%%%%%%%%
%
%
%%%%%%%%%%%%%%%%%%%%%% figure %%%%%%%%%%%%%%%%%%%%%%%%%%%%%%
\begin{figure}[]
\begin{center}
\includegraphics[width=0.990\columnwidth]{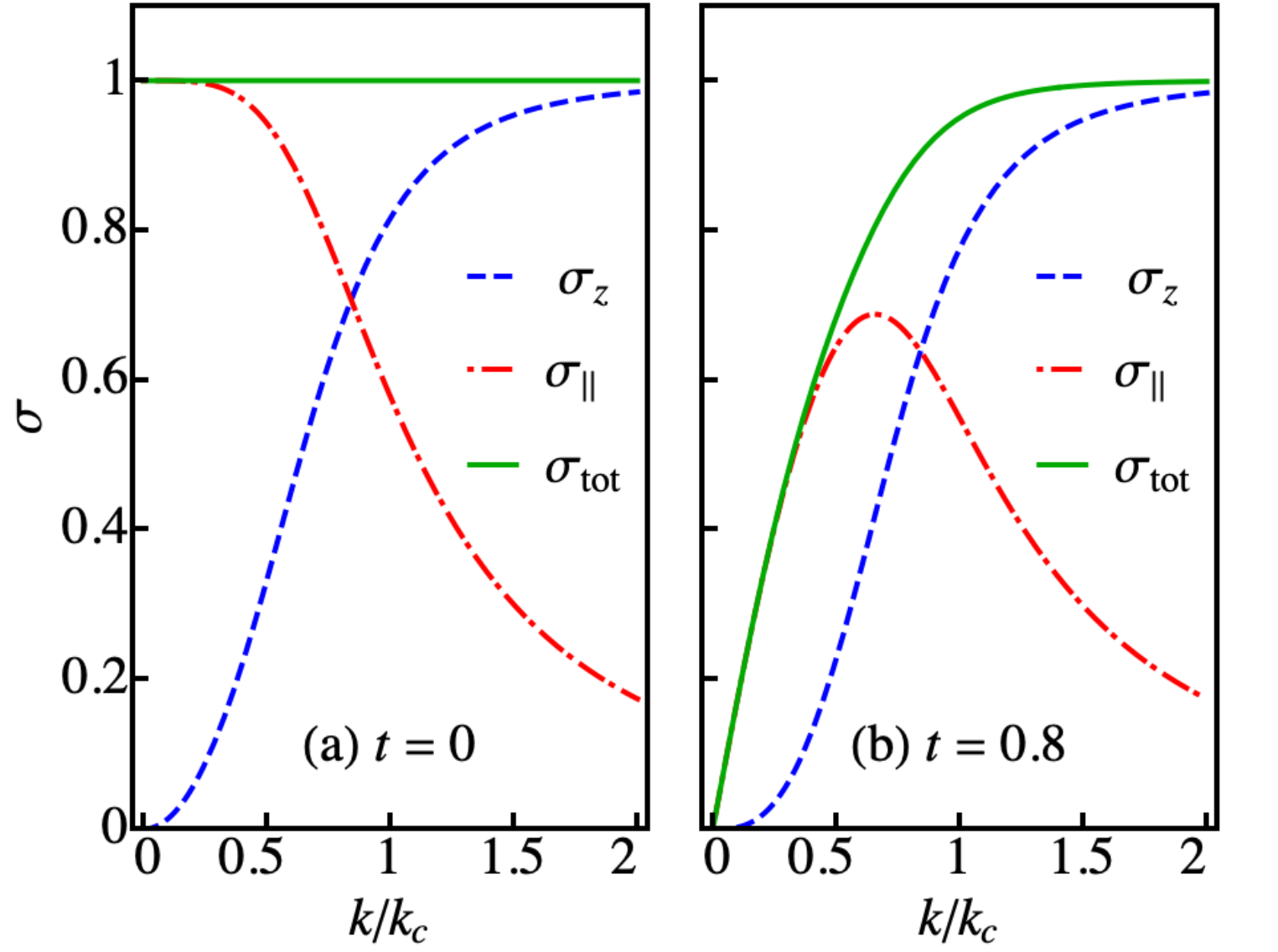}
\end{center}
\vspace{-0.5cm}
\caption{In-plane $(\parallel)$, out-of-plane $(z)$ and total (tot) spin component of surface states
for isolated $(\hatt=0)$ (left) and coupled $(\hatt=0.8)$ (right) T,B surfaces. For non-zero $\hatt$ both components
will be suppressed at the Dirac point due to the gapping. Here $\theta_\bk$=0, i.e. the wave vector points 
to the dents in Fig.~\ref{fig:FermiSurf}.}
\label{fig:spin}
\end{figure}
%%%%%%%%%%%%%%%%%%%%%%fig%%%%%%%%%%%%%%%%%%%%%%%%%%%%%%%%
%

The Dirac point of topological surface states may be viewed as a monopole in momentum space. It 
is connected with a nonvanishing Berry-phase when encircling the Dirac point on a closed path $C$ containing the origin. First we consider the case of warped Dirac cone of a single (T or B) surface. Then the  Berry phase is given by ($\kappa=\pm$)~\cite{xiao:10,taskin:11}
%
%---------------------------------------------------------------------------------------------------
\be
\bl
&
\gamma_{\kappa} 
=
 \oint_Cd\bk\cdot i\bra \psi^\bk_\kappa |\bigtriangledown_\bk |\psi^\bk_\kappa\ket,
 %\\
 %&
\label{eqn:Berrysurf}
\el
\ee
%---------------------------------------------------------------------------------------------------
%
where the Berry connection described by
$$
i\bra \psi^\bk_\kappa |\bigtriangledown_\bk |\psi^\bk_\kappa\ket
=
 \kappa[\cos\phi_\bk-1],
$$
 depends on the warping through $\phi_\bk$. Because of the anti- periodic property  
 $\cos[\phi(\theta_\bk+\frac{\pi}{3})]=-\cos[\phi(\theta_\bk)]$,
  the contour integral over the cosine vanishes and we get $\gamma_{\pm}=\mp\pi$. This means that the topologically non-trivial Berry phase for the Dirac cone states is not influenced by the warping effect since the latter does not destroy the $\bk=0$ singularity. Therefore when considering the effect of the inter-surface tunnelling $t$ on the Berry phase we may safely neglect the warping for simplicity. The tunnelling leads to film states determined by the T,B inter-surface mixing angle $\psi_\bk$ (Eq.~\ref{eqn:eigenvec}). The Berry connection for the isotropic $(\phi_\bk=\frac{\pi}{2})$ film states  $(\tau=1,2,\kappa=\pm)$,
 $|\tps^\bk_{\tau\kappa}\ket$ with a given bare Dirac cone energy $E=E_\bk=vk$ may be calculated as
%---------------------------------------------------------------------------------------------------
\be
\bl
&
\gamma_{\tau\kappa} 
=
 \oint_Cd\bk\cdot i\bra \psi^\bk_\kappa |\bigtriangledown_\bk |\psi^\bk_\kappa\ket
 =
(-1)^\tau\kappa\pi\cos  2\psi 
; \\
% \label{eqn:Berryfilm}
%\el
%\ee
%with 
%\be
%\bl
&
\cos 2\psi
\!=\!
\frac{E}{(E^2+t^2)^\fs}
\!=\!
\left\{
\begin{array}{ll}
\frac{E}{t}& \;E\ll t \;(k\ll t/v) \\
1-\fs(\frac{t}{E})^2& \; E\gg t \; (k\gg t/v)
\end{array}
\right.
.
\label{eqn:Berryfilm}
\el
\ee
%---------------------------------------------------------------------------------------------------
%
Therefore the Berry phase of states at the gap threshold which have parabolic dispersion vanishes while for energies much larger than the gap it approaches the values $\gamma_{\tau\kappa}=(-1)^\tau\kappa\pi$,
 $(\tau=1,2,\kappa=\pm 1)$ of the isolated surface Dirac cones. Here $\kappa=\pm$ correspond now to {\it degenerate} pairs (they become the $T+,B-$ states for $\tau =1$ and  $B+,T-$ states for $\tau =2$ when $t,\psi_\bk\rightarrow 0$). 
The reduction of the Berry phase close to the mass gap of  Dirac electrons  leads to a violation of topological protection. Therefore WAL due to destructive interference caused by the Berry phase $\mp\pi$ is suppressed and breaks down completely for large enough mass gap in ultrathin films with $d\leq 5{\rm QL}$~\cite{taskin:12}. Furthermore the acquired degeneracy of film states opens the backscattering channel for the gapped states reducing the surface mobility~\cite{taskin:12} and also influencing the QPI signatures.

\section{Selfconsistent t-matrix theory for impurity scattering}
\label{sec:tmat}

The STM-QPI method measures the electronic density fluctuations on the surface caused the interference of scattered and ingoing waves at an impurity site. It must be stressed that this is a single impurity effect~\cite{capriotti:03} although the amplitude of the density fluctuations will be proportional to the number of impurities. In the situation of a thin film another aspect is important. Due to the tunnelling the film states are eigenstates composed of surface states on both top and bottom surface. Therefore, even if the density fluctuations are measured on the top surface they will also be influenced by the scattering on the bottom surface due to the tunnelling. This effect which is illustrated in Fig.~\ref{fig:spintex} has to be included in the calculation. For the impurity scattering potential in spin $(\sigma)$ and T,B surface space $(\alpha)$ we assume the generic form
%
%---------------------------------------------------------------------------------------------------
\be
\bl
&
\hV=(V_c\sigma_0+V_m\sigma_z)\alpha_0=V\alpha_0;
\\[0.3cm]
&
V= 
\left(
\begin{array}{cc}
V_\ua& 0 
\\
0& V_\da
\end{array}
\right);\;\;\;
V_{\ua,\da}=V_c\pm V_m.
\label{eqn:Vscatt}
\el
\ee
%---------------------------------------------------------------------------------------------------
%
Here $V_c$ denotes charge and  V$_m$ exchange scattering by normal and magnetic impurities, respectively,  
which does {\it not} depend on the momentum transfer $\bq=\bk'-\bk$ of the scattering  process. In this case a closed solution for the scattering t-matrix necessary to compute the QPI spectrum is possible in spin representation which then will have to be transformed to the helical eigenstate basis of the surfaces. The second exchange scattering term in Eq.~(\ref{eqn:Vscatt}) corresponds to a frozen impurity spin (along z-direction),   neglecting any spin dynamics such as treated in Ref.~\onlinecite{derry:15}. This spin polarisation can be achieved either by tiny magnetic field or, since the impurity concentration is finite, may be the result of long-range exchange interactions among the magnetic impurities that lead to quasi-static behaviour (long relaxation times) for each impurity spin. 
 For the potential in Eq.(~\ref{eqn:Vscatt}) we assume only intra-surface scattering $(\sim \alpha_0)$,  i.e. cross scattering between opposite T, B surfaces is neglected. This is a reasonable simplification because, firstly the scattering in QPI experiments happens mostly via gas impurity atoms or molecules adsorbed above the surfaces. Therefore their perturbing potential should
be constrained to the top and bottom surfaces. Since the {\it overlap} of T,B surface state wave functions is small at each surface no sizable inter-surface (T-B) scattering  $V'$ for $d> 2$QL is expected. Secondly, for zero momentum transfer $\bq$ a scattering between equal  helicity states on  opposite surfaces vanishes because of opposite spins for initial and final states (due to the
reversed helicity sense on both surfaces) This means that the scattering $V'$ has to be odd in $\bq$ whereas
the intra-plane scattering $V$ is even, taken as constant. Because the Dirac cones have small radius 
($k_c\ll\pi/a$) all scattering processes on the cone have small momentum transfer and therefore $V'$ will also be small for them.\\
%
%%%%%%%%%%%%%%%%%%%%%% figure %%%%%%%%%%%%%%%%%%%%%%%%%%%%%%
\begin{figure}[t]
\begin{center}
\includegraphics[width=0.990\columnwidth]{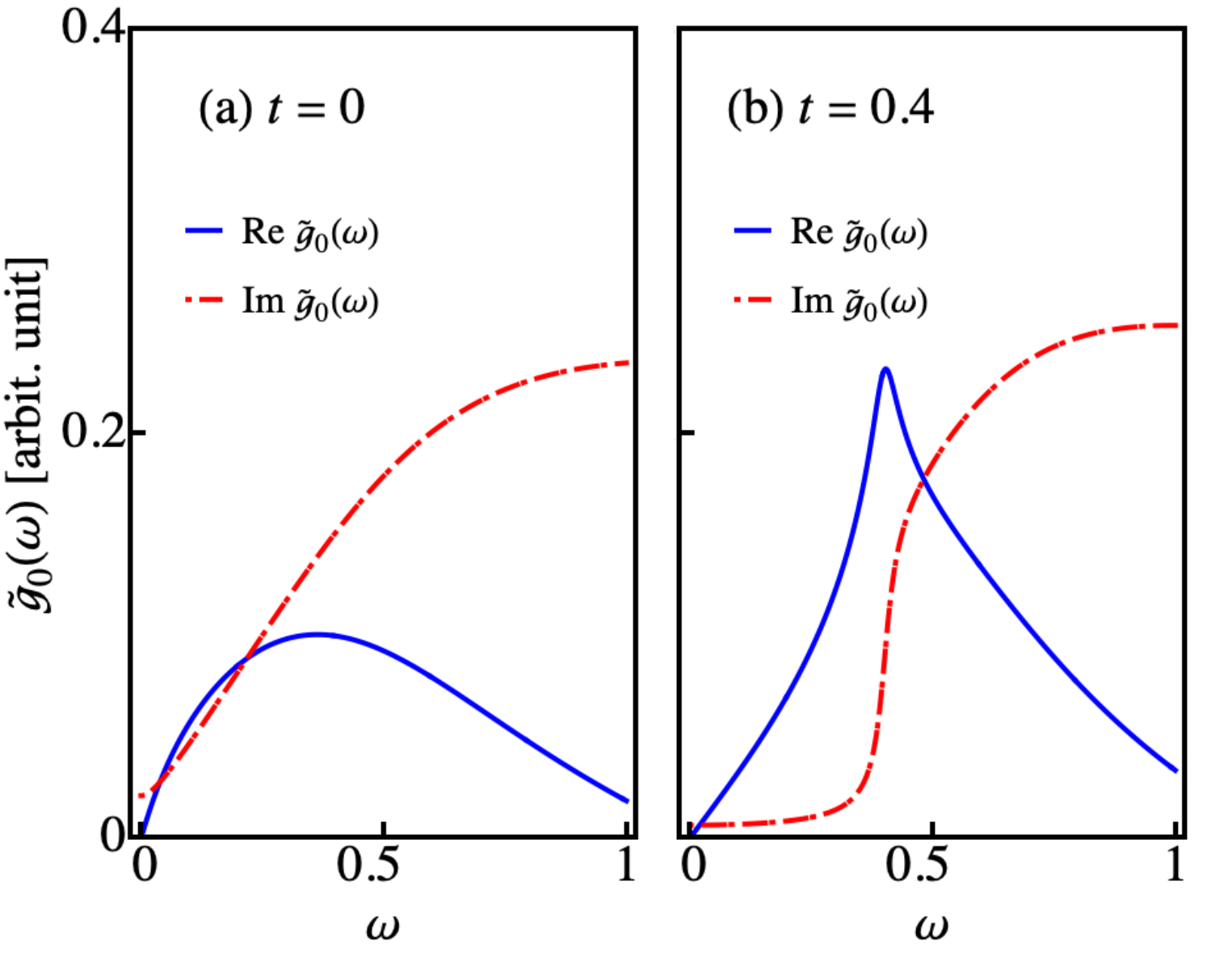}
\end{center}
\vspace{-0.5cm}
\caption{Momentum integrated Green's function $\tg_0(\omega)$ (Eq.~\ref{eqn:auxiliary}) which determines the frequency 
dependence of the full t-matrix elements in Eq.~(\ref{eqn:tmat_spin}). It becomes develops singular behaviour at the gap edge $\omega=t$.}
\label{fig:g0-function}
\end{figure}
%%%%%%%%%%%%%%%%%%%%%%fig%%%%%%%%%%%%%%%%%%%%%%%%%%%%%%%%
%
\subsection{Closed solution in spin basis}

For a momentum-independent scattering potential the t-matrix equation in spin basis may be solved as
%
%---------------------------------------------------------------------------------------------------
\be
\bl
\hat{t}_{\bk\bk'}(\om)
&=
\hat{t}(\om)=[1-\hV{\hg}(\om)]^{-1}\hV
\\
&
\equiv R^{-1}\hV=t_T\oplus t_B.
\label{eqn:tmatdef}
\el
\ee
%---------------------------------------------------------------------------------------------------
%
We made the approximation that impurities on T,B do {\it not} scatter {\it between} the surfaces but only {\it within} each of them, 
i.e. we approximate $R^{-1}=R^{-1}_T\oplus R^{-1}_B$  and therefore $t_{TB}=t_{BT}=0$. However the propagation, i.e. tunnelling  {\it between} surfaces is fully taken into account by the non-diagonal Green's function elements given below. Furthermore we assume that the type and scattering potential of impurities on T,~B are the  same and therefore set $\hV_T=\hV_B=\hV$. Then
%
%---------------------------------------------------------------------------------------------------
\be
\bl
&
t=
\left(
\begin{array}{cc}
t_{\ua\ua}&t_{\ua\da}  \\
t_{\da\ua}&t_{\da\da}
\end{array}
\right);
\;\;\;
%\\
%&
V=
\left(
\begin{array}{cc}
V_v+V_m& 0 \\
0&V_c-V_m
\end{array}
\right);
\;
\\
&
{\hg}(\om)=\frac{1}{N}\sum_\bk {\hG}_\bk(\om).
\label{eq:tdef}
\el
\ee
%---------------------------------------------------------------------------------------------------
%
The spin flip amplitudes $t_{\ua\da}$,  $t_{\ua\da}$ that appear formally vanish identically (Appendix \ref{sec:rmatrix}) so that we only have to consider the diagonal elements $t_{\ua\ua}$,  $t_{\da\da}$ in the following. Ultimately this is due to the absence of spin-flip terms in the magnetic scattering term of Eq.~(\ref{eqn:Vscatt}).
Furthermore the film Green's function ${\hG}_\bk(\om)$ in spin basis is given by
%
%---------------------------------------------------------------------------------------------------
\be
\hG_\bk(\om)=[\om-\hat{H}]^{-1};\;\;\; 
\tilH_\bk=
\left(
 \begin{array}{cc}
h_\bk& t\sigma_0 \\
 t\sigma_0& h_\bk
\end{array}
\right).
\ee
%---------------------------------------------------------------------------------------------------
%
With $h_\bk$ defined in Eqs.~(\ref{eqn:hamsig},~and~\ref{eqn:hamat}).
The individual t-matrix  elements $t_{\sigma\sigma}$ may be evaluated (see Appendix \ref{sec:rmatrix}) from Eq.~(\ref{eqn:tmatdef}). It is also useful to introduce (anti-)symmetric combinations by
$t_{s,a}=\fs(t_{\ua\ua}\pm t_{\da\da})$. Furthermore we need the determinants $D={\rm det} [R_{T,B}]$ which are given by
%
%---------------------------------------------------------------------------------------------------
\be
D=1-2V_c\tg_0+(V_c^2-V_m^2)\tg_0^2.
\label{eqn:determinant}
\ee
%---------------------------------------------------------------------------------------------------
%
The momentum- integrated Green's function in the above expressions which depend only on frequency $\om$ are defined by
(see Eq.~(\ref{eqn:filmdirac})):
%
%---------------------------------------------------------------------------------------------------
\be
\tg_0(\om)=
\frac{1}{N}\sum_\bk\frac{\om}{[(\om)^2-\tE_\bk^2]};\;\;\;\;\;
\label{eqn:auxiliary}
\ee
%---------------------------------------------------------------------------------------------------
%
the (anti-) symmetrized t-matrix elements are then obtained as
%
%---------------------------------------------------------------------------------------------------
\be
\bl
&
t_s(\om)=\frac{[V_c-(V_c^2-V_m^2)\tg_0(\om)]}{1-2V_c\tg_0(\om)+(V_c^2-V_m^2)\tg_0(\om)^2};
\\
&
t_a(\om)=\frac{V_m}{1-2V_c\tg_0(\om)+(V_c^2-V_m^2)\tg_0(\om)^2}.
\label{eqn:tmat_spin}
\el
\ee
%---------------------------------------------------------------------------------------------------
%
Obviously the anti-symmetric amplitudes require a magnetic scattering and  for normal scattering  $(V_m=0,~t_a=0)$, we only obtain the symmetric scattering amplitude $t_s=V_c/(1+V_c\tg_0)$.
\
For this calculation we used the spin basis because then the scattering potential is isotropic allowing for a closed solution. But for later calculation of the QPI spectrum we now have to transform back to the helical eigenbasis of the surfaces.
The dynamics of the scattering matrix is determined by that of the momentum-integrated $\tg_0(\om)$ which is shown in Fig.~\ref{fig:g0-function}.
The transparent closed form of the t-matrix obtained in Eq.~(\ref{eqn:tmat_spin}) which is essential for the following discussions was only possible due to the restrictions of the model mentioned in the introduction. A more general t-matrix formalism including momentum dependent scattering and the realistic layer dependence of surface state wave functions would require a high dimensional real space representation of the Green's functions and t-matrix as e.g. in Refs.~\cite{fu:12,derry:15,ruessmann:20}. This would require to solve the t- matrix equation and calculation of the resulting QPI spectrum within a fully numerical treatment that is beyond the present approach and intention of this work.

\subsection{transformation to helical basis}

The transformation to helical basis is accomplished by the unitary matrix $S_\bk$ in Eq.~(\ref{eqn:unmat}) which is identical for T,B.  The transformed scattering matrix
%
%---------------------------------------------------------------------------------------------------
\be
\tilt_{\bk\bk'}=S_\bk^\dg 
\;
t_{\bk\bk'}
\;
S_{\bk'},
\label{eqn:tmat_heli}
\ee
%---------------------------------------------------------------------------------------------------
%
is then explitcitly given in terms of the spin based solution Eq.~(\ref{eqn:tmat_spin}) by its elements for both T,B blocks as
%
%---------------------------------------------------------------------------------------------------
\be
\bl
\tilt_{++}&=\alpha^+_{++}t_s  +\alpha_{++}^-t_a;\;\;\; \tilt_{+-}=\alpha^-_{+-}t_s  +\alpha_{+-}^+t_a;
\\
\tilt_{--}&=\alpha^+_{--}t_s  -\alpha_{--}^-t_a;\;\;\; \tilt_{-+}=\alpha^-_{-+}t_s  +\alpha_{-+}^+t_a .
\label{eqn:tmatheli1}
\el
\ee
%---------------------------------------------------------------------------------------------------
%
The (T,B- independent) form factors $\alpha^\pm_{ij} (i,j=\pm)$  that are functions of  $(\phi_\bk,\theta_\bk,\phi_{\bk'},\theta_{\bk'})$ originating from the helical eigenstates are given explicitly  in Appendix \ref{sec:formfactor}.
Before proceeding to the QPI spectrum it is useful to discuss the general solution for the t-matrix 
in the simple case of the Born approximation (BA) corresponding to single scattering from the impurity, i.e. to first order in $V_c$ and $V_m$.
%
%%%%%%%%%%%%%%%%%%%%%% figure %%%%%%%%%%%%%%%%%%%%%%%%%%%%%%
\begin{figure}[t]
\begin{center}
\includegraphics[width=0.9960\columnwidth]{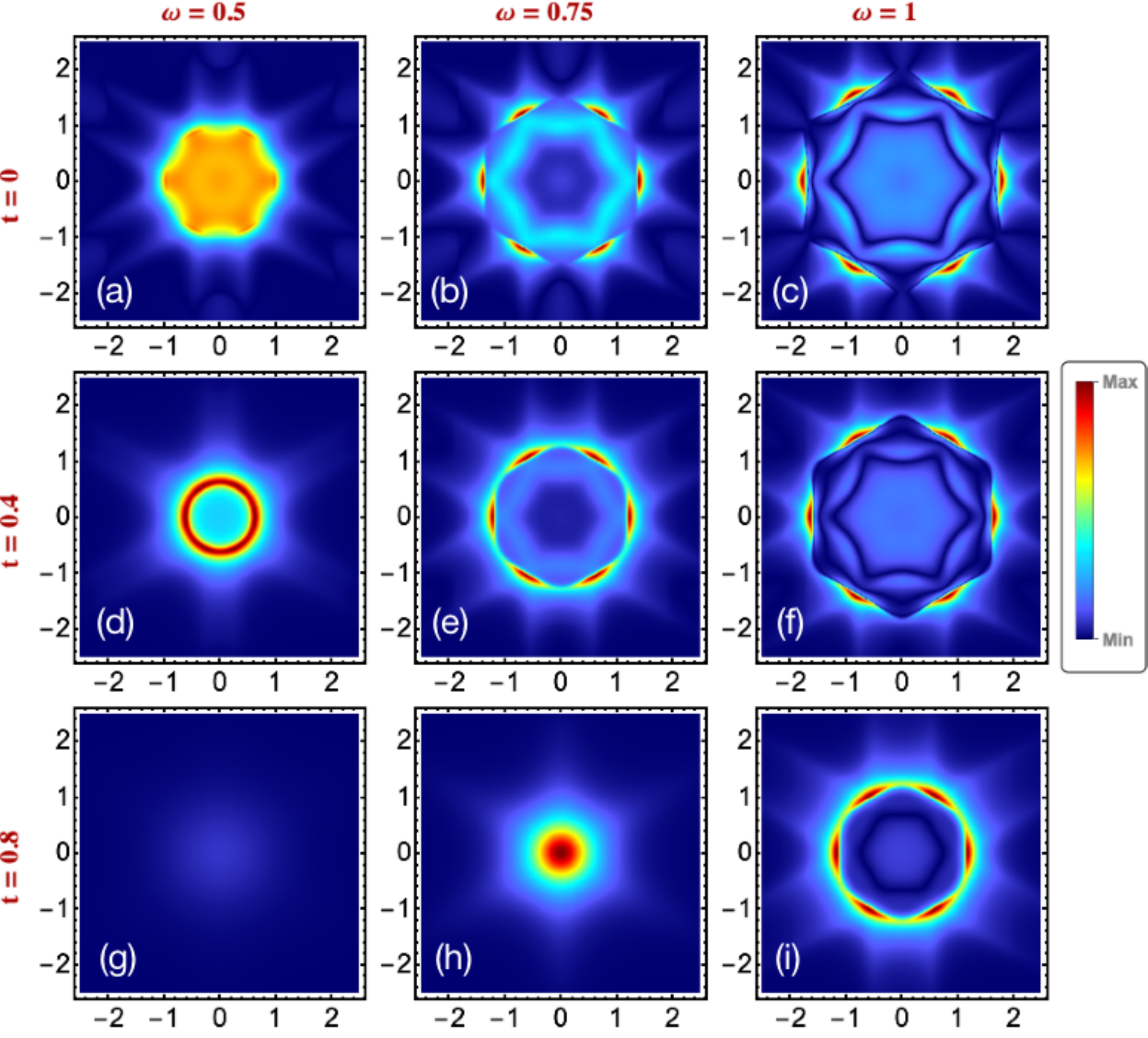}
\end{center}
\vspace{-0.5cm}
\caption{QPI spectrum in Born approximation $(Eq.~(\ref{eqn:Bornc}))$ for normal scattering $V_c\equiv 1$ only $(V_m=0)$. The images are for  for various bias voltages $eV=\omega$ (rows) and inter-surface tunnelling strength t (columns). When the latter is increased the typical QPI structure appear only for $\omega >t$ due to the vanishing of constant energy surfaces for smaller frequencies (Fig.~\ref{fig:FermiSurf}). The details of the QPI images are explained in Sec.~\ref{sec:numerical}.}
\label{fig:QPI_BAc}
\end{figure}
%%%%%%%%%%%%%%%%%%%%%%fig%%%%%%%%%%%%%%%%%%%%%%%%%%%%%%%%
%
\subsection{Born approximation and selection rules for scattering}
\label{sec:BA1}

In BA we have $t_s=V_c$ and $t_a=V_m$. Treating normal $(V_c)$ and magnetic $(V_m)$ scattering separately, we get the simple complementary results, respectively [c.f. Eq.~(\ref{eqn:formmat})]:
%
%---------------------------------------------------------------------------------------------------
\be
\bl
&
\tilt_{\bk\bk'}=V_c\tilde{\alpha}_{s\bk\bk'}=V_c
\left(
\begin{array}{cc}
\alpha_{++}^+& \alpha_{+-}^-\\[0.2cm]
\alpha_{-+}^-&\alpha_{--}^+
\end{array}
\right);\;\;\;
\text{and}\;\;\;
\\
&
\tilt_{\bk\bk'}=V_m\tilde{\alpha}_{a\bk\bk'}=V_m
\left(
\begin{array}{cc}
\alpha_{++}^-& \alpha_{+-}^+\\[0.2cm]
\alpha_{-+}^+&-\alpha_{--}^-
\end{array}
\right);\;\;\;
\el
\ee
%---------------------------------------------------------------------------------------------------
%
First we consider the limiting case of isotropic Dirac cones, i.e. vanishing warping term $\lambda\rightarrow 0$. Then $\phi_\bk=\frac{\pi}{2}$ and $\cph=\sph=\frac{\sqrt{2}}{2}$ and similar for $\bk'$. Explicitly we have
%
%---------------------------------------------------------------------------------------------------
\be
\bl
&
\tilt_{\bk\bk'}=\frac{V_c}{2}
\left(
\begin{array}{cc}
1+e^{i(\theta_\bk-\theta_{\bk'})}& -i(e^{i\theta_{\bk}}-e^{i\theta_{\bk'}})\\[0.2cm]
-i(e^{-i\theta_{\bk}}-e^{-i\theta_{\bk'}})&1+e^{-i(\theta_\bk-\theta_{\bk'})}
\end{array}
\right);
\\&
%\text{and}\;\;\;
\tilt_{\bk\bk'}=\frac{V_m}{2}
\left(
\begin{array}{cc}
1-e^{i(\theta_\bk-\theta_{\bk'})}& i(e^{i\theta_{\bk}}+e^{i\theta_{\bk'}})\\[0.2cm]
-i(e^{-i\theta_{\bk}}+e^{-i\theta_{\bk'}})&-(1-e^{-i(\theta_\bk-\theta_{\bk'})})
\end{array}
\right)
.
\el
\ee
%---------------------------------------------------------------------------------------------------
%
In both cases the symmetry relation $\tilt^\dg_{\bk\bk'}=\tilt_{\bk'\bk}$ is fulfilled. It is instructive to look at the special examples of forward 
scattering $(f): \theta_{\bk'}=\theta_{\bk}$ and backward scattering $(b): \theta_{\bk'}=\theta_{\bk}\pm\pi$.  For charge scattering this means
%
%---------------------------------------------------------------------------------------------------
\be
\bl
&
(f): \tilt_{\bk\bk'}=V_c
\left(
\begin{array}{cc}
1& 0\\
0 &1
\end{array}
\right);\;\;\;
\text{and}
\\&
(b): \tilt_{\bk\bk'}=V_c
\left(
\begin{array}{cc}
0 & -ie^{i\theta_{\bk}}\\[0.2cm]
ie^{-i\theta_{\bk'}}&0
\end{array}
\right)
\label{eqn:BAcfb}
\el
\ee
%---------------------------------------------------------------------------------------------------
%
while for purely magnetic scattering the complementary result is 
%
%---------------------------------------------------------------------------------------------------
\be
\bl
&
(f): \tilt_{\bk\bk}=V_m
\left(
\begin{array}{cc}
0& ie^{i\theta_{\bk}}\\
 -ie^{-i\theta_{\bk}}&0
\end{array}
\right);\;\;\;
\text{and}
\\&
(b): \tilt_{\bk\bk'}=V_m
\left(
\begin{array}{cc}
1 & 0\\[0.2cm]
0&-1
\end{array}
\right).
\label{eqn:BAmfb}
\el
\ee
%---------------------------------------------------------------------------------------------------
%
This demonstrates that backward scattering between same helicity states is forbidden for normal impurity scattering while it is allowed in the case of magnetic impurities. This will have direct influence on the QPI spectrum (Sec.\ref{sec:QPI}).\\

It is also useful to derive the explicit expressions of form factors for forward 
$(f;\; \theta_\bk=\theta_{\bk'})$ and backward 
$(b;\; \theta_\bk=\theta_{\bk'}+\pi)$
 scattering including the effect of finite warping $(\lambda\neq0)$ when $\phi_\bk\neq 0$ in general. Using $\phi_{\bk'}=\phi_\bk$ 
 in case $(f)$ and the identities $\spr=\cph;\;\; \cpr=\sph$ in case $(b)$  
we nevertheless get for normal scattering $V_c$ $(V_m=0)$ the {\it identical} result as in Eq.~(\ref{eqn:BAcfb}).
Thus the spin mixing angle $\phi_\bk$ of the warped case does not enter in the $f,b$ amplitudes and the result is identical to the isotropic Dirac case. In particular this means that the backscattering for equal helicity states remains forbidden even for the eigenstates in the warped Dirac cone. This is natural since the normal scattering does not react to the changed spin texture caused by the warping term.\\

On the other hand for exchange scattering $V_m$ $(V_c=0)$ the change in spin texture will be important and therefore the result for the scattering matrix  will depend on the  spin mixing angle $\phi_\bk$ caused by warping. In BA we obtain from the only nonvanishing second term in Eq.~(\ref{eqn:tmat_short})
%
%---------------------------------------------------------------------------------------------------
\be\no
(f): \tilt_{\bk\bk}=V_m
\left(
\begin{array}{cc}
\cos\phi_\bk& ie^{i\theta_{\bk}}\sin\phi_\bk\\
 -ie^{-i\theta_{\bk}}\sin\phi_\bk&\cos\phi_\bk
\end{array}
\right);\;\;\;
\ee
%---------------------------------------------------------------------------------------------------
and
%---------------------------------------------------------------------------------------------------
\be
(b): \tilt_{\bk\bk'}=V_m
\left(
\begin{array}{cc}
\sin\phi_\bk & -ie^{i\theta_\bk}\cos\phi_\bk\\[0.2cm]
ie^{i\theta_{\bk'}}\cos\phi_\bk&-\sin\phi_\bk
\end{array}
\right).
\label{eqn:BAmfb_warp}
\ee
%---------------------------------------------------------------------------------------------------
%
In the isotropic cone limit $(\lambda\rightarrow 0,\;\; \phi_\bk=\frac{\pi}{2})$ we recover the result of Eq.~(\ref{eqn:BAmfb}). We note that in the warped case when $\phi_\bk\neq \frac{\pi}{2}$ in general the forward scattering for equal helicity states no longer vanishes as in Eq.~(\ref{eqn:BAmfb}) due to the presence of the perpendicular spin component  $\sim\cos\phi_\bk$   of the eigenstates [Eq.~(\ref{eqn:spinpol})].
%
%%%%%%%%%%%%%%%%%%%%%% figure %%%%%%%%%%%%%%%%%%%%%%%%%%%%%%
\begin{figure}[t]
%\vspace{0.5cm}
\begin{center}
\includegraphics[width=0.970\columnwidth]{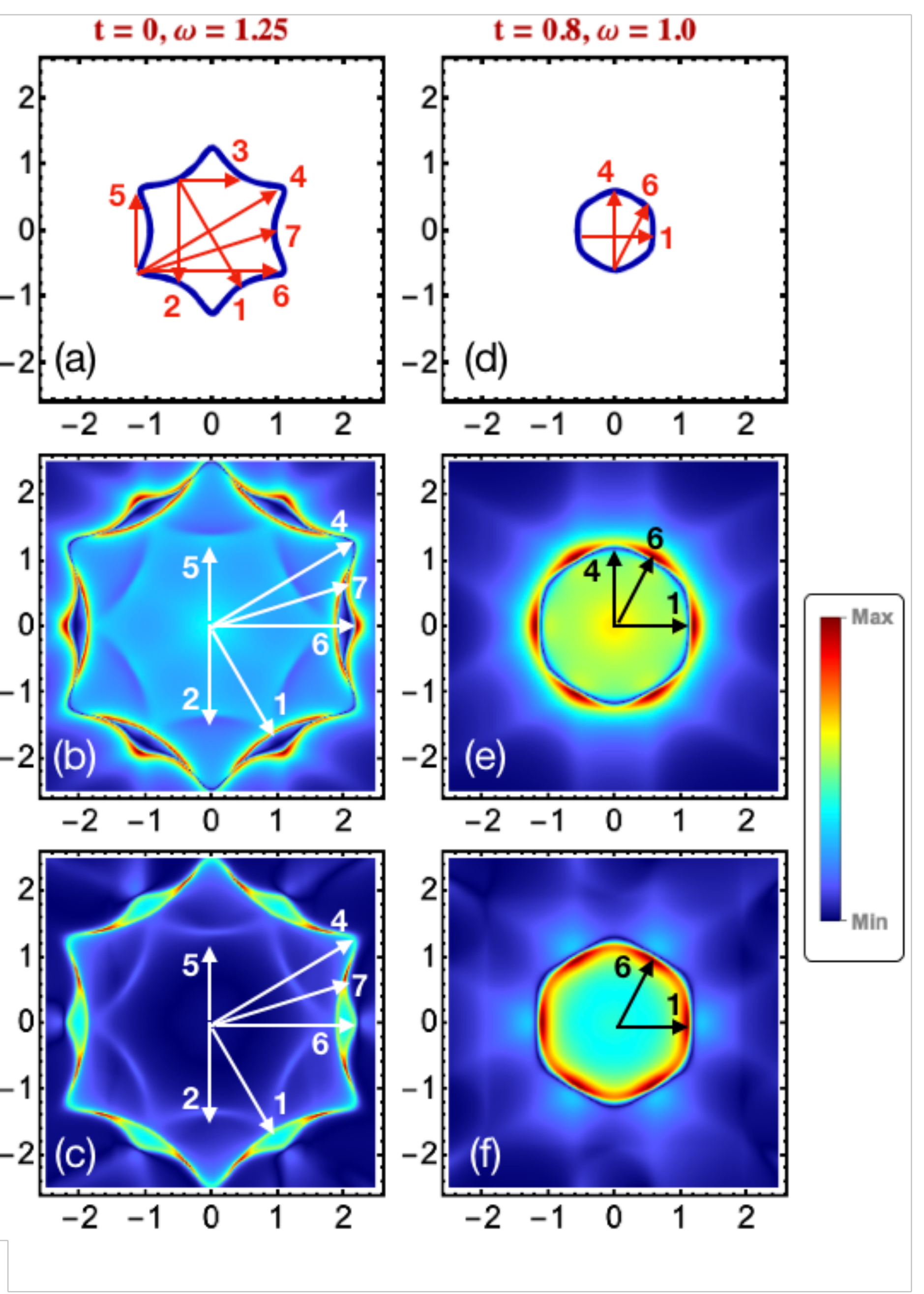}
\end{center}
\vspace{-0.75cm}
\caption{Comparison of characteristic wave vectors $\bq_i (i=1,7)$ conntecting Fermi surface tips and dents (a,d) with 
prominent features in the QPI spectrum for normal charge $(V_c)$ (b,e) and magnetic $(V_m)$ (c,f) impurity scattering. Most $\bq_i$ can be clearly associated in FS and QPI images, only $\bq_3$ has too weak intensity. Note that intensity for back-scattering vectors $\bq_1,\bq_4$ passing through the origin is suppressed for normal impurity scattering (b,e) while it is finite for magnetic scattering (c,f).}
\label{fig:FS-QPI-comp}
\end{figure}
%%%%%%%%%%%%%%%%%%%%%%fig%%%%%%%%%%%%%%%%%%%%%%%%%%%%%%%%
%
\section{Quasiparticle interference spectrum}
\label{sec:QPI}

The spectral  Fourier component of the surface density modulation visible by STM-QPI is given by~\cite{capriotti:03, akbari:11,akbari:13,*akbari:13b}
%
%---------------------------------------------------------------------------------------------------
\be
\bl
\delta N_T(\bq,\omega)
&=
-\frac{1}{\pi}
{\rm Im}
\Big[
\Lambda_T(\bq,\om)
\Big]_{\om\rightarrow \omega+i\delta};
\\
\Lambda_T(\bq,\om)
&=\frac{1}{N}\sum_\bk 
{\rm tr}_\sigma
\Big[
G_\bk t_{\bk\bk'} G_{\bk'}
\Big]_{TT}
\\
&=:
 \frac{1}{N}\sum_\bk 
 {\rm tr}_\sigma
 \Big[
 X_{\bk\bk'}
 \Big]_{TT}.
 \hspace{1cm}
\label{eqn:QPI}
\el
\ee
%---------------------------------------------------------------------------------------------------
%
Here $\omega=eV$ with $V$ denoting the variable STM-tip bias voltage (Fig.~\ref{fig:film}). It is assumed that the tip is placed to
the top (T) surface, therefore only the spin-trace over the T-block of t-matrix and Green's function product has to
be performed. Explicitly we have two contributions given by
%
%---------------------------------------------------------------------------------------------------
\be
[X_{\bk\bk'}]_{TT}
=G_\bk^{TT}t_{\bk\bk'}G_{\bk'}^{TT}+
G_\bk^{TB}t_{\bk\bk'}G_{\bk'}^{BT}.
\label{eqn:QPI_kernel}
\ee
%---------------------------------------------------------------------------------------------------
%

 As it stands everything is still written on spin basis. Since the Green's functions are diagonal
in the helical bases one should transform to the latter, using Eq.~(\ref{eqn:tmat_heli}) and the similar transformation for the Green's function:
\bea
\tG^{\alpha\alpha'}_{\bk\bk'}=S_\bk^\dg G^{\alpha,\alpha'}_{\bk\bk'}S_{\bk'},
\label{eqn:Green_heli}
\eea
where $\alpha,\alpha'=T,B$. For the t-matrix we restricted to T,B diagonal elements only, neglecting the impurity {\it scattering}
between the bottom and top surfaces. This cannot be done for the Green's functions because the inter-surface tunnelling or hybridisation is essential for the low energy surface states of the TI film. Therefore the T,B nondiagonal Green's function elements have to be kept. This leads to two terms in the kernel Eq.~(\ref{eqn:QPI_kernel}) of the QPI spectrum involving scattering at the top (first term), and via hybridisation of surface states also at the bottom (second term) surface. They are schematically shown in Fig.~\ref{fig:film}.\\
%
%%%%%%%%%%%%%%%%%%%%%% figure %%%%%%%%%%%%%%%%%%%%%%%%%%%%%%
\begin{figure}[t]
\begin{center}
\includegraphics[width=0.990\columnwidth]{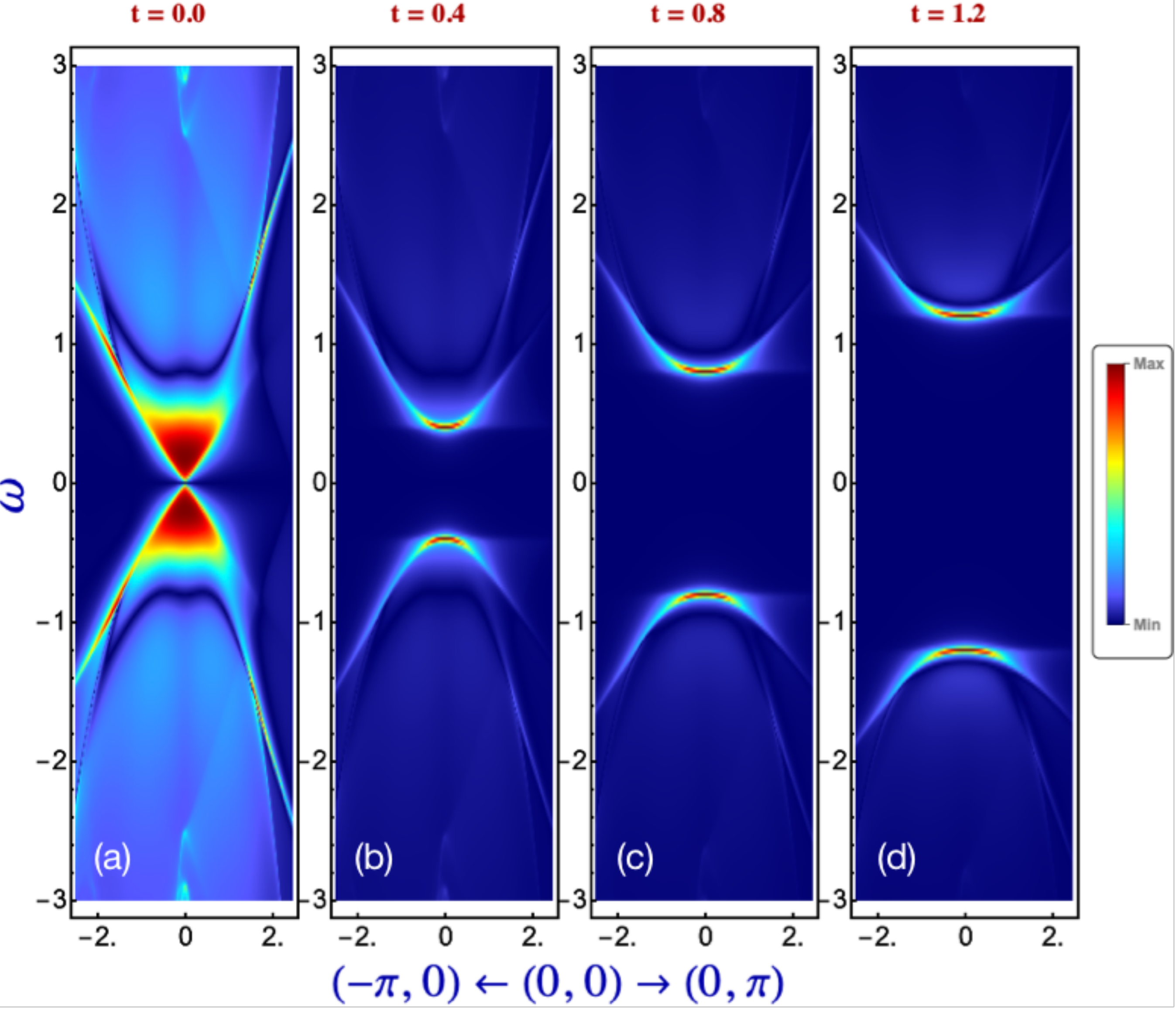}
\end{center}
\vspace{-0.35cm}
\caption{The dispersion characteristics of QPI spectrum along symmetry directions of the SBZ $\bar{\text{M}}\bar{\Gamma}$ $(-\pi,0)$ and $\bar{\Gamma}\bar{\text{K}}$ $(0,\pi)$ as function of inter-surface tunnelling strength. The gap opening of the Dirac cones can clearly be followed in the QPI dispersion.}
\label{fig:QPIdisp}
\end{figure}
%%%%%%%%%%%%%%%%%%%%%%fig%%%%%%%%%%%%%%%%%%%%%%%%%%%%%%%%
%

The T,B Green's function blocks are all diagonal in the helical eigenstate basis. Therefore it is economic to transform Eq.~(\ref{eqn:QPI}) to this basis using Eqs.~(\ref{eqn:tmat_heli},~and~\ref{eqn:Green_heli}): The latter leads to
%
%---------------------------------------------------------------------------------------------------
\be
\bl
&
\tG^{TT,BB}_{\bk}=
\left(
\begin{array}{cc}
\tA_{\pm\bk} & 0\\[0.2cm]
0&\tA_{\mp\bk}
\end{array}
\right);\;\;\;
\text{and}
\\
&
\tG^{TB}_{\bk}=\tG^{BT}_{\bk}=
\left(
\begin{array}{cc}
\tB_{0\bk} & 0\\[0.2cm]
0&\tB_{0\bk}
\end{array}
\right).
\el
\ee
%---------------------------------------------------------------------------------------------------
%
Where TT and BB correspond to upper and lower sign, respectively. Furthermore 
%
%---------------------------------------------------------------------------------------------------
\bea
\tA_{\pm\bk}=\frac{\om\pm E_\bk}{[(\om)^2-\tE^2_\bk]};\;\;\;
\tB_{0\bk}=\frac{t}{[(\om)^2-\tE^2_\bk]}.
\eea
%---------------------------------------------------------------------------------------------------
%
Using the diagonal Green's functions the kernel in Eq.~(\ref{eqn:QPI_kernel}) may be transformed to helical basis as
%
%---------------------------------------------------------------------------------------------------
\bea
\no
[X_{\bk\bk'}]_{TT}
&=&
S_\bk[\tG^{TT}_{\bk}\tilt_{\bk\bk'}\tG^{TT}_{\bk'}]S^\dg_{\bk'}+
S_\bk[\tG^{TB}_{\bk}\tilt_{\bk\bk'}\tG^{BT}_{\bk'}]S^\dg_{\bk'}
\\
&=&
[X_{\bk\bk'}]^{(1)}_{TT}+[X_{\bk\bk'}]^{(2)}_{TT}
.
\eea
%---------------------------------------------------------------------------------------------------
%
Evaluating the two terms
and inserting into Eq.~(\ref{eqn:QPI}) we obtain, correspondingly for $(1)$ and $(2)$:
%
%---------------------------------------------------------------------------------------------------
\bea
\bl
\nonumber
\Lambda_T^{(1)}(\bq,\om)
=&
\\
&\hspace{-1.1cm}
\frac{1}{N}
\sum_\bk
\Big[
\alpha_{--}^+\tilt_{++}\tA_{+\bk}\tA_{+\bk'}
+
\alpha_{++}^+\tilt_{--}\tA_{-\bk}\tA_{-\bk'}
\\
&
-
\alpha_{-+}^-\tilt_{+-}\tA_{+\bk}\tA_{-\bk'}-
\alpha_{+-}^-\tilt_{-+}\tA_{-\bk}\tA_{+\bk'}
\Big];
\\
\Lambda_T^{(2)}(\bq,\om)
=
&\frac{1}{N}\sum_\bk\tB_{0\bk}\tB_{0\bk'}
\times
\\
&
\hspace{-1cm}
\Big[
\alpha_{--}^+\tilt_{++}+
\alpha_{++}^+\tilt_{--} -
\alpha_{-+}^-\tilt_{+-}-
\alpha_{+-}^-\tilt_{-+}
\Big],
\label{eqn:QPI_general1}
\el
\\
\eea
%---------------------------------------------------------------------------------------------------
%
%
%---------------------------------------------------------------------------------------------------
%\begin{widetext}
%\bea
%\bl
%\Lambda_T^{(1)}(\bq,\om)
%&=\frac{1}{N}\sum_\bk
%[\alpha_{--}^+\tilt_{++}\tA_{+\bk}\tA_{+\bk'}+
%\alpha_{++}^+\tilt_{--}\tA_{-\bk}\tA_{-\bk'}-
%\alpha_{-+}^-\tilt_{+-}\tA_{+\bk}\tA_{-\bk'}-
%\alpha_{+-}^-\tilt_{-+}\tA_{-\bk}\tA_{+\bk'}];
%\\
%\Lambda_T^{(2)}(\bq,\om)
%&=\frac{1}{N}\sum_\bk\tB_{0\bk}\tB_{0\bk'}
%[\alpha_{--}^+\tilt_{++}+
%\alpha_{++}^+\tilt_{--} -
%\alpha_{-+}^-\tilt_{+-}-
%\alpha_{+-}^-\tilt_{-+}]
%\label{eqn:QPI_general1}.
%\el
%\eea
%\end{widetext}
%---------------------------------------------------------------------------------------------------
%
with the total result  
$$\Lambda_T(\bq,\om)=\Lambda_T^{(1)}(\bq,\om)+\Lambda_T^{(2)}(\bq,\om).$$
Here $\tilt_{\kappa\kappa'}(\om)$ is given by Eqs.~(\ref{eqn:tmatheli1}~and~\ref{eqn:tmat_spin}).

%---------------------------------------------------------------------------------------------------
%---------------------------------------------------------------------------------------------------
%---------------------------------------------------------------------------------------------------
\subsection{Born approximation for QPI}
\label{sec:BA2}

To obtain a better insight in the general expression for the QPI spectrum we first simplify to the case of Born approximation
with only single scattering events at the impurity included. In this case we have $t^{T,B}_s=V_c$ and $t^{T,B}_a=V_m$.  Again we treat normal (c) and magnetic (m) scattering separately. Inserting into Eq.~(\ref{eqn:QPI_general1}) and using the explicit form factors and expressions for $\tA_{\pm\bk}$ we obtain for normal scattering, after considerable algebra:
%
%---------------------------------------------------------------------------------------------------
\be
\bl
&\Lambda_c(\bq,\om)
=
\\
&\hspace{1cm}
\frac{2V_c}{N}\sum_\bk
\frac{m_+^c(\bk\bk')[(\om)^2+t^2]+m_-^c(\bk\bk')E_\bk E_{\bk'}}
{[(\om)^2-\tE^2_\bk][(\om)^2-\tE^2_{\bk'}]},
\el
\ee
with
\be
\bl\nonumber
m^c_+(\bk\bk')
&=
1;
\\
m^c_-(\bk\bk')&=\cos\phi_\bk\cos\phi_{\bk'}+\sin\phi_\bk\sin\phi_{\bk'}\cos(\theta_\bk-\theta_{\bk'}),
\label{eqn:Bornc}
\el
\ee
%---------------------------------------------------------------------------------------------------
%
and likewise for magnetic scattering
%\begin{widetext}
%
%---------------------------------------------------------------------------------------------------
\be
\bl
&
\Lambda_m(\bq,\om)=
\\
&\hspace{0.cm}
\frac{2V_m}{N}\sum_\bk
\frac{\om[m^m_+(\bk)E_\bk +m^m_+(\bk')E_{\bk'}]+m^m_-(\bk\bk')E_\bk E_{\bk'}}
{[(\om)^2-\tE^2_\bk][(\om)^2-\tE^2_{\bk'}]},
\el
\ee
with
\be\nonumber
\bl
&
m^m_+(\bk)=
\cos\phi_\bk;\;\;\;m^m_+(\bk')=\cos\phi_{\bk'};
\\
&m^m_-(\bk\bk')
=-i\sin\phi_\bk\sin\phi_{\bk'}\sin(\theta_\bk-\theta_{\bk'}).
\el
\ee
%---------------------------------------------------------------------------------------------------
%\end{widetext}
%
Note that in these expression the gapped cone energies $\tE_\bk$ of the film appear in
the denominator while the ungapped energies $E_\bk$ of isolated surfaces remain in the numerator.
It is instructive to consider the limit of the isotropic Dirac cone when 
$\lambda\rightarrow 0$ and then $\phi_\bk=\frac{\pi}{2}$. Inserting  this into the form factors $m^\pm_{c,m}(\bk,\bk')$ the above formulas then simplify to 
%
%---------------------------------------------------------------------------------------------------
\be
\bl
\Lambda_c(\bq,\om)=
&
\frac{2V_c}{N}\sum_\bk
\frac{(\om)^2+t^2+\cos(\theta_\bk-\theta_{\bk'})E_\bk E_{\bk'}}
{[(\om)^2-\tE^2_\bk][(\om)^2-\tE^2_{\bk'}]};
\\
\Lambda_m(\bq,\om)=
&
i
\frac{2V_m}{N}\sum_\bk
\frac{\sin(\theta_\bk-\theta_{\bk'})E_\bk E_{\bk'}}
{[(\om)^2-\tE^2_\bk][(\om)^2-\tE^2_{\bk'}]}.
\hspace{0.8cm}
\el
\ee
%---------------------------------------------------------------------------------------------------
%
For the decoupled case $(t=0,\; \tE_\bk=E_\bk)$ these expressions can also be directly obtained  from Eq.~(\ref{eqn:QPI}) by using the single- surface Green's functions of Eq.~(\ref{eqn:Green_single2}) for the isotropic case. Due to the sign change of the numerator of $\Lambda_m$  under $\bk\leftrightarrow \bk'$ which results from the helical spin-locking we have $\Lambda_m\equiv 0$ in Born approximation. For a finite result one has to use the full t-matrix theory.
%
%%%%%%%%%%%%%%%%%%%%%% figure %%%%%%%%%%%%%%%%%%%%%%%%%%%%%%
\begin{figure}[t]
\begin{center}
\includegraphics[width=0.990\columnwidth]{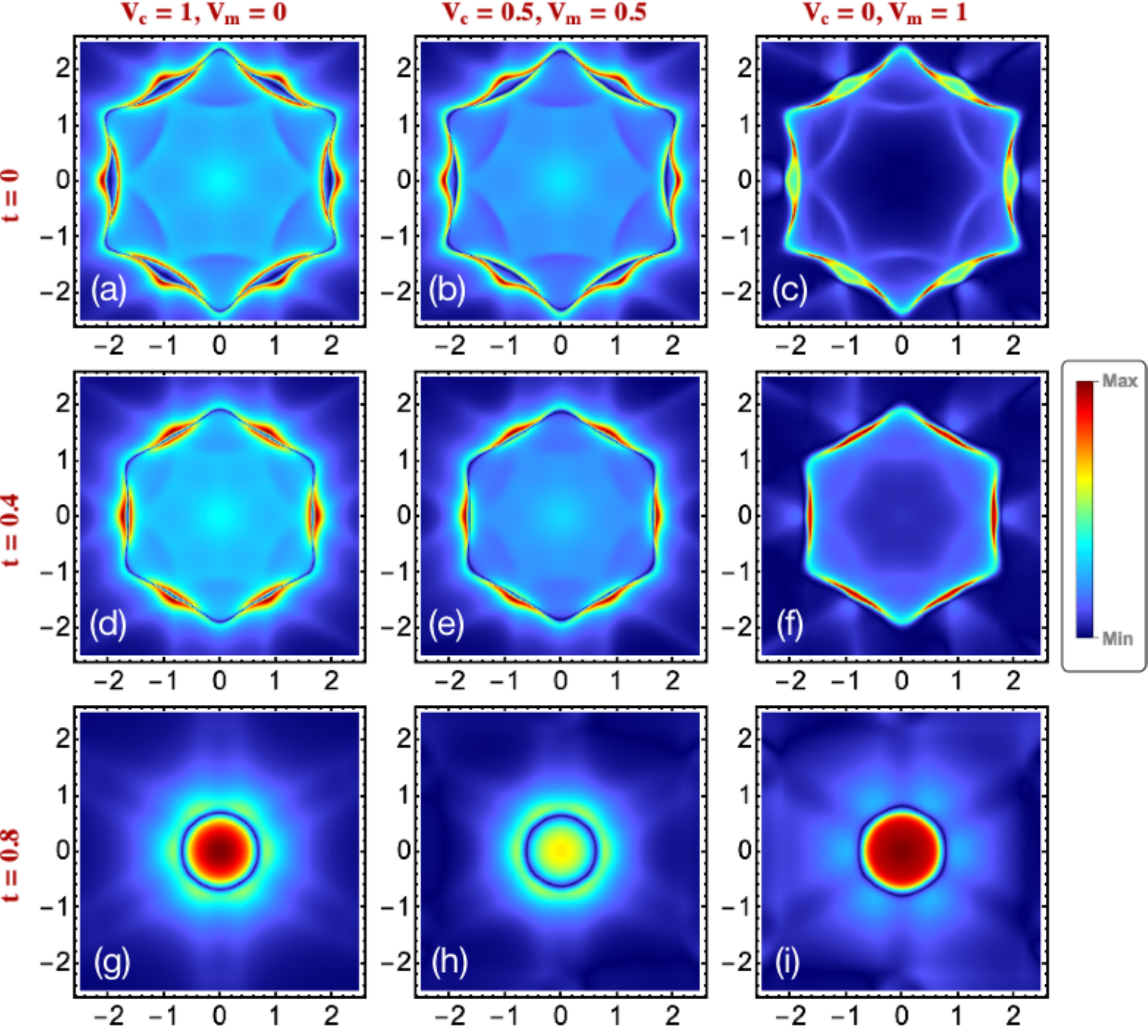}
\end{center}
\vspace{-0.35cm}
\caption{QPI spectrum of thin film for frequency $\omega =1.25$ for various tunnelling 
strengths $t$ (columns) and impurity scattering strengths $(V_c,V_m)$ (rows). }
\label{fig:QPI_tmat}
\end{figure}
%%%%%%%%%%%%%%%%%%%%%%fig%%%%%%%%%%%%%%%%%%%%%%%%%%%%%%%%
%
\subsection{Full t-matrix expressions QPI spectrum}

For computational convenience we also give here the rather lengthy explicit expression for the QPI spectrum including the general t-matrix elements that will mostly be used in the numerical calculations. The total QPI spectrum, equal to the sum $\Lambda_T(\bq,\om)=\Lambda_T^{(1)}(\bq,\om)+\Lambda_T^{(2)}(\bq,\om)$ from Eq.~(\ref{eqn:QPI_general1}) is then obtained as
%
%---------------------------------------------------------------------------------------------------
%\begin{widetext}
%\bea
%\Lambda_T(\bq,\om)&&=
%\frac{1}{N}\sum_\bk\tilt_{++}(\om)\frac
%{\alpha^+_{--}[(\om+E_\bk)(\om+E_{\bk'})+t^2]}
%{[(\om)^2-\tE^2_\bk][(\om)^2-\tE^2_{\bk'}]}+
%\frac{1}{N}\sum_\bk\tilt_{--}(\om)\frac
%{\alpha^+_{++}[(\om-E_\bk)(\om-E_{\bk'})+t^2]}
%{[(\om)^2-\tE^2_\bk][(\om)^2-\tE^2_{\bk'}]}\nonumber\\
%&&
%-\frac{1}{N}\sum_\bk\tilt_{+-}(\om)\frac
%{\alpha^-_{-+}[(\om+E_\bk)(\om-E_{\bk'})+t^2]}
%{[(\om)^2-\tE^2_\bk][(\om)^2-\tE^2_{\bk'}]}-
%\frac{1}{N}\sum_\bk\tilt_{-+}(\om)\frac
%{\alpha^-_{+-}[(\om-E_\bk)(\om+E_{\bk'})+t^2]}
%{[(\om)^2-\tE^2_\bk][(\om)^2-\tE^2_{\bk'}]}.
%\nonumber\\
%\eea
%\end{widetext}
%---------------------------------------------------------------------------------------------------
%
%---------------------------------------------------------------------------------------------------
\be
\bl
&\Lambda_T(\bq,\om)
=
\frac{1}{N}\sum_\bk
\frac
{1}
{[(\om)^2-\tE^2_\bk][(\om)^2-\tE^2_{\bk'}]}
\times
\hspace{0.5cm}
\\
&\hspace{2.cm}
\Bigg[
\tilt_{++} %(\om)
\alpha^+_{--}
\Big(
(\om+E_\bk)(\om+E_{\bk'})+t^2
\Big)
\\
&\hspace{2cm}
+
\tilt_{--} %(\om)
\alpha^+_{++}
\Big(
(\om-E_\bk)(\om-E_{\bk'})+t^2
\Big)
\\
&\hspace{2cm}
-
\tilt_{+-} %(\om)
\alpha^-_{-+}
\Big(
(\om+E_\bk)(\om-E_{\bk'})+t^2
\Big)
\\
&\hspace{2cm}
-
\tilt_{-+} %(\om)
\alpha^-_{+-}
\Big(
(\om-E_\bk)(\om+E_{\bk'})+t^2
\Big)
\Bigg].
\label{eqn:QPI_closed}
\el
\ee
%---------------------------------------------------------------------------------------------------
%
%
In each term the first part describes the QPI contribution on the top surface due to scattering on top surface while the second part  proportional to $t^2$ represents the QPI contribution on the top surface due to scattering on the {\it bottom} surface. Obviously this term can only be present when there is tunnelling (described by $G_{TB},G_{BT}$ in Fig.~\ref{fig:film}) between the T,B  surface states, therefore it vanishes for decoupled surfaces. In these equations the expressions for the full t-matrix elements $\tilt_{\kappa\kappa'}(\om)$  in helical presentation are obtained from Eq.~(\ref{eqn:tmatheli1}). They are composed of the irreducible $t_{s,a}$ matrix elements in Eq.~(\ref{eqn:tmat_spin}) and the form factors $\alpha_{\kappa\kappa'}^\pm$ of Appendix \ref{sec:formfactor}.

\section{Numerical results and discussion}
\label{sec:numerical}

We now discuss the numerical results for the QPI spectrum under systematic variation of bias voltage $\omega=eV$, inter-surface tunnelling $t$ of the thin film and normal $(V_c)$ and magnetic $(V_m)$ impurity scattering potential (assumed identical on both surfaces). We mostly use the full t-matrix approximation as given in closed form by Eq.~(\ref{eqn:QPI_closed}) in  this section except when stated otherwise.  In Fig.~\ref{fig:QPI_BAc}, we show an overview over the images for normal $V_c$- type scattering  where the inter-surface tunnelling $t$ varies along the columns and the bias voltage $eV=\omega$ (measured from the Dirac point) along the rows. Here for once the Born approximation [Eq.~(\ref{eqn:Bornc})] is employed for comparison.  Generally in QPI images the Fermi surface for a given $\omega$ is reproduced , with a doubling of the FS radius. However, the intensities at wave vectors connecting special Fermi surface points like tips and dents may be strongly enhanced or depressed. The overall extension of the QPI images increases with $\omega$, the distance from the Dirac point,  according to the diameter of the cut $\omega=E_\bk$ (or $\omega=\tE_\bk$)  through the warped cone which gives the snowflake FS.
The  shape of the pattern and its increasing radius with $\omega$ is clearly seen in the first row representing the isolated surfaces $(t=0)$.  When the tunnelling is turned on (2$^{\rm nd}$ and 3$^{\rm rd}$ row) the low energy spectrum is gapped and therefore around $\omega\simeq t$ (i.f. we use $\omega,t\geq 0$)  the QPI image will be strongly modified: The radius shrinks and the anisotropic `snowflakes' character is reduced leading to a more isotropic image. This is completely in accordance with the evolution of the Fermi surfaces in Fig.~\ref{fig:FermiSurf}. Therefore QPI investigation of thin films can give full information on the thickness (tunnelling) dependence of low energy quasiparticles close to the gap threshold. For larger energies (voltages) $\omega \gg t$ above the gap the QPI image approaches that of the isolated surfaces (c.f. the two top right figures).

A detailed comparison of Fermi surface shape (first row a,d) and according QPI image is presented in Fig.~\ref{fig:FS-QPI-comp}(a-f) for normal (second row b,e) as well as magnetic (third row c,f) scattering mechanism. Similar as in Refs.~\cite{lee:09,zhang:09} we can identify characteristic wave vectors 
$\bq_i$ (i=1-7) in Fig.~\ref{fig:FS-QPI-comp}(a,d) connecting special points where the azimuthal group velocity vanishes (tips and dents of the snowflake) and should therefore figure prominently in the QPI image. Indeed most of the $\bq_i$ vectors (with the exception of $\bq_3$) can be identified in Figs.~\ref{fig:FS-QPI-comp}(b,c) for the isolated surface corresponding either to
large (dark red) intensity or low and vanishing (deep blue) intensity areas with arrows representing  the characteristic wave vectors pointing to them. Some of the latter in Fig.~\ref{fig:FS-QPI-comp} correspond to the forbidden backscattering as they are connected by   $\bq_1, \bq_4$ scattering vectors that pass through the origin such that $\bk'=-\bk$. These regions are, however, quite narrow  because of the vicinity of (equivalent) closeby  allowed $\bq_6,\bq_7$ vectors that are associated with large intensities. The  $\bq_2,\bq_5$ scattering vectors can also be identified though less prominently. The overall pattern for magnetic scattering in (c) looks quite similar  although the pointwise intensities are largely different. In particular for magnetic impurities the backscattering vectors  $\bq_1, \bq_4$  are allowed (Secs.~\ref{sec:BA1},\ref{sec:BA2}) and are also associated with finite QPI spectral intensity.
%
%%%%%%%%% figure %%%%%%%%%%%%%%%%%%%%%%%%%%%%%%
\begin{figure}[t]
%\vspace{0.5cm}
\begin{center}
\includegraphics[width=1.02\linewidth]{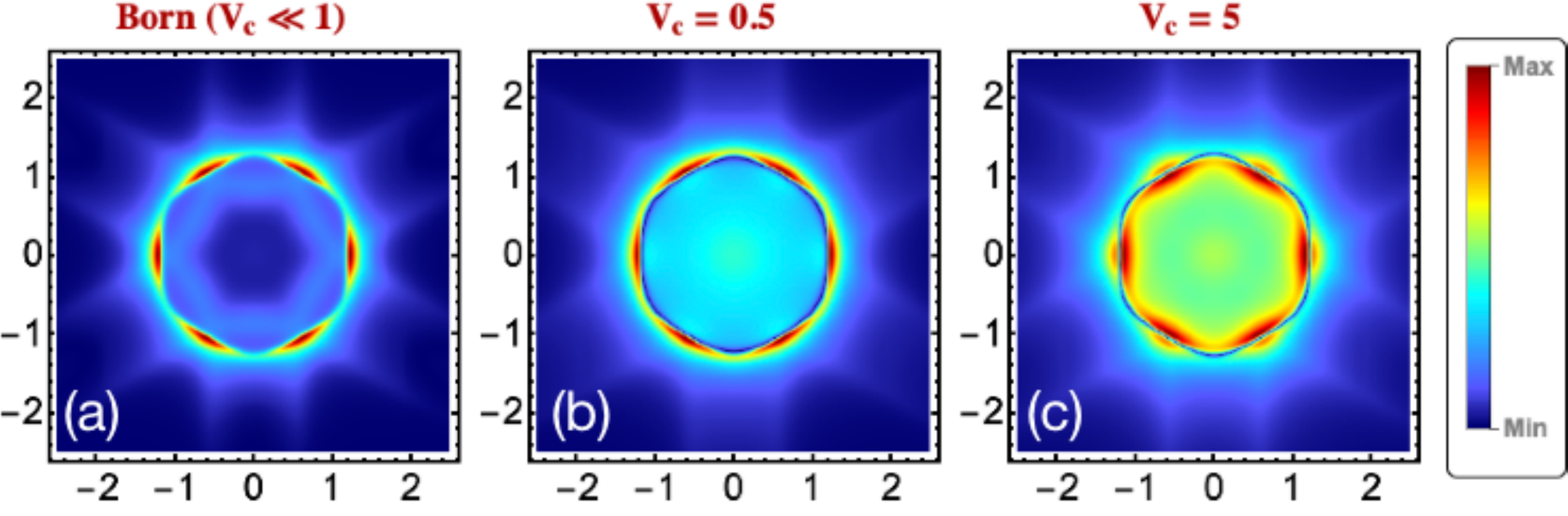}
\end{center}
\vspace{-0.5cm}
\caption{The evolution of QPI pattern from small to large normal scattering potential
 $V_c$ at the energy $\omega=0.75$ for the case of $t=0.4$.}
\label{fig:Vc-evolution}
\end{figure}
%%%%%%%%%%%%%%%%%%%%%%fig%%%%%%%%%%%%%%%%%%%%%%%%%%%%%%%%
%
There is another weak, though noteworthy peculiar feature of the QPI image: The single snowflake of the FS experiences a doubling in the QPI image. The outer one has the same orientation as the FS and is associated mostly with scattering vectors close to the group $\bq_1,\bq_4,\bq_6,\bq_7$ and their equivalents while the inner smaller image is rotated by $\pi/3$ with respect to the FS and is mostly associated with scattering vectors close to $\bq_2,\bq_5$ and equivalents.

When the inter-surface tunnelling $t$ is turned on and the gap is opened at the Dirac point, at first for $\omega \gg |t|$ there is no drastic change in QPI image characteristics as seen from the upper right corner of Fig.~\ref{fig:QPI_BAc}, except that the characteristic scattering vector lenghts $|\bq_i|$ are shrinking. This becomes more dramatic when $\omega$ approaches the (half-) gap size $|t|$ and the FS structure becomes more circular (i). Furthermore the tunnelling opens backscattering channels close to the gap edge increasing the relative size of the high intensity areas (dark red).  As a result the QPI snowflake image gradually `melts' into a circular  drop shape where only the most prominent scattering vectors $\bq_6$,~and~$\bq_7$ can still be discerned. Their length is now close to $2k_F$ of the nearly spherical Fermi surface image [Figs.~\ref{fig:QPI_BAc}(d,i)~and~\ref{fig:FS-QPI-comp}(e,f)] corresponding to the parabolic surface dispersion close to the gap edge (c.f. Fig.~\ref{fig:FermiSurf}).
This gradual transformation of the QPI image from the six-pronged (double) `snowflakes' at zero or small tunnelling (isolated surfaces) to the  more isotropic, almost circular structures at large tunnelling (few quintuple layer films) should be worthwhile to investigate experimentally with STM-QPI technique for a number of reasons: Firstly it has not yet been observed in ARPES experiments \cite{zhang:10}. Furthermore in QPI this transition contains direct visual evidence for the reappearance of backscattering, as function of increasing $t$, i.e. decreasing film thickness $d$ as clearly seen in Figs.~\ref{fig:QPI_BAc},\ref{fig:FS-QPI-comp}. This observation is central to the change of  surface state topology with thickness (witnessed also by the Berry phase of Eq.~(\ref{eqn:Berryfilm})).  For $d > 5$ QL the backscattering is largely forbidden and  QPI intensity  is strongly suppressed for backscattering vectors. For $d < 5$ QL it reappears and as a consequence the QPI image becomes more circular  (due to the 
gap opening) with more isotropic intensity distribution. Therefore this change demonstrates pointwise in k-space how
the backscattering reappears as function of film thickness. This is not directly possible by ARPES which  probes just single particle spectral function and also not by magnetotransport which can only indirectly conclude (via destruction of weak antilocalization \cite{taskin:12}) on this mechanism.\\

Instead of looking at the QPI image of $|\Lambda(q_x,q_y,\omega)|$ in the 2D SBZ as before we may consider the QPI `quasiparticle dispersion' defined by the complementary image $|\Lambda(\bq=\hbq,\omega)|$ in the $(q,~\omega)$-plane for fixed momentum direction $\hbq\parallel$ $\bar{\Gamma}\bar{\text{K}}$ or  $\bar{\Gamma}\bar{\text{M}}$  in the SBZ which should directly demonstrate the change of surface quasiparticle spectrum with tunnelling strength. The dispersion results for normal scattering are shown in Fig.~\ref{fig:QPIdisp}. For isolated surfaces ($t=0$) the Dirac cone dispersion can clearly be identified by the envelope high intensity region. For T,B surfaces connected by the tunnelling the gap $|t|$ opens progressively and the parabolic dispersion for $\omega\geq |t|$ is again seen in by a sharp prominent high intensity envelope. The destruction of the Dirac cone by the gap opening via inter-surface tunnelling has sofar been observed directly only in ARPES experiments for \bse~thin films between $1-6$ QL thickness~\cite{zhang:10} and it would be highly desirable to investigate this also by the complementary QPI method as proposed here. 
The existence of the Dirac cone dispersion for the isolated surfaces ($t=0$  in Fig.~\ref{fig:QPIdisp})  has indeed been demonstrated before with the QPI method~\cite{gomes:09,alpichshev:10, okada:11}.\\

In an alternative constant-$\omega$ presentation of QPI images $|\Lambda(q_x,q_y,\omega=eV)|$,
 we again keep the bias voltage fixed and change the tunnelling $t$ and scattering strengths $V_c$, and $V_m$.
This is presented in Fig.~\ref{fig:QPI_tmat}, where we show the images for a constant $\omega =1.25$ and as function of $t$ (columns) and relative normal $(V_c)$ to magnetic $(V_m)$ scattering strength (rows).
While the overall image features are similar, the substructure of the large intensity regions (dents) depends in a subtle way on the relative size of  $V_c$, and $V_m$.
Finally in Fig.~\ref{fig:Vc-evolution}, we show an example of the evolution of QPI spectrum for normal scattering from very low (BA) to very strong scattering strength. The global structure of the image is preserved but details show again subtle changes: The high intensity spots for small scattering merge into a ring for large $V_c$, but still keeping low intensity at the backscattering vectors. Furthermore the interior of the structure acquires more intensity presumably due to the increasing importance of multiple scattering processes with growing $V_c$ contained in the full t-matrix approach.\\
%\begin{widetext}
%
%%%%%%%%% figure %%%%%%%%%%%%%%%%%%%%%%%%%%%%%%
\begin{figure}[t]
\begin{center}
\includegraphics[width=0.990\columnwidth]{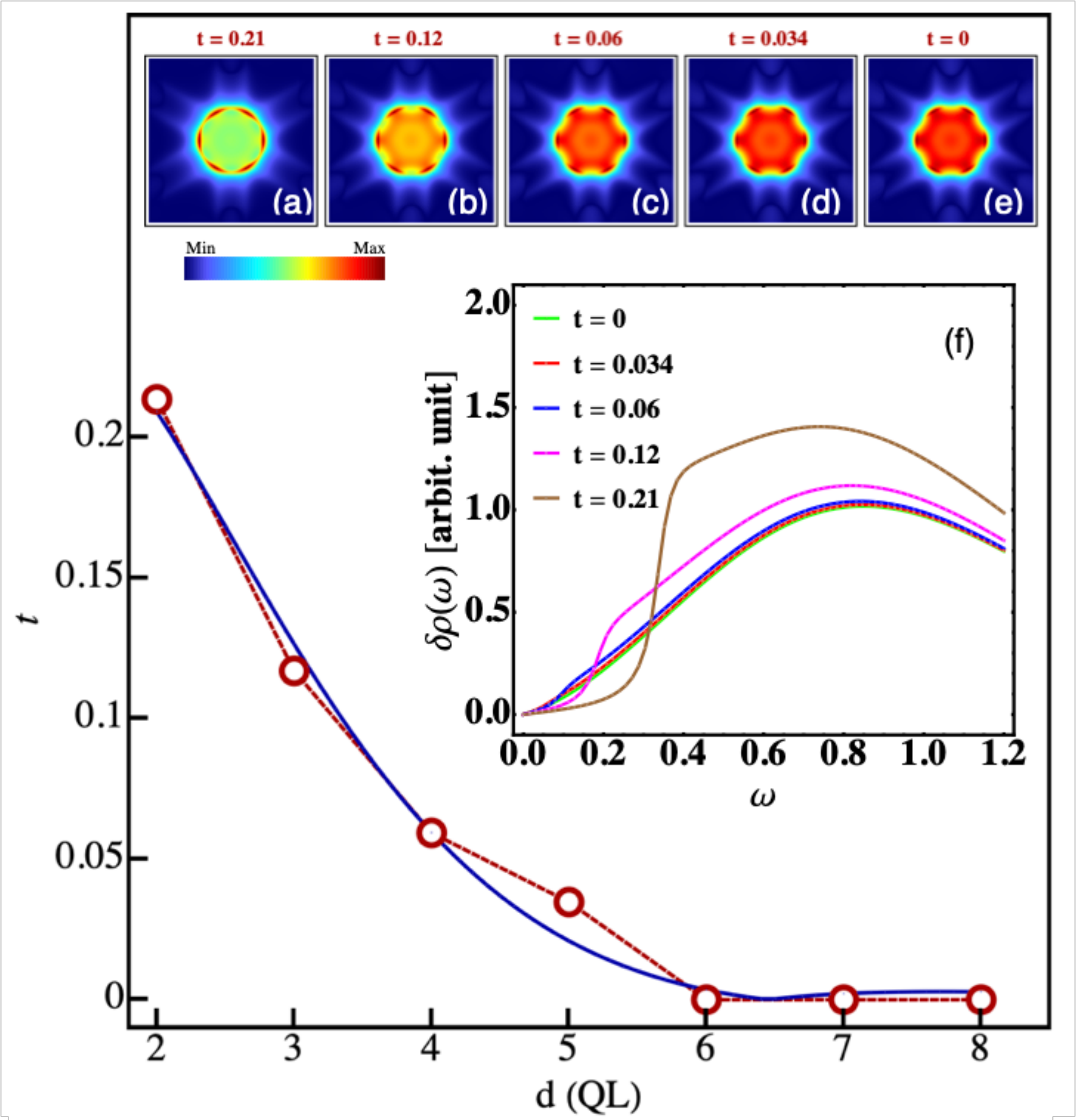}
\end{center}
\vspace{-0.5cm}
\caption{Monotonic (half-) gap $|t(d)|$ (units $E^*$) dependence  (red line and symbols) on film thickness $d$ (units QL) obtained for \bse~by ARPES (adapted from Zhang et al~\cite{zhang:10}). The blue line is a fit with Eq.~(\ref{eqn:td}) (see text). Upper inset  shows evolution of QPI pattern $|\Lambda_T(\bq,\omega=0.5)|$ for t-values  corresponding to the thickness sequence $d$~(2QL - 6QL). For large/small gaps $|t|$  isotropic circular/anisotropic snowflake patterns are seen. Bias voltage $eV=\omega$ is kept constant with respect to the Dirac point corresponding to each $d$. Lower inset: Total spectral surplus density $\delta\rho(\omega)=(-1/\pi N)\sum_\bk {\rm Im}[\Lambda_T(\bq,\omega)]$ at the impurity scattering site which exhibits the Dirac cone gap.}
\label{fig:t-thickness-se}
\end{figure}
%%%%%%%%%%%%%%%%%%%%%%fig%%%%%%%%%%%%%%%%%%%%%%%%%%%%%%%%
%
%\end{widetext}

Sofar the tunnelling matrix element $t$ has been treated as an arbitrary but a fixed parameter. It is, however, an effective (film-) surface states parameter that derives from the solution of the true boundary value problem~\cite{lu:10,asmar:18} starting from the valence (VB) and conduction band (CB) states of the bulk Hamiltoninan given in $\bk\cdot\bp$~-~parametrized form~\cite{liu:10,asmar:17}.  Using these bulk parameters it was shown in Ref.~\onlinecite{asmar:18} that the dependence of $t(d)$ on film thickness $d$ should be given by 
\bea
t(d)=t_0\exp(-\frac{d}{d_0})\sin(\frac{d}{d'_0})
\label{eqn:td}
\eea
where the reference energy $t_0[E^*]$ and length $d_0$ [QL], $d_0'$ [QL] scales are given in terms of the $\bk\cdot\bp$ - parameters of the true bulk~\cite{liu:10,asmar:17}.  The $t(d)$ dependence contains an overall exponential decrease with increasing $d$ characterized by the decay length $d_0$ but also an oscillatory term governed by $d_0'$ (oscillation period $2\pi d'_0$). The oscillatory term leads to a closing of the gap for intermediate d-values~\cite{asmar:18} $d=n\pi d'_0 (n=1,2..)$. From Ref.~\onlinecite{asmar:18} we derive the following set of {\it theoretically predicted} energy $[E^*]$ and length [QL]  scales for \bse: $(t_0,d_0,d'_0)_{th}=( 2.68, 0.64, 0.9)$ and \bte: $(t_0,d_0,d'_0)_{th}=( 0.80, 1.79, 0.3)$.
%
%\bea
%\text{\bse:}\;\;
%(t_0,d_0,d'_0)_{th}&=&( 2.68, 0.64, 0.9)\nonumber\\
%\text{\bte:}\;\;
%(t_0,d_0,d'_0)_{th}&=&( 0.80, 1.79, 0.3)
%\label{eqn:thpar}
%\eea
%
For the first compound this would lead to a vanishing of t(d) for intermediate $d\simeq 3$ due to a single oscillation in the range of sizable $t(d)$. This is not observed in ARPES experiments in \bse~which show no gap oscillation or closing~\cite{zhang:10} but rather a monotonous decrease of $|t(d)|$ in the range $2{\rm QL}<d<7{\rm QL}$ where it vanishes on the upper value, restoring the isolated Dirac surface states. This is shown by the symbols connected by a red line in Fig.~\ref{fig:t-thickness-se}. There is a natural source for the origin of the discrepancy: The bulk $\bk\cdot\bp$ parameters which depend on the position of bulk CB and VB edge are assumed as independent of thickness d for the film. However, this is unrealistic because band-bending effects due to both surfaces lead to thickness-dependent CB and VB edges as indeed observed in ARPES \cite{zhang:09} in addition to a strongly $d$-dependent energy position of the Dirac point of surface states. Therefore one has to expect that realistic parameters for $t(d)$ may be quite different from the theoretical ones given above. 

In a more practical strategy we take the observed monotonic exponential decay of $t(d)$ as real and derive the parameter set by fitting to the ARPES results for \bse~\cite{zhang:09} corresponding to the blue line in Fig.~\ref{fig:t-thickness-se}. This leads to the partial experimental set for \bse~: $(t_0,d_0)_{exp}=( 1.0, 1.46)$. Assuming the theoretical ratio $d'_0/d_0=1.41$ we then obtain $d'_0=2.05$. Because $d'_0$ is now considerably larger than the purely theoretical value the first zero of $t(d)$ due to the oscillatory term is shifted to $d$-values where the exponential term has already reduced the gap already close to zero. This means the oscillatory dependence of $t(d)$ is no longer observable. Therefore we suggest there is no fundamental contradiction between ARPES results and theoretical model for the inter-surface tunnelling but rather the effective theoretical parameter values do not correspond to the real ones, possibly due to the effect of the observed strong band bending effects found for VB and CB in this compound.  \\
For \bte~sofar no systematic ARPES results for thin films are available that could be used to constrain the parameter set for $t(d)$.
The  theoretical  values given above predict a much smaller oscillation period $2\pi d'_0$ of $t(d)$ for the \bte~compound which would lead to many zeroes of the tunnelling or gap in the thickness range $d=2-8$ as discussed below. Therefore even if the real parameter set is again different the chances to observe oscillatory behaviour should be much higher in \bte.\\
%
%%%%%%%%% figure %%%%%%%%%%%%%%%%%%%%%%%%%%%%%%

\begin{figure}[t]
\begin{center}
\includegraphics[width=0.990\columnwidth]{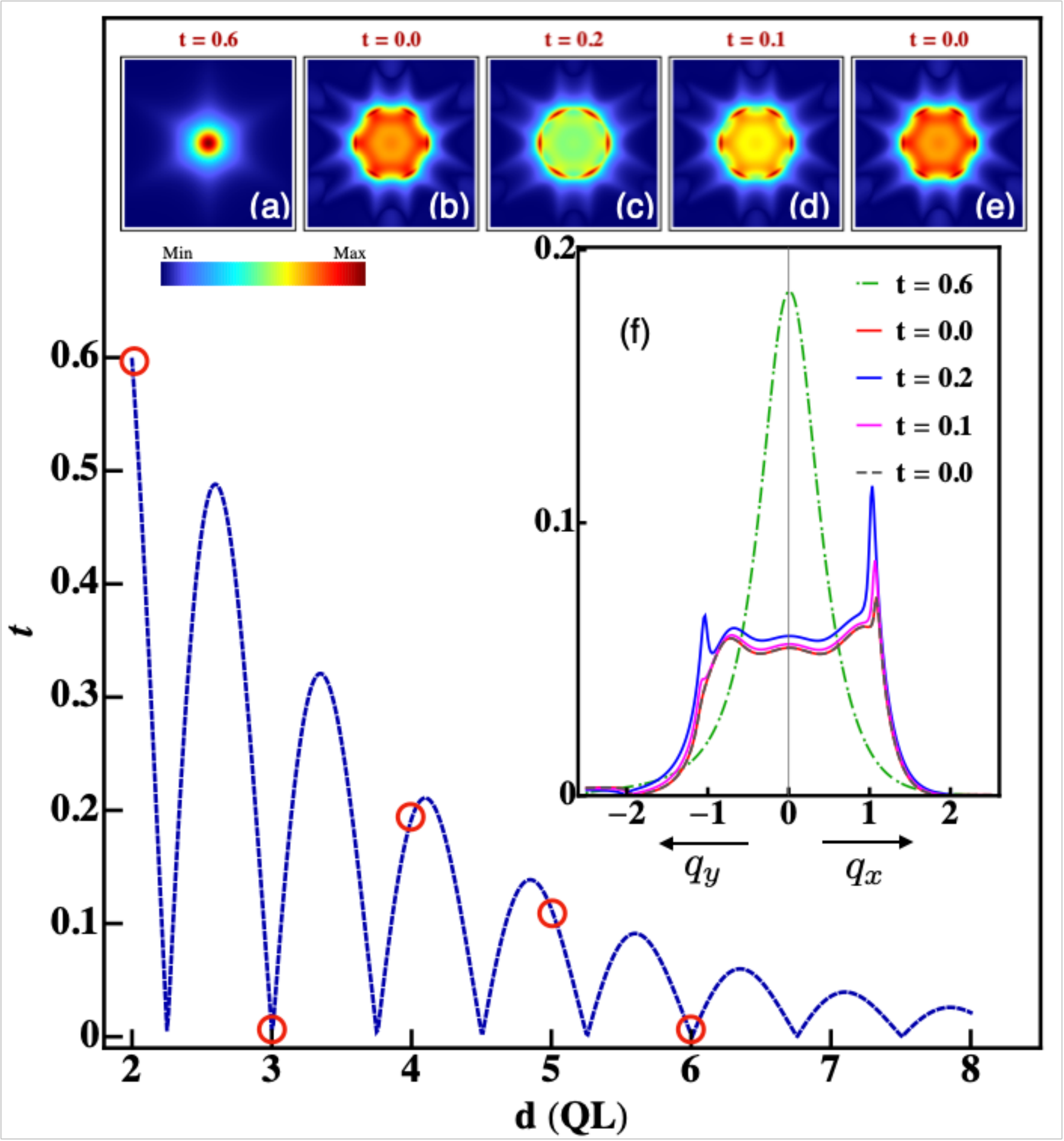}
\end{center}
\vspace{-0.5cm}
\caption{Oscillatory (half-) gap $|t(d)|$ (units $E^*$) dependence on film thickness $d$ (units of QL) 
obtained from Eq.~(\ref{eqn:td})  corresponding to a case of rapid oscillation of $|t(d)|$.
Concommitant oscillations in QPI spectra (upper row inset, $\omega=0.5$) between snowflake (b,d,e) and isotropic shape (a,c) depends 
critically on the positions of zeroes and maxima with respect to the possible film thickness $d=2,3,..$. Red circles correspond to actual values  of $|t|$ in the row images. The central image (f) shows the distinct intensity $|\Lambda(\bq,\omega)|$ behaviour for two directions due to different backscattering rules as function of thickness. For the highest $t=0.6>\omega$ it collapses to a bell shape curve because there are no more scattering processes on the equal-energy surface $\tE_\bk=\omega$.}
\label{fig:t-thickness-te}
\end{figure}

%%%%%%%%%%%%%%%%%%%%%%fig%%%%%%%%%%%%%%%%%%%%%%%%%%%%%%%%
%

Now we discuss the expected QPI pattern dependence on thickness which may show dramatic effects depending on the hybridisation gap or  $t(d)$ behaviour. Let us assume for each thickness the bias voltage is kept at a constant value with respect to the Dirac point in each film because the absolute value of the latter itself depends on $d$ as shown in Ref.~\onlinecite{zhang:09}.
 Then, as the film thickness is changed the gap strongly varies and the according QPI pattern should change between relatively isotropic circular pattern for large $|t|$ and strongly six-pronged snowflake patterns for small $|t|$. One may have the two principal cases:
 %--------------------
 \begin{itemize}
\item[] i)  For monotonic gap dependence as in \bse~the QPI pattern should also change monotonically from the almost circular shape with isotropic intensity to the snowflake shape with sixfold peaked intensities due to the forbidden backscattering in the latter.\\
\item[] ii) For oscillatory behaviour of the gap, as possibly realized in \bte,~the QPI images can also oscillate between more isotropic and snowflake type patterns. This depends sensitively on the precise positions of the zeroes and maxima of $t(d)$. If the former happen to lie close to an integer value of d (the ones achievable for real films) the gap will be zero and the snowflake structure will appear although $t_0$ is finite. If, on the other hand the maxima  of $t(d)$ come to lie close to one of these $d-$values the QPI pattern shows more isotropic form. In the general case when zeroes and maximum positions of $t(d)$ shift with respect to the integer $d$- values  oscillatory behaviour of QPI pattern between isotropic and  snowflake type may be expected. Observation of such sequence in STM-QPI experiments should be taken as direct proof for the gap oscillation.\\
 \end{itemize}
 %--------------------

First we discuss case (i) which is realized in \bse~(Fig.~\ref{fig:t-thickness-se}). The symbols present the tunneling $|t|[E^*]$ or gap size $(\Delta=2|t|)$ as function of thickness $d$[QL] as determined by ARPES \cite{zhang:09}. The red line is a guide to the eye and the blue line is a fit with Eq.~(\ref{eqn:td})
taking $t_0,d_0$ as free parameters and assuming the theoretical value $d'_0/d_0=1.41$ for \bse. This shifts the zero due to the oscillatory term to a thickness $d=6-7$ (arrow) where $|t|$ has already been exponentially suppressed and cannot be observed. For the theoretical
values of $t_0,d_0$ the zero would have appeared at $d=2.8$.
 On the top row of Fig.~\ref{fig:t-thickness-se}  we show the expected QPI image at $\omega=0.5$ for the five measured (half-) gap sizes on the red line in corresponding sequence. For increasing $d$ one observes the gradual appearance of the snow-flake features out of the isotropic QPI image for maximum gap at 2QL caused by the concomitant decrease in $|t|$.  If this behaviour could be observed in STM-QPI it would support the suggestion of ARPES that the gap decays indeed monotonically with increasing $d$ due to the large oscillation period $2\pi d'_0$ in this compound.\\
 
The second case (ii) with oscillations in $t(d)$ playing a role may qualitatively  correspond to \bte~although there are no ARPES results yet that allow to constrain the parameter set.  In the  model calculation of in  Fig.~\ref{fig:t-thickness-te} we use the parameter set $(t_0,d_0,d'_0)=( 2.1, 1.79, 0.24)$, slightly different from the theoretical one but has similar oscillation pattern as in Ref.~\onlinecite{asmar:18}. It is shown in Fig.~\ref{fig:t-thickness-te} as the blue curve with actual values of $|t(d)|$ for $d=2,3,\cdots$[QL] indicated by red circles which also oscillate. The corresponding sequence of QPI patterns is shown in the top row and it clearly exhibits an oscillation between snowflake pattern for small $|t|$(b,d,e) and a more istropic pattern for larger $|t|$(a,c). Experimental observation of this effect would give a direct evidence for the oscillation of the inter-surface hybrization strength $t(d)$. A detailed analysis of the QPI intensity given in the central inset provides even more insight. Here $|\Lambda(\bq,\omega=0.5)|$ is plotted for two directions $\bq=(q_x,0)$ and  $\bq=(0,q_y)$. From Fig.~\ref{fig:FS-QPI-comp}(b) one can see that $q_x$ direction corresponds to allowed scattering vectors $\bq_6$ whereas $q_y$ direction corresponds to  (equivalent) $\bq_4$ vectors forbidden due to backscattering. Therefore for small $|t|$ in the inset (red curves)  the intensity along $q_y$ shows a steep drop when crossing the $2k_F$ value while it has a sharp spike along $q_x$. For thickness values with larger $|t|$ the forbidden backscattering reappears and therefore the spike is now seen also for $q_y$ direction (blue curve).  Therefore the detailed intensity analysis of QPI pattern as function of film thickness can give a pointwise (in \bk -space) information on the backscattering variation with inter-surface tunneling strength and the associated topological character of near-gap hybridized surface states.

\section{Summary and Conclusion}
\label{sec:conc}

In this work we proposed a theory of quasiparticle interference for topological insulator thin films. We start from a model of single Dirac cone surface states with a warping term included to reproduce the realistic snowflake FS shape of the isolated TI surface in compounds like \bse~and \bte. The main physical effect of thin film geometry is implemented within a model where equal helicity states can tunnel from one surface to the other, described by the tunnelling energy $t$. This has two consequences: A gap opening is observed due to the (momentum-dependent) mixing of top/bottom states and the helical spin texture and Berry phase of Dirac states is strongly modified close to the gap edge at $\omega= |t|$, leading to a vanishing of the in-plane spin component and Berry phase.

The warped surface cones have been previousy found for isolated surfaces both by ARPES and QPI methods. The gap opening of surface states in thin films as function of thickness has sofar only been directly seen with the former method. Here we investigated what is to be expected for QPI experiments. 
To calculate the QPI signal we use a completely analytical t-matrix calculation with a full summation to infinite order in the scattering strength as well as the Born approximation. To carry out this approach it is necessary to restrict to a tractable model of impurity scattering and four surface basis states connected by the inter-surface tunnelling.
For the former we employ a generic potential consisting of momentum-independent intra-plane scalar and exchange scattering whose strengths are model parameters. We find a closed solution of the t-matrix including the effect of inter-plane propagation.
The impurity scattering and the tunnelling to the STM tip is expressed in the spin basis whereas the eigenstates of the warped cones have a complicated helical spin texture. Therefore the analytical espressions involve a summation over terms with different azimuthal form factors which also control the behaviour under backward scattering.

The QPI images can be interpreted in terms of characteristic scattering vectors connecting tips/dents in the snowflake FS for a given bias voltage. The allowed scattering and forbidden backscattering can be identified for normal impurities as long as the inter-plane tunnelling is moderate. For constant $\omega$ the latter, when the gap size increases, shrinks the dimension of the QPI image and its anisotropic features. This turns the six- pronged snowflakes of QPI with strong intensity anisotropy into a more isotropic circular $2k_F$ image of the Fermi surface of the gapped Dirac cones, also due to reopening of backscattering. Once $\omega < |t|$ falls below the (half-) gap size all specific QPI features are suppressed. 
 Such point-wise (in \bk-space) observation of the changing backscattering properties which characterize the topological nature of hybridised gap threshold states evidenced by the Berry phase is a unique ability of the QPI method and would certainly be worthwhile to discover.
A further strong incentive to carry out such investigation is suggested  by previous surface state calculations in \bse~and \bte~that have shown $t(d)$ to behave non-monotonically or even oscillates with film thickness $d$. This would imply that as thickness is reduced continuously the QPI image of TI surface state will oscillate between anisotropic snowflake and nearly isotropic circular patterns. 
Previous ARPES experiments in \bse~have, however, found only a monotonous reduction of the gap with increasing film thickness. As discussed in Sec.~\ref{sec:numerical} this advocates that the empirical parameters that describe the tunnelling energy $t(d)$ are different from theoretical ones such that the oscillatory period is much longer (Fig.~\ref{fig:t-thickness-se}), possibly due to band bending effects. However, \bte~may be a more favorable case because it is predicted to have a higher number of $t(d)$ oscillations, and some of them might survive even when again the empirical parameters for the tunnelling are different from the calculated ones. As shown in Fig.~\ref{fig:t-thickness-te} the sequence of QPI patterns for various $d =2,3, \cdots$ depends sensitively on the oscillation period and the possible coincidence of film thickness with the zeroes or maxima of $t(d)$. Therefore QPI method would be able to identify the predicted oscillatory behaviour of tunnelling induced gap in \bte.
Furthermore the dispersion of surface state energies may be extracted from QPI thin film images in $(q,\omega)$ plane for fixed direction of scattering vector $\hbq$. It shows the breaking of the Dirac cone by gapping and appearance of  quadratic low energy dispersion which is anisotropic due to the effect of warping. Their observation would give a further instructive comparison to ARPES results  as function of film thickness. Finally we want to mention that a first indication of the thin film effects discussed in this work may have recently been observed in a different context \cite{wu:20}. The gap opening of Dirac cones was clearly identified by QPI method in the multilayer MnBi$_4$Te$_7$ where it is induced by  hybridisation of surface states with adjacent inequivalent layer rather than the opposite surface as is the case in the thin films discussed here.

\section*{ACKNOWLEDGMENTS}
A.A. acknowledges financial support from the National Research Foundation (NRF) funded by the Ministry of Science of Korea (Grants: No. 2016K1A4A01922028, No. 2017R1D1A1B03033465, and No. 2019R1H1A2039733).

%\newpage

\appendix

\section{Explicit helicity eigenstates of the film}
\label{sec:heleigen}

The eigenvectors for coupled surface states given in Eq.~(\ref{eqn:eigenvec}) form the columns
of the unitary transformation $W_\bk$ that diagonalizes the coupled layer $4\times 4$  spin  Hamiltonian $H_\bk$ according to
$\tilde{H}_\bk=W_\bk^\dagger H_\bk W_\bk=\tE_\bk\tau_z\otimes\kappa_0$ where $\tau=1,2$ is the upper/lower band index and $\kappa=\pm$ the helicity. It is given explicitly by\\
%
%---------------------------------------------------------------------------------------------------
\begin{widetext}
\be
W_\bk=
\left(
 \begin{array}{cccc}
\cos\psi_\bk\cph& i\sin\psi_\bk\sph e^{-i\theta_\bk}&-\sin\psi_\bk\cph &i\cos\psi_\bk\sph e^{-i\theta_\bk}  \\
i\cos\psi_\bk\sph e^{i\theta_\bk}&\sin\psi_\bk\cph&- i\sin\psi_\bk\sph e^{i\theta_\bk} &  \cos\psi_\bk\cph \\
\sin\psi_\bk\cph&i\cos\psi_\bk\sph e^{-i\theta_\bk}& \cos\psi_\bk\cph & - i\sin\psi_\bk\sph e^{-i\theta_\bk} \\
i\sin\psi_\bk\sph e^{i\theta_\bk} &\cos \psi_\bk\cph&i\cos\psi_\bk\sph e^{i\theta_\bk}  & -\sin\psi_\bk\cph
\end{array}
\right),
\label{eqn:Wunitary}
\ee
\end{widetext}
%---------------------------------------------------------------------------------------------------
%
where the sequence of columns corresponds to $W_\bk=( |\tilde{\psi}_{1+}\ket , |\tilde{\psi}_{1-}\ket , |\tilde{\psi}_{2+}\ket, |\tilde{\psi}_{2-}\ket $) ordering of coupled surface eigenstates $|\tilde{\psi}_{\tau\kappa}\ket$ with energies $\tE_\bk,\tE_\bk,-\tE_\bk,-\tE_\bk$ respectively.

\section{Form factors for the t-matrix}
\label{sec:formfactor}

In this appendix we list the explicit expressions for the form factors $\alpha_{\kappa\kappa'}^\pm$ where $\kappa~(\kappa')$ denotes the helicity quantum numbers. We have
%
%---------------------------------------------------------------------------------------------------
\be
\bl
\alpha_{++}^\pm
&=\cph\cpr \pm\sph\spr e^{i(\theta_\bk-\theta_{\bk'})},
\\
\alpha_{--}^\pm
&=\cph\cpr \pm\sph\spr e^{-i(\theta_\bk-\theta_{\bk'})},
\\
\alpha_{+-}^\pm
&=i(\cph\spr e^{i\theta_{\bk'}} \pm\cpr\sph e^{i\theta_\bk}),
\\
\alpha_{-+}^\pm
&=-i(\cpr\sph e^{-i\theta_{\bk}} \pm\cph\spr e^{-i\theta_{\bk'}}).
\el
\ee
%---------------------------------------------------------------------------------------------------
%
Here $\theta_\bk$ is the azimuthal angle of the \bk-vector and $\phi_\bk$ is defined in Eq.~(\ref{eqn:phiangle}).
In these expressions the {\it upper} indices $\pm$ on the l.h.s. are correlated with the sign choice on the r.h.s..
We can arrange the form factors in $2\times 2$ matrices $\tilde{\alpha}_s$, $\tilde{\alpha}_a$ according to
%
%---------------------------------------------------------------------------------------------------
\be
\bl
\tilde{\alpha}_{s\bk\bk'}=
\left(
\begin{array}{cc}
\alpha_{++}^+ & \alpha_{+-}^-\\[0.1cm]
\alpha_{-+}^-&\alpha_{--}^+
\end{array}
\right);\;\;\;
%\text{and}\;\;\;
\tilde{\alpha}_{a\bk\bk'}=
\left(
\begin{array}{cc}
\alpha_{++}^- & \alpha_{+-}^+\\[0.1cm]
\alpha_{-+}^+&-\alpha_{--}^-
\end{array}
\right).
\label{eqn:formmat}
\el
\ee
%---------------------------------------------------------------------------------------------------
%
In terms of these form factor matrices the scattering matrix of Eq.~(\ref{eqn:tmatheli1}) is then given in shorthand notation by
%
%---------------------------------------------------------------------------------------------------
\be
\tilt_{\bk\bk'}=
\left(
\begin{array}{cc}
\tilt_{++} & \tilt_{+-} \\
\tilt_{-+}&\tilt_{--}
\end{array}
\right)
=
t_s(\om)\tilde{\alpha}_{s\bk\bk'}+t_a(\om)\tilde{\alpha}_{a\bk\bk'},
\label{eqn:tmat_short}
\ee
%---------------------------------------------------------------------------------------------------
%
where the (equal) top (T) and bottom B) (anti-) symmetrized t-matrix elements $t_{s,a}(\om)$  that depend only on frequency are given in Eq.~(\ref{eqn:tmat_spin}) for the general case. \\

%---------------------------------------------------------------------------------------------------
%---------------------------------------------------------------------------------------------------
\section{Form of the R-and t- matrix in  spin representation}
\label{sec:rmatrix}
For the calculation of the t-matrix elements in spin basis we need to calculate $R^{-1}=[1-\hat{V}{\hg}(\om)]^{-1}$ in Eq.~(\ref{eqn:tmatdef}). 
Restricting to intra-surface $(T\leftrightarrow T, B\leftrightarrow B)$  scattering (which is exact in Born approximation) we only have to consider the diagonal blocks 
(T/B correspond to upper/lower signs, respectively). Using the auxiliary functions defined in Eq.~(\ref{eqn:auxiliary1}) we obtain for the R-matrix
%
%---------------------------------------------------------------------------------------------------
\be
R_{T,B}=
\left(
\begin{array}{cc}
[1-V_\ua\hg_0\mp V_\ua f_c]  &-V_\ua f_s\\
- V_\da f^*_s& [1-V_\da\hg_0\pm V_\da f_c]
\end{array}
\right),
\ee
%---------------------------------------------------------------------------------------------------
%
where we defined $V_{\ua\da}=V_c\pm V_m$. The (complex) auxiliary function $f_s$, $f_c$ are defined by
%
%---------------------------------------------------------------------------------------------------
\be
\bl
&
f_s(\om)=\frac{i}{N}\sum_\bk F_s(\om,\theta_\bk)e^{-i\theta_\bk};
\\
&
 F_s(\om,\theta_\bk)
 =\frac{\sin\phi_\bk E_\bk}{[(\om)^2-\tE_\bk^2]};
 \el
\ee
and
\be
\bl
&
f_c(\om)=\frac{1}{N}\sum_\bk F_c(\om,\theta_\bk);
\\
&
F_c(\om,\theta_\bk)
=\frac{\cos\phi_\bk E_\bk}{[(\om)^2-\tE_\bk^2]}.
\label{eqn:auxiliary1}
\el
\ee
%---------------------------------------------------------------------------------------------------
%
From the expressions for $\phi_\bk,E_\bk,\tE_\bk$ given in Sec.~\ref{sec:model} one may show that $F_s(\om,\theta_\bk)$ is periodic with period $\frac{\pi}{3}$, i.e. $F_s(\om,\theta_\bk+\frac{\pi}{3})=F_s(\om,\theta_\bk)$, whereas $F_c(\om,\theta_\bk)$ is anti-periodic with period $\frac{\pi}{3}$, i.e. $F_c(\om,\theta_\bk+\frac{\pi}{3})=-F_c(\om,\theta_\bk)$. From this property it follows immediately that $f_c(\om)\equiv 0$, and due to the different periodicities in the integral for $f_s$ also $f_s\equiv 0$. Therefore $R^{-1}$ may be easily obtained and finally from Eq.~(\ref{eqn:tmatdef}) the (diagonal) t-matrix elements in  spin representation are given by
%
%---------------------------------------------------------------------------------------------------
\be
\bl
t_{\ua\ua}=\frac{V_c+V_m}{1-(V_c+V_m)\tg_0};\;\;\;
t_{\da\da}=\frac{V_c-V_m}{1-(V_c-V_m)\tg_0},
\el
\ee
%---------------------------------------------------------------------------------------------------
%
where $\tg_0(\om)$ is given in Eq.~(\ref{eqn:tmat_spin}). For the t-matrix in helical basis representation [Eq.~(\ref{eqn:tmatheli1})], it is then more convenient to use (anti-)symmetric combinations defined by $t_{s,a}=\fs(t_{\ua\ua}\pm t_{\da\da})$, which are explicitly written in Eq.~(\ref{eqn:tmat_spin}).
%\vspace{3cm}

%\vspace{2cm}
\bibliography{References}

\begin{thebibliography}{43}%
\makeatletter
\providecommand \@ifxundefined [1]{%
 \@ifx{#1\undefined}
}%
\providecommand \@ifnum [1]{%
 \ifnum #1\expandafter \@firstoftwo
 \else \expandafter \@secondoftwo
 \fi
}%
\providecommand \@ifx [1]{%
 \ifx #1\expandafter \@firstoftwo
 \else \expandafter \@secondoftwo
 \fi
}%
\providecommand \natexlab [1]{#1}%
\providecommand \enquote  [1]{``#1''}%
\providecommand \bibnamefont  [1]{#1}%
\providecommand \bibfnamefont [1]{#1}%
\providecommand \citenamefont [1]{#1}%
\providecommand \href@noop [0]{\@secondoftwo}%
\providecommand \href [0]{\begingroup \@sanitize@url \@href}%
\providecommand \@href[1]{\@@startlink{#1}\@@href}%
\providecommand \@@href[1]{\endgroup#1\@@endlink}%
\providecommand \@sanitize@url [0]{\catcode `\\12\catcode `\$12\catcode
  `\&12\catcode `\#12\catcode `\^12\catcode `\_12\catcode `\%12\relax}%
\providecommand \@@startlink[1]{}%
\providecommand \@@endlink[0]{}%
\providecommand \url  [0]{\begingroup\@sanitize@url \@url }%
\providecommand \@url [1]{\endgroup\@href {#1}{\urlprefix }}%
\providecommand \urlprefix  [0]{URL }%
\providecommand \Eprint [0]{\href }%
\providecommand \doibase [0]{http://dx.doi.org/}%
\providecommand \selectlanguage [0]{\@gobble}%
\providecommand \bibinfo  [0]{\@secondoftwo}%
\providecommand \bibfield  [0]{\@secondoftwo}%
\providecommand \translation [1]{[#1]}%
\providecommand \BibitemOpen [0]{}%
\providecommand \bibitemStop [0]{}%
\providecommand \bibitemNoStop [0]{.\EOS\space}%
\providecommand \EOS [0]{\spacefactor3000\relax}%
\providecommand \BibitemShut  [1]{\csname bibitem#1\endcsname}%
\let\auto@bib@innerbib\@empty
%</preamble>
\bibitem [{\citenamefont {Hsieh}\ \emph {et~al.}(2009)\citenamefont {Hsieh},
  \citenamefont {Xia}, \citenamefont {Qian}, \citenamefont {Wray},
  \citenamefont {Dil}, \citenamefont {Meier}, \citenamefont {Osterwalder},
  \citenamefont {Patthey}, \citenamefont {Checkelsky}, \citenamefont {Ong},
  \citenamefont {Fedorov}, \citenamefont {Lin}, \citenamefont {Bansil},
  \citenamefont {Grauer}, \citenamefont {Hor}, \citenamefont {Cava},\ and\
  \citenamefont {Hasan}}]{hsieh:09}%
  \BibitemOpen
  \bibfield  {author} {\bibinfo {author} {\bibfnamefont {D.}~\bibnamefont
  {Hsieh}}, \bibinfo {author} {\bibfnamefont {Y.}~\bibnamefont {Xia}}, \bibinfo
  {author} {\bibfnamefont {D.}~\bibnamefont {Qian}}, \bibinfo {author}
  {\bibfnamefont {L.}~\bibnamefont {Wray}}, \bibinfo {author} {\bibfnamefont
  {J.~H.}\ \bibnamefont {Dil}}, \bibinfo {author} {\bibfnamefont
  {F.}~\bibnamefont {Meier}}, \bibinfo {author} {\bibfnamefont
  {J.}~\bibnamefont {Osterwalder}}, \bibinfo {author} {\bibfnamefont
  {L.}~\bibnamefont {Patthey}}, \bibinfo {author} {\bibfnamefont {J.~G.}\
  \bibnamefont {Checkelsky}}, \bibinfo {author} {\bibfnamefont {N.~P.}\
  \bibnamefont {Ong}}, \bibinfo {author} {\bibfnamefont {A.~V.}\ \bibnamefont
  {Fedorov}}, \bibinfo {author} {\bibfnamefont {H.}~\bibnamefont {Lin}},
  \bibinfo {author} {\bibfnamefont {A.}~\bibnamefont {Bansil}}, \bibinfo
  {author} {\bibfnamefont {D.}~\bibnamefont {Grauer}}, \bibinfo {author}
  {\bibfnamefont {Y.~S.}\ \bibnamefont {Hor}}, \bibinfo {author} {\bibfnamefont
  {R.~J.}\ \bibnamefont {Cava}}, \ and\ \bibinfo {author} {\bibfnamefont
  {M.~Z.}\ \bibnamefont {Hasan}},\ }\href {\doibase 10.1038/nature08234}
  {\bibfield  {journal} {\bibinfo  {journal} {Nature}\ }\textbf {\bibinfo
  {volume} {460}},\ \bibinfo {pages} {1101} (\bibinfo {year}
  {2009})}\BibitemShut {NoStop}%
\bibitem [{\citenamefont {Kuroda}\ \emph {et~al.}(2010)\citenamefont {Kuroda},
  \citenamefont {Arita}, \citenamefont {Miyamoto}, \citenamefont {Ye},
  \citenamefont {Jiang}, \citenamefont {Kimura}, \citenamefont {Krasovskii},
  \citenamefont {Chulkov}, \citenamefont {Iwasawa}, \citenamefont {Okuda},
  \citenamefont {Shimada}, \citenamefont {Ueda}, \citenamefont {Namatame},\
  and\ \citenamefont {Taniguchi}}]{kuroda:10}%
  \BibitemOpen
  \bibfield  {author} {\bibinfo {author} {\bibfnamefont {K.}~\bibnamefont
  {Kuroda}}, \bibinfo {author} {\bibfnamefont {M.}~\bibnamefont {Arita}},
  \bibinfo {author} {\bibfnamefont {K.}~\bibnamefont {Miyamoto}}, \bibinfo
  {author} {\bibfnamefont {M.}~\bibnamefont {Ye}}, \bibinfo {author}
  {\bibfnamefont {J.}~\bibnamefont {Jiang}}, \bibinfo {author} {\bibfnamefont
  {A.}~\bibnamefont {Kimura}}, \bibinfo {author} {\bibfnamefont {E.~E.}\
  \bibnamefont {Krasovskii}}, \bibinfo {author} {\bibfnamefont {E.~V.}\
  \bibnamefont {Chulkov}}, \bibinfo {author} {\bibfnamefont {H.}~\bibnamefont
  {Iwasawa}}, \bibinfo {author} {\bibfnamefont {T.}~\bibnamefont {Okuda}},
  \bibinfo {author} {\bibfnamefont {K.}~\bibnamefont {Shimada}}, \bibinfo
  {author} {\bibfnamefont {Y.}~\bibnamefont {Ueda}}, \bibinfo {author}
  {\bibfnamefont {H.}~\bibnamefont {Namatame}}, \ and\ \bibinfo {author}
  {\bibfnamefont {M.}~\bibnamefont {Taniguchi}},\ }\href {\doibase
  10.1103/PhysRevLett.105.076802} {\bibfield  {journal} {\bibinfo  {journal}
  {Phys. Rev. Lett.}\ }\textbf {\bibinfo {volume} {105}},\ \bibinfo {pages}
  {076802} (\bibinfo {year} {2010})}\BibitemShut {NoStop}%
\bibitem [{\citenamefont {Hoefer}\ \emph {et~al.}(2014)\citenamefont {Hoefer},
  \citenamefont {Becker}, \citenamefont {Rata}, \citenamefont {Swanson},
  \citenamefont {Thalmeier},\ and\ \citenamefont {Tjeng}}]{hoefer:14}%
  \BibitemOpen
  \bibfield  {author} {\bibinfo {author} {\bibfnamefont {K.}~\bibnamefont
  {Hoefer}}, \bibinfo {author} {\bibfnamefont {C.}~\bibnamefont {Becker}},
  \bibinfo {author} {\bibfnamefont {D.}~\bibnamefont {Rata}}, \bibinfo {author}
  {\bibfnamefont {J.}~\bibnamefont {Swanson}}, \bibinfo {author} {\bibfnamefont
  {P.}~\bibnamefont {Thalmeier}}, \ and\ \bibinfo {author} {\bibfnamefont
  {L.~H.}\ \bibnamefont {Tjeng}},\ }\href {\doibase 10.1073/pnas.1410591111}
  {\bibfield  {journal} {\bibinfo  {journal} {Proceedings of the National
  Academy of Sciences}\ }\textbf {\bibinfo {volume} {111}},\ \bibinfo {pages}
  {14979} (\bibinfo {year} {2014})}\BibitemShut {NoStop}%
\bibitem [{\citenamefont {Roushan}\ \emph {et~al.}(2009)\citenamefont
  {Roushan}, \citenamefont {Seo}, \citenamefont {Parker}, \citenamefont {Hor},
  \citenamefont {Hsieh}, \citenamefont {Qian}, \citenamefont {Richardella},
  \citenamefont {Hasan}, \citenamefont {Cava},\ and\ \citenamefont
  {Yazdani}}]{roushan:09}%
  \BibitemOpen
  \bibfield  {author} {\bibinfo {author} {\bibfnamefont {P.}~\bibnamefont
  {Roushan}}, \bibinfo {author} {\bibfnamefont {J.}~\bibnamefont {Seo}},
  \bibinfo {author} {\bibfnamefont {C.~V.}\ \bibnamefont {Parker}}, \bibinfo
  {author} {\bibfnamefont {Y.~S.}\ \bibnamefont {Hor}}, \bibinfo {author}
  {\bibfnamefont {D.}~\bibnamefont {Hsieh}}, \bibinfo {author} {\bibfnamefont
  {D.}~\bibnamefont {Qian}}, \bibinfo {author} {\bibfnamefont {A.}~\bibnamefont
  {Richardella}}, \bibinfo {author} {\bibfnamefont {M.~Z.}\ \bibnamefont
  {Hasan}}, \bibinfo {author} {\bibfnamefont {R.~J.}\ \bibnamefont {Cava}}, \
  and\ \bibinfo {author} {\bibfnamefont {A.}~\bibnamefont {Yazdani}},\ }\href
  {\doibase 10.1038/nature08308} {\bibfield  {journal} {\bibinfo  {journal}
  {Nature}\ }\textbf {\bibinfo {volume} {460}},\ \bibinfo {pages} {1106}
  (\bibinfo {year} {2009})}\BibitemShut {NoStop}%
\bibitem [{\citenamefont {Zhang}\ \emph
  {et~al.}(2009{\natexlab{a}})\citenamefont {Zhang}, \citenamefont {Cheng},
  \citenamefont {Chen}, \citenamefont {Jia}, \citenamefont {Ma}, \citenamefont
  {He}, \citenamefont {Wang}, \citenamefont {Zhang}, \citenamefont {Dai},
  \citenamefont {Fang}, \citenamefont {Xie},\ and\ \citenamefont
  {Xue}}]{zhang:09}%
  \BibitemOpen
  \bibfield  {author} {\bibinfo {author} {\bibfnamefont {T.}~\bibnamefont
  {Zhang}}, \bibinfo {author} {\bibfnamefont {P.}~\bibnamefont {Cheng}},
  \bibinfo {author} {\bibfnamefont {X.}~\bibnamefont {Chen}}, \bibinfo {author}
  {\bibfnamefont {J.-F.}\ \bibnamefont {Jia}}, \bibinfo {author} {\bibfnamefont
  {X.}~\bibnamefont {Ma}}, \bibinfo {author} {\bibfnamefont {K.}~\bibnamefont
  {He}}, \bibinfo {author} {\bibfnamefont {L.}~\bibnamefont {Wang}}, \bibinfo
  {author} {\bibfnamefont {H.}~\bibnamefont {Zhang}}, \bibinfo {author}
  {\bibfnamefont {X.}~\bibnamefont {Dai}}, \bibinfo {author} {\bibfnamefont
  {Z.}~\bibnamefont {Fang}}, \bibinfo {author} {\bibfnamefont {X.}~\bibnamefont
  {Xie}}, \ and\ \bibinfo {author} {\bibfnamefont {Q.-K.}\ \bibnamefont
  {Xue}},\ }\href {\doibase 10.1103/PhysRevLett.103.266803} {\bibfield
  {journal} {\bibinfo  {journal} {Phys. Rev. Lett.}\ }\textbf {\bibinfo
  {volume} {103}},\ \bibinfo {pages} {266803} (\bibinfo {year}
  {2009}{\natexlab{a}})}\BibitemShut {NoStop}%
\bibitem [{\citenamefont {{Gomes}}\ \emph {et~al.}(2009)\citenamefont
  {{Gomes}}, \citenamefont {{Ko}}, \citenamefont {{Mar}}, \citenamefont
  {{Chen}}, \citenamefont {{Shen}},\ and\ \citenamefont
  {{Manoharan}}}]{gomes:09}%
  \BibitemOpen
  \bibfield  {author} {\bibinfo {author} {\bibfnamefont {K.~K.}\ \bibnamefont
  {{Gomes}}}, \bibinfo {author} {\bibfnamefont {W.}~\bibnamefont {{Ko}}},
  \bibinfo {author} {\bibfnamefont {W.}~\bibnamefont {{Mar}}}, \bibinfo
  {author} {\bibfnamefont {Y.}~\bibnamefont {{Chen}}}, \bibinfo {author}
  {\bibfnamefont {Z.-X.}\ \bibnamefont {{Shen}}}, \ and\ \bibinfo {author}
  {\bibfnamefont {H.~C.}\ \bibnamefont {{Manoharan}}},\ }\href@noop {}
  {\bibfield  {journal} {\bibinfo  {journal} {arXiv e-prints}\ ,\ \bibinfo
  {eid} {arXiv:0909.0921}} (\bibinfo {year} {2009})},\ \Eprint
  {http://arxiv.org/abs/0909.0921} {arXiv:0909.0921 [cond-mat.mes-hall]}
  \BibitemShut {NoStop}%
\bibitem [{\citenamefont {Alpichshev}\ \emph {et~al.}(2010)\citenamefont
  {Alpichshev}, \citenamefont {Analytis}, \citenamefont {Chu}, \citenamefont
  {Fisher}, \citenamefont {Chen}, \citenamefont {Shen}, \citenamefont {Fang},\
  and\ \citenamefont {Kapitulnik}}]{alpichshev:10}%
  \BibitemOpen
  \bibfield  {author} {\bibinfo {author} {\bibfnamefont {Z.}~\bibnamefont
  {Alpichshev}}, \bibinfo {author} {\bibfnamefont {J.~G.}\ \bibnamefont
  {Analytis}}, \bibinfo {author} {\bibfnamefont {J.-H.}\ \bibnamefont {Chu}},
  \bibinfo {author} {\bibfnamefont {I.~R.}\ \bibnamefont {Fisher}}, \bibinfo
  {author} {\bibfnamefont {Y.~L.}\ \bibnamefont {Chen}}, \bibinfo {author}
  {\bibfnamefont {Z.~X.}\ \bibnamefont {Shen}}, \bibinfo {author}
  {\bibfnamefont {A.}~\bibnamefont {Fang}}, \ and\ \bibinfo {author}
  {\bibfnamefont {A.}~\bibnamefont {Kapitulnik}},\ }\href {\doibase
  10.1103/PhysRevLett.104.016401} {\bibfield  {journal} {\bibinfo  {journal}
  {Phys. Rev. Lett.}\ }\textbf {\bibinfo {volume} {104}},\ \bibinfo {pages}
  {016401} (\bibinfo {year} {2010})}\BibitemShut {NoStop}%
\bibitem [{\citenamefont {Okada}\ \emph {et~al.}(2011)\citenamefont {Okada},
  \citenamefont {Dhital}, \citenamefont {Zhou}, \citenamefont {Huemiller},
  \citenamefont {Lin}, \citenamefont {Basak}, \citenamefont {Bansil},
  \citenamefont {Huang}, \citenamefont {Ding}, \citenamefont {Wang},
  \citenamefont {Wilson},\ and\ \citenamefont {Madhavan}}]{okada:11}%
  \BibitemOpen
  \bibfield  {author} {\bibinfo {author} {\bibfnamefont {Y.}~\bibnamefont
  {Okada}}, \bibinfo {author} {\bibfnamefont {C.}~\bibnamefont {Dhital}},
  \bibinfo {author} {\bibfnamefont {W.}~\bibnamefont {Zhou}}, \bibinfo {author}
  {\bibfnamefont {E.~D.}\ \bibnamefont {Huemiller}}, \bibinfo {author}
  {\bibfnamefont {H.}~\bibnamefont {Lin}}, \bibinfo {author} {\bibfnamefont
  {S.}~\bibnamefont {Basak}}, \bibinfo {author} {\bibfnamefont
  {A.}~\bibnamefont {Bansil}}, \bibinfo {author} {\bibfnamefont {Y.-B.}\
  \bibnamefont {Huang}}, \bibinfo {author} {\bibfnamefont {H.}~\bibnamefont
  {Ding}}, \bibinfo {author} {\bibfnamefont {Z.}~\bibnamefont {Wang}}, \bibinfo
  {author} {\bibfnamefont {S.~D.}\ \bibnamefont {Wilson}}, \ and\ \bibinfo
  {author} {\bibfnamefont {V.}~\bibnamefont {Madhavan}},\ }\href {\doibase
  10.1103/PhysRevLett.106.206805} {\bibfield  {journal} {\bibinfo  {journal}
  {Phys. Rev. Lett.}\ }\textbf {\bibinfo {volume} {106}},\ \bibinfo {pages}
  {206805} (\bibinfo {year} {2011})}\BibitemShut {NoStop}%
\bibitem [{\citenamefont {Cheng}\ \emph {et~al.}(2012)\citenamefont {Cheng},
  \citenamefont {Zhang}, \citenamefont {He}, \citenamefont {Chen},
  \citenamefont {Ma},\ and\ \citenamefont {Xue}}]{cheng:12}%
  \BibitemOpen
  \bibfield  {author} {\bibinfo {author} {\bibfnamefont {P.}~\bibnamefont
  {Cheng}}, \bibinfo {author} {\bibfnamefont {T.}~\bibnamefont {Zhang}},
  \bibinfo {author} {\bibfnamefont {K.}~\bibnamefont {He}}, \bibinfo {author}
  {\bibfnamefont {X.}~\bibnamefont {Chen}}, \bibinfo {author} {\bibfnamefont
  {X.}~\bibnamefont {Ma}}, \ and\ \bibinfo {author} {\bibfnamefont
  {Q.}~\bibnamefont {Xue}},\ }\href {\doibase
  https://doi.org/10.1016/j.physe.2011.11.004} {\bibfield  {journal} {\bibinfo
  {journal} {Physica E: Low-dimensional Systems and Nanostructures}\ }\textbf
  {\bibinfo {volume} {44}},\ \bibinfo {pages} {912 } (\bibinfo {year}
  {2012})}\BibitemShut {NoStop}%
\bibitem [{\citenamefont {Kohsaka}\ \emph {et~al.}(2017)\citenamefont
  {Kohsaka}, \citenamefont {Machida}, \citenamefont {Iwaya}, \citenamefont
  {Kanou}, \citenamefont {Hanaguri},\ and\ \citenamefont
  {Sasagawa}}]{kohsaka:17}%
  \BibitemOpen
  \bibfield  {author} {\bibinfo {author} {\bibfnamefont {Y.}~\bibnamefont
  {Kohsaka}}, \bibinfo {author} {\bibfnamefont {T.}~\bibnamefont {Machida}},
  \bibinfo {author} {\bibfnamefont {K.}~\bibnamefont {Iwaya}}, \bibinfo
  {author} {\bibfnamefont {M.}~\bibnamefont {Kanou}}, \bibinfo {author}
  {\bibfnamefont {T.}~\bibnamefont {Hanaguri}}, \ and\ \bibinfo {author}
  {\bibfnamefont {T.}~\bibnamefont {Sasagawa}},\ }\href {\doibase
  10.1103/PhysRevB.95.115307} {\bibfield  {journal} {\bibinfo  {journal} {Phys.
  Rev. B}\ }\textbf {\bibinfo {volume} {95}},\ \bibinfo {pages} {115307}
  (\bibinfo {year} {2017})}\BibitemShut {NoStop}%
\bibitem [{\citenamefont {Lee}\ \emph {et~al.}(2009)\citenamefont {Lee},
  \citenamefont {Wu}, \citenamefont {Arovas},\ and\ \citenamefont
  {Zhang}}]{lee:09}%
  \BibitemOpen
  \bibfield  {author} {\bibinfo {author} {\bibfnamefont {W.-C.}\ \bibnamefont
  {Lee}}, \bibinfo {author} {\bibfnamefont {C.}~\bibnamefont {Wu}}, \bibinfo
  {author} {\bibfnamefont {D.~P.}\ \bibnamefont {Arovas}}, \ and\ \bibinfo
  {author} {\bibfnamefont {S.-C.}\ \bibnamefont {Zhang}},\ }\href {\doibase
  10.1103/PhysRevB.80.245439} {\bibfield  {journal} {\bibinfo  {journal} {Phys.
  Rev. B}\ }\textbf {\bibinfo {volume} {80}},\ \bibinfo {pages} {245439}
  (\bibinfo {year} {2009})}\BibitemShut {NoStop}%
\bibitem [{\citenamefont {Zhou}\ \emph {et~al.}(2009)\citenamefont {Zhou},
  \citenamefont {Fang}, \citenamefont {Tsai},\ and\ \citenamefont
  {Hu}}]{zhou:09}%
  \BibitemOpen
  \bibfield  {author} {\bibinfo {author} {\bibfnamefont {X.}~\bibnamefont
  {Zhou}}, \bibinfo {author} {\bibfnamefont {C.}~\bibnamefont {Fang}}, \bibinfo
  {author} {\bibfnamefont {W.-F.}\ \bibnamefont {Tsai}}, \ and\ \bibinfo
  {author} {\bibfnamefont {J.}~\bibnamefont {Hu}},\ }\href {\doibase
  10.1103/PhysRevB.80.245317} {\bibfield  {journal} {\bibinfo  {journal} {Phys.
  Rev. B}\ }\textbf {\bibinfo {volume} {80}},\ \bibinfo {pages} {245317}
  (\bibinfo {year} {2009})}\BibitemShut {NoStop}%
\bibitem [{\citenamefont {Guo}\ and\ \citenamefont {Franz}(2010)}]{guo:10}%
  \BibitemOpen
  \bibfield  {author} {\bibinfo {author} {\bibfnamefont {H.-M.}\ \bibnamefont
  {Guo}}\ and\ \bibinfo {author} {\bibfnamefont {M.}~\bibnamefont {Franz}},\
  }\href {\doibase 10.1103/PhysRevB.81.041102} {\bibfield  {journal} {\bibinfo
  {journal} {Phys. Rev. B}\ }\textbf {\bibinfo {volume} {81}},\ \bibinfo
  {pages} {041102} (\bibinfo {year} {2010})}\BibitemShut {NoStop}%
\bibitem [{\citenamefont {Liu}\ \emph {et~al.}(2012)\citenamefont {Liu},
  \citenamefont {Qi},\ and\ \citenamefont {Zhang}}]{Liu:2012aa}%
  \BibitemOpen
  \bibfield  {author} {\bibinfo {author} {\bibfnamefont {Q.}~\bibnamefont
  {Liu}}, \bibinfo {author} {\bibfnamefont {X.-L.}\ \bibnamefont {Qi}}, \ and\
  \bibinfo {author} {\bibfnamefont {S.-C.}\ \bibnamefont {Zhang}},\ }\href
  {\doibase 10.1103/PhysRevB.85.125314} {\bibfield  {journal} {\bibinfo
  {journal} {Phys. Rev. B}\ }\textbf {\bibinfo {volume} {85}},\ \bibinfo
  {pages} {125314} (\bibinfo {year} {2012})}\BibitemShut {NoStop}%
\bibitem [{\citenamefont {{R{\"u}{\ss}mann}}\ \emph {et~al.}(2020)\citenamefont
  {{R{\"u}{\ss}mann}}, \citenamefont {{Mavropoulos}},\ and\ \citenamefont
  {{Bl{\"u}gel}}}]{ruessmann:20}%
  \BibitemOpen
  \bibfield  {author} {\bibinfo {author} {\bibfnamefont {P.}~\bibnamefont
  {{R{\"u}{\ss}mann}}}, \bibinfo {author} {\bibfnamefont {P.}~\bibnamefont
  {{Mavropoulos}}}, \ and\ \bibinfo {author} {\bibfnamefont {S.}~\bibnamefont
  {{Bl{\"u}gel}}},\ }\href@noop {} {\bibfield  {journal} {\bibinfo  {journal}
  {arXiv e-prints}\ ,\ \bibinfo {eid} {arXiv:2001.06189}} (\bibinfo {year}
  {2020})},\ \Eprint {http://arxiv.org/abs/2001.06189} {arXiv:2001.06189
  [cond-mat.mes-hall]} \BibitemShut {NoStop}%
\bibitem [{\citenamefont {Taskin}\ \emph {et~al.}(2012)\citenamefont {Taskin},
  \citenamefont {Sasaki}, \citenamefont {Segawa},\ and\ \citenamefont
  {Ando}}]{taskin:12}%
  \BibitemOpen
  \bibfield  {author} {\bibinfo {author} {\bibfnamefont {A.~A.}\ \bibnamefont
  {Taskin}}, \bibinfo {author} {\bibfnamefont {S.}~\bibnamefont {Sasaki}},
  \bibinfo {author} {\bibfnamefont {K.}~\bibnamefont {Segawa}}, \ and\ \bibinfo
  {author} {\bibfnamefont {Y.}~\bibnamefont {Ando}},\ }\href {\doibase
  10.1103/PhysRevLett.109.066803} {\bibfield  {journal} {\bibinfo  {journal}
  {Phys. Rev. Lett.}\ }\textbf {\bibinfo {volume} {109}},\ \bibinfo {pages}
  {066803} (\bibinfo {year} {2012})}\BibitemShut {NoStop}%
\bibitem [{\citenamefont {Lu}\ \emph {et~al.}(2011)\citenamefont {Lu},
  \citenamefont {Shi},\ and\ \citenamefont {Shen}}]{lu:11}%
  \BibitemOpen
  \bibfield  {author} {\bibinfo {author} {\bibfnamefont {H.-Z.}\ \bibnamefont
  {Lu}}, \bibinfo {author} {\bibfnamefont {J.}~\bibnamefont {Shi}}, \ and\
  \bibinfo {author} {\bibfnamefont {S.-Q.}\ \bibnamefont {Shen}},\ }\href
  {\doibase 10.1103/PhysRevLett.107.076801} {\bibfield  {journal} {\bibinfo
  {journal} {Phys. Rev. Lett.}\ }\textbf {\bibinfo {volume} {107}},\ \bibinfo
  {pages} {076801} (\bibinfo {year} {2011})}\BibitemShut {NoStop}%
\bibitem [{\citenamefont {Lu}\ and\ \citenamefont {Shen}(2011)}]{lu:11b}%
  \BibitemOpen
  \bibfield  {author} {\bibinfo {author} {\bibfnamefont {H.-Z.}\ \bibnamefont
  {Lu}}\ and\ \bibinfo {author} {\bibfnamefont {S.-Q.}\ \bibnamefont {Shen}},\
  }\href {\doibase 10.1103/PhysRevB.84.125138} {\bibfield  {journal} {\bibinfo
  {journal} {Phys. Rev. B}\ }\textbf {\bibinfo {volume} {84}},\ \bibinfo
  {pages} {125138} (\bibinfo {year} {2011})}\BibitemShut {NoStop}%
\bibitem [{\citenamefont {Zhang}\ \emph {et~al.}(2010)\citenamefont {Zhang},
  \citenamefont {He}, \citenamefont {Chang}, \citenamefont {Song},
  \citenamefont {Wang}, \citenamefont {Chen}, \citenamefont {Jia},
  \citenamefont {Fang}, \citenamefont {Dai}, \citenamefont {Shan},
  \citenamefont {Shen}, \citenamefont {Niu}, \citenamefont {Qi}, \citenamefont
  {Zhang}, \citenamefont {Ma},\ and\ \citenamefont {Xue}}]{zhang:10}%
  \BibitemOpen
  \bibfield  {author} {\bibinfo {author} {\bibfnamefont {Y.}~\bibnamefont
  {Zhang}}, \bibinfo {author} {\bibfnamefont {K.}~\bibnamefont {He}}, \bibinfo
  {author} {\bibfnamefont {C.-Z.}\ \bibnamefont {Chang}}, \bibinfo {author}
  {\bibfnamefont {C.-L.}\ \bibnamefont {Song}}, \bibinfo {author}
  {\bibfnamefont {L.-L.}\ \bibnamefont {Wang}}, \bibinfo {author}
  {\bibfnamefont {X.}~\bibnamefont {Chen}}, \bibinfo {author} {\bibfnamefont
  {J.-F.}\ \bibnamefont {Jia}}, \bibinfo {author} {\bibfnamefont
  {Z.}~\bibnamefont {Fang}}, \bibinfo {author} {\bibfnamefont {X.}~\bibnamefont
  {Dai}}, \bibinfo {author} {\bibfnamefont {W.-Y.}\ \bibnamefont {Shan}},
  \bibinfo {author} {\bibfnamefont {S.-Q.}\ \bibnamefont {Shen}}, \bibinfo
  {author} {\bibfnamefont {Q.}~\bibnamefont {Niu}}, \bibinfo {author}
  {\bibfnamefont {X.-L.}\ \bibnamefont {Qi}}, \bibinfo {author} {\bibfnamefont
  {S.-C.}\ \bibnamefont {Zhang}}, \bibinfo {author} {\bibfnamefont {X.-C.}\
  \bibnamefont {Ma}}, \ and\ \bibinfo {author} {\bibfnamefont {Q.-K.}\
  \bibnamefont {Xue}},\ }\href {\doibase 10.1038/nphys1689} {\bibfield
  {journal} {\bibinfo  {journal} {Nature Physics}\ }\textbf {\bibinfo {volume}
  {6}},\ \bibinfo {pages} {584} (\bibinfo {year} {2010})}\BibitemShut {NoStop}%
\bibitem [{\citenamefont {Neupane}\ \emph {et~al.}(2014)\citenamefont
  {Neupane}, \citenamefont {Richardella}, \citenamefont {S{\'a}nchez-Barriga},
  \citenamefont {Xu}, \citenamefont {Alidoust}, \citenamefont {Belopolski},
  \citenamefont {Liu}, \citenamefont {Bian}, \citenamefont {Zhang},
  \citenamefont {Marchenko}, \citenamefont {Varykhalov}, \citenamefont {Rader},
  \citenamefont {Leandersson}, \citenamefont {Balasubramanian}, \citenamefont
  {Chang}, \citenamefont {Jeng}, \citenamefont {Basak}, \citenamefont {Lin},
  \citenamefont {Bansil}, \citenamefont {Samarth},\ and\ \citenamefont
  {Hasan}}]{neupane:14}%
  \BibitemOpen
  \bibfield  {author} {\bibinfo {author} {\bibfnamefont {M.}~\bibnamefont
  {Neupane}}, \bibinfo {author} {\bibfnamefont {A.}~\bibnamefont
  {Richardella}}, \bibinfo {author} {\bibfnamefont {J.}~\bibnamefont
  {S{\'a}nchez-Barriga}}, \bibinfo {author} {\bibfnamefont {S.}~\bibnamefont
  {Xu}}, \bibinfo {author} {\bibfnamefont {N.}~\bibnamefont {Alidoust}},
  \bibinfo {author} {\bibfnamefont {I.}~\bibnamefont {Belopolski}}, \bibinfo
  {author} {\bibfnamefont {C.}~\bibnamefont {Liu}}, \bibinfo {author}
  {\bibfnamefont {G.}~\bibnamefont {Bian}}, \bibinfo {author} {\bibfnamefont
  {D.}~\bibnamefont {Zhang}}, \bibinfo {author} {\bibfnamefont
  {D.}~\bibnamefont {Marchenko}}, \bibinfo {author} {\bibfnamefont
  {A.}~\bibnamefont {Varykhalov}}, \bibinfo {author} {\bibfnamefont
  {O.}~\bibnamefont {Rader}}, \bibinfo {author} {\bibfnamefont
  {M.}~\bibnamefont {Leandersson}}, \bibinfo {author} {\bibfnamefont
  {T.}~\bibnamefont {Balasubramanian}}, \bibinfo {author} {\bibfnamefont
  {T.-R.}\ \bibnamefont {Chang}}, \bibinfo {author} {\bibfnamefont {H.-T.}\
  \bibnamefont {Jeng}}, \bibinfo {author} {\bibfnamefont {S.}~\bibnamefont
  {Basak}}, \bibinfo {author} {\bibfnamefont {H.}~\bibnamefont {Lin}}, \bibinfo
  {author} {\bibfnamefont {A.}~\bibnamefont {Bansil}}, \bibinfo {author}
  {\bibfnamefont {N.}~\bibnamefont {Samarth}}, \ and\ \bibinfo {author}
  {\bibfnamefont {M.~Z.}\ \bibnamefont {Hasan}},\ }\href {\doibase
  10.1038/ncomms4841} {\bibfield  {journal} {\bibinfo  {journal} {Nature
  Communications}\ }\textbf {\bibinfo {volume} {5}},\ \bibinfo {pages} {3841}
  (\bibinfo {year} {2014})}\BibitemShut {NoStop}%
\bibitem [{\citenamefont {Capriotti}\ \emph {et~al.}(2003)\citenamefont
  {Capriotti}, \citenamefont {Scalapino},\ and\ \citenamefont
  {Sedgewick}}]{capriotti:03}%
  \BibitemOpen
  \bibfield  {author} {\bibinfo {author} {\bibfnamefont {L.}~\bibnamefont
  {Capriotti}}, \bibinfo {author} {\bibfnamefont {D.~J.}\ \bibnamefont
  {Scalapino}}, \ and\ \bibinfo {author} {\bibfnamefont {R.~D.}\ \bibnamefont
  {Sedgewick}},\ }\href {\doibase 10.1103/PhysRevB.68.014508} {\bibfield
  {journal} {\bibinfo  {journal} {Phys. Rev. B}\ }\textbf {\bibinfo {volume}
  {68}},\ \bibinfo {pages} {014508} (\bibinfo {year} {2003})}\BibitemShut
  {NoStop}%
\bibitem [{\citenamefont {Balatsky}\ \emph {et~al.}(2006)\citenamefont
  {Balatsky}, \citenamefont {Vekhter},\ and\ \citenamefont
  {Zhu}}]{balatsky:06}%
  \BibitemOpen
  \bibfield  {author} {\bibinfo {author} {\bibfnamefont {A.~V.}\ \bibnamefont
  {Balatsky}}, \bibinfo {author} {\bibfnamefont {I.}~\bibnamefont {Vekhter}}, \
  and\ \bibinfo {author} {\bibfnamefont {J.-X.}\ \bibnamefont {Zhu}},\ }\href
  {\doibase 10.1103/RevModPhys.78.373} {\bibfield  {journal} {\bibinfo
  {journal} {Rev. Mod. Phys.}\ }\textbf {\bibinfo {volume} {78}},\ \bibinfo
  {pages} {373} (\bibinfo {year} {2006})}\BibitemShut {NoStop}%
\bibitem [{\citenamefont {Lu}\ \emph {et~al.}(2010)\citenamefont {Lu},
  \citenamefont {Shan}, \citenamefont {Yao}, \citenamefont {Niu},\ and\
  \citenamefont {Shen}}]{lu:10}%
  \BibitemOpen
  \bibfield  {author} {\bibinfo {author} {\bibfnamefont {H.-Z.}\ \bibnamefont
  {Lu}}, \bibinfo {author} {\bibfnamefont {W.-Y.}\ \bibnamefont {Shan}},
  \bibinfo {author} {\bibfnamefont {W.}~\bibnamefont {Yao}}, \bibinfo {author}
  {\bibfnamefont {Q.}~\bibnamefont {Niu}}, \ and\ \bibinfo {author}
  {\bibfnamefont {S.-Q.}\ \bibnamefont {Shen}},\ }\href {\doibase
  10.1103/PhysRevB.81.115407} {\bibfield  {journal} {\bibinfo  {journal} {Phys.
  Rev. B}\ }\textbf {\bibinfo {volume} {81}},\ \bibinfo {pages} {115407}
  (\bibinfo {year} {2010})}\BibitemShut {NoStop}%
\bibitem [{\citenamefont {Asmar}\ \emph {et~al.}(2018)\citenamefont {Asmar},
  \citenamefont {Sheehy},\ and\ \citenamefont {Vekhter}}]{asmar:18}%
  \BibitemOpen
  \bibfield  {author} {\bibinfo {author} {\bibfnamefont {M.~M.}\ \bibnamefont
  {Asmar}}, \bibinfo {author} {\bibfnamefont {D.~E.}\ \bibnamefont {Sheehy}}, \
  and\ \bibinfo {author} {\bibfnamefont {I.}~\bibnamefont {Vekhter}},\ }\href
  {\doibase 10.1103/PhysRevB.97.075419} {\bibfield  {journal} {\bibinfo
  {journal} {Phys. Rev. B}\ }\textbf {\bibinfo {volume} {97}},\ \bibinfo
  {pages} {075419} (\bibinfo {year} {2018})}\BibitemShut {NoStop}%
\bibitem [{\citenamefont {Zhang}\ \emph
  {et~al.}(2009{\natexlab{b}})\citenamefont {Zhang}, \citenamefont {Liu},
  \citenamefont {Qi}, \citenamefont {Dai}, \citenamefont {Fang},\ and\
  \citenamefont {Zhang}}]{zhang:09a}%
  \BibitemOpen
  \bibfield  {author} {\bibinfo {author} {\bibfnamefont {H.}~\bibnamefont
  {Zhang}}, \bibinfo {author} {\bibfnamefont {C.-X.}\ \bibnamefont {Liu}},
  \bibinfo {author} {\bibfnamefont {X.-L.}\ \bibnamefont {Qi}}, \bibinfo
  {author} {\bibfnamefont {X.}~\bibnamefont {Dai}}, \bibinfo {author}
  {\bibfnamefont {Z.}~\bibnamefont {Fang}}, \ and\ \bibinfo {author}
  {\bibfnamefont {S.-C.}\ \bibnamefont {Zhang}},\ }\href@noop {} {\bibfield
  {journal} {\bibinfo  {journal} {Nature Physics}\ }\textbf {\bibinfo {volume}
  {5}},\ \bibinfo {pages} {438} (\bibinfo {year}
  {2009}{\natexlab{b}})}\BibitemShut {NoStop}%
\bibitem [{\citenamefont {Zhang}\ \emph {et~al.}(2012)\citenamefont {Zhang},
  \citenamefont {Kane},\ and\ \citenamefont {Mele}}]{zhang:12}%
  \BibitemOpen
  \bibfield  {author} {\bibinfo {author} {\bibfnamefont {F.}~\bibnamefont
  {Zhang}}, \bibinfo {author} {\bibfnamefont {C.~L.}\ \bibnamefont {Kane}}, \
  and\ \bibinfo {author} {\bibfnamefont {E.~J.}\ \bibnamefont {Mele}},\
  }\href@noop {} {\bibfield  {journal} {\bibinfo  {journal} {Phys. Rev. B}\
  }\textbf {\bibinfo {volume} {86}},\ \bibinfo {pages} {081303} (\bibinfo
  {year} {2012})}\BibitemShut {NoStop}%
\bibitem [{\citenamefont {Fu}(2009)}]{fu:09}%
  \BibitemOpen
  \bibfield  {author} {\bibinfo {author} {\bibfnamefont {L.}~\bibnamefont
  {Fu}},\ }\href {\doibase 10.1103/PhysRevLett.103.266801} {\bibfield
  {journal} {\bibinfo  {journal} {Phys. Rev. Lett.}\ }\textbf {\bibinfo
  {volume} {103}},\ \bibinfo {pages} {266801} (\bibinfo {year}
  {2009})}\BibitemShut {NoStop}%
\bibitem [{\citenamefont {Thalmeier}(2011)}]{thalmeier:11}%
  \BibitemOpen
  \bibfield  {author} {\bibinfo {author} {\bibfnamefont {P.}~\bibnamefont
  {Thalmeier}},\ }\href {\doibase 10.1103/PhysRevB.84.155102} {\bibfield
  {journal} {\bibinfo  {journal} {Phys. Rev. B}\ }\textbf {\bibinfo {volume}
  {84}},\ \bibinfo {pages} {155102} (\bibinfo {year} {2011})}\BibitemShut
  {NoStop}%
\bibitem [{\citenamefont {Wang}\ and\ \citenamefont {Zhu}(2013)}]{wang:13}%
  \BibitemOpen
  \bibfield  {author} {\bibinfo {author} {\bibfnamefont {J.}~\bibnamefont
  {Wang}}\ and\ \bibinfo {author} {\bibfnamefont {B.-F.}\ \bibnamefont {Zhu}},\
  }\href@noop {} {\bibfield  {journal} {\bibinfo  {journal} {Chin. Phys. B}\
  }\textbf {\bibinfo {volume} {22}},\ \bibinfo {pages} {067301} (\bibinfo
  {year} {2013})}\BibitemShut {NoStop}%
\bibitem [{\citenamefont {Bansil}\ \emph {et~al.}(2016)\citenamefont {Bansil},
  \citenamefont {Lin},\ and\ \citenamefont {Das}}]{bansil:16}%
  \BibitemOpen
  \bibfield  {author} {\bibinfo {author} {\bibfnamefont {A.}~\bibnamefont
  {Bansil}}, \bibinfo {author} {\bibfnamefont {H.}~\bibnamefont {Lin}}, \ and\
  \bibinfo {author} {\bibfnamefont {T.}~\bibnamefont {Das}},\ }\href@noop {}
  {\bibfield  {journal} {\bibinfo  {journal} {Rev. Mod. Phys.}\ }\textbf
  {\bibinfo {volume} {88}},\ \bibinfo {pages} {021004} (\bibinfo {year}
  {2016})}\BibitemShut {NoStop}%
\bibitem [{\citenamefont {Hasan}\ and\ \citenamefont {Moore}(2011)}]{hasan:11}%
  \BibitemOpen
  \bibfield  {author} {\bibinfo {author} {\bibfnamefont {M.~Z.}\ \bibnamefont
  {Hasan}}\ and\ \bibinfo {author} {\bibfnamefont {J.~E.}\ \bibnamefont
  {Moore}},\ }\href {\doibase 10.1146/annurev-conmatphys-062910-140432}
  {\bibfield  {journal} {\bibinfo  {journal} {Annual Review of Condensed Matter
  Physics}\ }\textbf {\bibinfo {volume} {2}},\ \bibinfo {pages} {55} (\bibinfo
  {year} {2011})}\BibitemShut {NoStop}%
\bibitem [{\citenamefont {Ando}(2013)}]{ando:13}%
  \BibitemOpen
  \bibfield  {author} {\bibinfo {author} {\bibfnamefont {Y.}~\bibnamefont
  {Ando}},\ }\href {\doibase 10.7566/JPSJ.82.102001} {\bibfield  {journal}
  {\bibinfo  {journal} {Journal of the Physical Society of Japan}\ }\textbf
  {\bibinfo {volume} {82}},\ \bibinfo {pages} {102001} (\bibinfo {year}
  {2013})}\BibitemShut {NoStop}%
\bibitem [{\citenamefont {Liu}\ \emph {et~al.}(2010)\citenamefont {Liu},
  \citenamefont {Zhang}, \citenamefont {Yan}, \citenamefont {Qi}, \citenamefont
  {Frauenheim}, \citenamefont {Dai}, \citenamefont {Fang},\ and\ \citenamefont
  {Zhang}}]{liu:10}%
  \BibitemOpen
  \bibfield  {author} {\bibinfo {author} {\bibfnamefont {C.-X.}\ \bibnamefont
  {Liu}}, \bibinfo {author} {\bibfnamefont {H.}~\bibnamefont {Zhang}}, \bibinfo
  {author} {\bibfnamefont {B.}~\bibnamefont {Yan}}, \bibinfo {author}
  {\bibfnamefont {X.-L.}\ \bibnamefont {Qi}}, \bibinfo {author} {\bibfnamefont
  {T.}~\bibnamefont {Frauenheim}}, \bibinfo {author} {\bibfnamefont
  {X.}~\bibnamefont {Dai}}, \bibinfo {author} {\bibfnamefont {Z.}~\bibnamefont
  {Fang}}, \ and\ \bibinfo {author} {\bibfnamefont {S.-C.}\ \bibnamefont
  {Zhang}},\ }\href {\doibase 10.1103/PhysRevB.81.041307} {\bibfield  {journal}
  {\bibinfo  {journal} {Phys. Rev. B}\ }\textbf {\bibinfo {volume} {81}},\
  \bibinfo {pages} {041307} (\bibinfo {year} {2010})}\BibitemShut {NoStop}%
\bibitem [{\citenamefont {Zhu}\ \emph {et~al.}(2013)\citenamefont {Zhu},
  \citenamefont {Veenstra}, \citenamefont {Levy}, \citenamefont {Ubaldini},
  \citenamefont {Syers}, \citenamefont {Butch}, \citenamefont {Paglione},
  \citenamefont {Haverkort}, \citenamefont {Elfimov},\ and\ \citenamefont
  {Damascelli}}]{zhu:13}%
  \BibitemOpen
  \bibfield  {author} {\bibinfo {author} {\bibfnamefont {Z.-H.}\ \bibnamefont
  {Zhu}}, \bibinfo {author} {\bibfnamefont {C.~N.}\ \bibnamefont {Veenstra}},
  \bibinfo {author} {\bibfnamefont {G.}~\bibnamefont {Levy}}, \bibinfo {author}
  {\bibfnamefont {A.}~\bibnamefont {Ubaldini}}, \bibinfo {author}
  {\bibfnamefont {P.}~\bibnamefont {Syers}}, \bibinfo {author} {\bibfnamefont
  {N.~P.}\ \bibnamefont {Butch}}, \bibinfo {author} {\bibfnamefont
  {J.}~\bibnamefont {Paglione}}, \bibinfo {author} {\bibfnamefont {M.~W.}\
  \bibnamefont {Haverkort}}, \bibinfo {author} {\bibfnamefont {I.~S.}\
  \bibnamefont {Elfimov}}, \ and\ \bibinfo {author} {\bibfnamefont
  {A.}~\bibnamefont {Damascelli}},\ }\href {\doibase
  10.1103/PhysRevLett.110.216401} {\bibfield  {journal} {\bibinfo  {journal}
  {Phys. Rev. Lett.}\ }\textbf {\bibinfo {volume} {110}},\ \bibinfo {pages}
  {216401} (\bibinfo {year} {2013})}\BibitemShut {NoStop}%
\bibitem [{\citenamefont {Xiao}\ \emph {et~al.}(2010)\citenamefont {Xiao},
  \citenamefont {Chang},\ and\ \citenamefont {Niu}}]{xiao:10}%
  \BibitemOpen
  \bibfield  {author} {\bibinfo {author} {\bibfnamefont {D.}~\bibnamefont
  {Xiao}}, \bibinfo {author} {\bibfnamefont {M.-C.}\ \bibnamefont {Chang}}, \
  and\ \bibinfo {author} {\bibfnamefont {Q.}~\bibnamefont {Niu}},\ }\href
  {\doibase 10.1103/RevModPhys.82.1959} {\bibfield  {journal} {\bibinfo
  {journal} {Rev. Mod. Phys.}\ }\textbf {\bibinfo {volume} {82}},\ \bibinfo
  {pages} {1959} (\bibinfo {year} {2010})}\BibitemShut {NoStop}%
\bibitem [{\citenamefont {Taskin}\ and\ \citenamefont
  {Ando}(2011)}]{taskin:11}%
  \BibitemOpen
  \bibfield  {author} {\bibinfo {author} {\bibfnamefont {A.~A.}\ \bibnamefont
  {Taskin}}\ and\ \bibinfo {author} {\bibfnamefont {Y.}~\bibnamefont {Ando}},\
  }\href {\doibase 10.1103/PhysRevB.84.035301} {\bibfield  {journal} {\bibinfo
  {journal} {Phys. Rev. B}\ }\textbf {\bibinfo {volume} {84}},\ \bibinfo
  {pages} {035301} (\bibinfo {year} {2011})}\BibitemShut {NoStop}%
\bibitem [{\citenamefont {Derry}\ \emph {et~al.}(2015)\citenamefont {Derry},
  \citenamefont {Mitchell},\ and\ \citenamefont {Logan}}]{derry:15}%
  \BibitemOpen
  \bibfield  {author} {\bibinfo {author} {\bibfnamefont {P.~G.}\ \bibnamefont
  {Derry}}, \bibinfo {author} {\bibfnamefont {A.~K.}\ \bibnamefont {Mitchell}},
  \ and\ \bibinfo {author} {\bibfnamefont {D.~E.}\ \bibnamefont {Logan}},\
  }\href@noop {} {\bibfield  {journal} {\bibinfo  {journal} {Phys. Rev. B}\
  }\textbf {\bibinfo {volume} {92}},\ \bibinfo {pages} {035126} (\bibinfo
  {year} {2015})}\BibitemShut {NoStop}%
\bibitem [{\citenamefont {Fu}\ \emph {et~al.}(2012)\citenamefont {Fu},
  \citenamefont {Zhang}, \citenamefont {Wang},\ and\ \citenamefont
  {Li}}]{fu:12}%
  \BibitemOpen
  \bibfield  {author} {\bibinfo {author} {\bibfnamefont {Z.-G.}\ \bibnamefont
  {Fu}}, \bibinfo {author} {\bibfnamefont {P.}~\bibnamefont {Zhang}}, \bibinfo
  {author} {\bibfnamefont {Z.}~\bibnamefont {Wang}}, \ and\ \bibinfo {author}
  {\bibfnamefont {S.-S.}\ \bibnamefont {Li}},\ }\href@noop {} {\bibfield
  {journal} {\bibinfo  {journal} {J. Phys. Condens. Matter}\ }\textbf {\bibinfo
  {volume} {24}},\ \bibinfo {pages} {145502} (\bibinfo {year}
  {2012})}\BibitemShut {NoStop}%
\bibitem [{\citenamefont {Akbari}\ \emph {et~al.}(2011)\citenamefont {Akbari},
  \citenamefont {Thalmeier},\ and\ \citenamefont {Eremin}}]{akbari:11}%
  \BibitemOpen
  \bibfield  {author} {\bibinfo {author} {\bibfnamefont {A.}~\bibnamefont
  {Akbari}}, \bibinfo {author} {\bibfnamefont {P.}~\bibnamefont {Thalmeier}}, \
  and\ \bibinfo {author} {\bibfnamefont {I.}~\bibnamefont {Eremin}},\ }\href
  {\doibase 10.1103/PhysRevB.84.134505} {\bibfield  {journal} {\bibinfo
  {journal} {Phys. Rev. B}\ }\textbf {\bibinfo {volume} {84}},\ \bibinfo
  {pages} {134505} (\bibinfo {year} {2011})}\BibitemShut {NoStop}%
\bibitem [{\citenamefont {Akbari}\ and\ \citenamefont
  {Thalmeier}(2013{\natexlab{a}})}]{akbari:13}%
  \BibitemOpen
  \bibfield  {author} {\bibinfo {author} {\bibfnamefont {A.}~\bibnamefont
  {Akbari}}\ and\ \bibinfo {author} {\bibfnamefont {P.}~\bibnamefont
  {Thalmeier}},\ }\href {\doibase 10.1209/0295-5075/102/57008} {\bibfield
  {journal} {\bibinfo  {journal} {{EPL} (Europhysics Letters)}\ }\textbf
  {\bibinfo {volume} {102}},\ \bibinfo {pages} {57008} (\bibinfo {year}
  {2013}{\natexlab{a}})}\BibitemShut {NoStop}%
\bibitem [{\citenamefont {Akbari}\ and\ \citenamefont
  {Thalmeier}(2013{\natexlab{b}})}]{akbari:13b}%
  \BibitemOpen
  \bibfield  {author} {\bibinfo {author} {\bibfnamefont {A.}~\bibnamefont
  {Akbari}}\ and\ \bibinfo {author} {\bibfnamefont {P.}~\bibnamefont
  {Thalmeier}},\ }\href {\doibase 10.1140/epjb/e2013-40859-6} {\bibfield
  {journal} {\bibinfo  {journal} {The European Physical Journal B}\ }\textbf
  {\bibinfo {volume} {86}},\ \bibinfo {pages} {495} (\bibinfo {year}
  {2013}{\natexlab{b}})}\BibitemShut {NoStop}%
\bibitem [{\citenamefont {Asmar}\ \emph {et~al.}(2017)\citenamefont {Asmar},
  \citenamefont {Sheehy},\ and\ \citenamefont {Vekhter}}]{asmar:17}%
  \BibitemOpen
  \bibfield  {author} {\bibinfo {author} {\bibfnamefont {M.~M.}\ \bibnamefont
  {Asmar}}, \bibinfo {author} {\bibfnamefont {D.~E.}\ \bibnamefont {Sheehy}}, \
  and\ \bibinfo {author} {\bibfnamefont {I.}~\bibnamefont {Vekhter}},\ }\href
  {\doibase 10.1103/PhysRevB.95.241115} {\bibfield  {journal} {\bibinfo
  {journal} {Phys. Rev. B}\ }\textbf {\bibinfo {volume} {95}},\ \bibinfo
  {pages} {241115} (\bibinfo {year} {2017})}\BibitemShut {NoStop}%
\bibitem [{\citenamefont {Wu}\ \emph {et~al.}()\citenamefont {Wu},
  \citenamefont {Li}, \citenamefont {Ma}, \citenamefont {Zhang}, \citenamefont
  {Liu}, \citenamefont {Zhou}, \citenamefont {Shao}, \citenamefont {Wang},
  \citenamefont {Hao}, \citenamefont {Feng}, \citenamefont {Schwier},
  \citenamefont {Kumar}, \citenamefont {Sun}, \citenamefont {Liu},
  \citenamefont {Shimada}, \citenamefont {Miyamoto}, \citenamefont {Okuda},
  \citenamefont {Wang}, \citenamefont {Xie}, \citenamefont {Chen},
  \citenamefont {Liu}, \citenamefont {Liu},\ and\ \citenamefont
  {Zhao}}]{wu:20}%
  \BibitemOpen
  \bibfield  {author} {\bibinfo {author} {\bibfnamefont {X.}~\bibnamefont
  {Wu}}, \bibinfo {author} {\bibfnamefont {J.}~\bibnamefont {Li}}, \bibinfo
  {author} {\bibfnamefont {X.-M.}\ \bibnamefont {Ma}}, \bibinfo {author}
  {\bibfnamefont {Y.}~\bibnamefont {Zhang}}, \bibinfo {author} {\bibfnamefont
  {Y.}~\bibnamefont {Liu}}, \bibinfo {author} {\bibfnamefont {C.-S.}\
  \bibnamefont {Zhou}}, \bibinfo {author} {\bibfnamefont {J.}~\bibnamefont
  {Shao}}, \bibinfo {author} {\bibfnamefont {Q.}~\bibnamefont {Wang}}, \bibinfo
  {author} {\bibfnamefont {Y.-J.}\ \bibnamefont {Hao}}, \bibinfo {author}
  {\bibfnamefont {Y.}~\bibnamefont {Feng}}, \bibinfo {author} {\bibfnamefont
  {E.~F.}\ \bibnamefont {Schwier}}, \bibinfo {author} {\bibfnamefont
  {S.}~\bibnamefont {Kumar}}, \bibinfo {author} {\bibfnamefont
  {H.}~\bibnamefont {Sun}}, \bibinfo {author} {\bibfnamefont {P.}~\bibnamefont
  {Liu}}, \bibinfo {author} {\bibfnamefont {K.}~\bibnamefont {Shimada}},
  \bibinfo {author} {\bibfnamefont {K.}~\bibnamefont {Miyamoto}}, \bibinfo
  {author} {\bibfnamefont {T.}~\bibnamefont {Okuda}}, \bibinfo {author}
  {\bibfnamefont {K.}~\bibnamefont {Wang}}, \bibinfo {author} {\bibfnamefont
  {M.}~\bibnamefont {Xie}}, \bibinfo {author} {\bibfnamefont {C.}~\bibnamefont
  {Chen}}, \bibinfo {author} {\bibfnamefont {Q.}~\bibnamefont {Liu}}, \bibinfo
  {author} {\bibfnamefont {C.}~\bibnamefont {Liu}}, \ and\ \bibinfo {author}
  {\bibfnamefont {Y.}~\bibnamefont {Zhao}},\ }\href@noop {} {\bibinfo
  {journal} {arXiv:2002.00320}\ }\BibitemShut {NoStop}%
\end{thebibliography}%

\end{document}